\documentclass[a4paper,11pt]{article}

\pdfoutput=1 % if your are submitting a pdflatex (i.e. if you have
\usepackage{cite}
\usepackage{graphicx}
\newcommand*{\email}[1]{\href{mailto:#1}{\nolinkurl{#1}} } 
\usepackage{hyperref}
\usepackage[affil-it]{authblk}
\usepackage{enumerate}

\usepackage{newtxtext}

\usepackage{slashed}
\usepackage{wrapfig}
\usepackage{textcomp}
\usepackage{amsmath,amssymb,fontenc, times}
\usepackage{float}
\usepackage{verbatim}

\usepackage{bm}
\usepackage{esvect}

\usepackage{diagbox}
\usepackage{adjustbox}
\usepackage{multirow}

\newcommand{\be}{\begin{equation}}
\newcommand{\ee}{\end{equation}}
\newcommand{\bea}{\begin{eqnarray}}
\newcommand{\eea}{\end{eqnarray}}
\newcommand{\bes}{\begin{subequations}}
\newcommand{\ees}{\end{subequations}}
\newcommand{\bear}{\begin{equation}\begin{array}}
\newcommand{\eear}[1]{\end{array}\label{#1}\end{equation}}

\newcommand{\beg}{\begin{equation}\begin{gathered}}
\newcommand{\eeg}{\end{gathered}\end{equation}}
\newcommand{\beal}{\begin{equation}\begin{aligned}}
\newcommand{\eeal}{\end{aligned}\end{equation}}
\newcommand{\begg}{\begin{gather*}}
\newcommand{\eegg}{\end{gather*}}
\newcommand{\vb}[1]{\vv{\bm{#1}}} 

\newcommand{\fr}[2]{\dfrac{{ #1}}{{ #2}}}

\renewcommand{\le}{\leqslant}

\newcommand{\epe}{\mbox{$e^+e^-\,$}}

{\end{list}}
\newcounter{enumct}

\newcommand{\MET}{\slash{\hspace{-2.4mm}E}_T}
\usepackage{color}    
\usepackage{lipsum}
\usepackage{soul}
\definecolor{darkgreen}{RGB}{11,150,35}

\usepackage{ulem}

\newcommand{\eps}{\varepsilon}

\begin{document}

\title{Decoding Dark Matter at future $e^+e^-$ colliders}

\author[a,b]{Alexander Belyaev\footnote{\email{ a.belyaev@soton.ac.uk}}}
\author[a]{Arran Freegard\footnote{\email{acf1g14@soton.ac.uk}}}
\author[c]{Ilya F. Ginzburg\footnote{\email{ginzburg@math.nsc.ru}}}
\author[a]{Daniel Locke\footnote{\email{D.R.Locke@soton.ac.uk}}}
\author[a,d]{Alexander Pukhov\footnote{\email{pukhov@lapp.in2p3.fr}}}

\affil[a]{School of Physics \& Astronomy, University of Southampton, Southampton SO17 1BJ, UK}
\affil[b]{Particle Physics Department, Rutherford Appleton Laboratory, Chilton, Didcot, Oxon OX11 0QX, UK}
\affil[c]{Sobolev Institute of Mathematics, Novosibirsk, 630090, Russia;\\
Novosibirsk State University, Novosibirsk, 630090, Russia}
\affil[d]{Skobeltsyn Inst. of Nuclear Physics, Moscow State Univ., Moscow 119992, Russia}

\maketitle

\date{\today}
%%%%%%%%%%%%%%%%%%%%%%%%%%%%%%%%%%%%%%%%%%%%%%%%%%%%%%%%%%%%%%%%%%%%%

\begin{abstract}
We explore the potential of the $\epe$ colliders
to discover dark matter and determine its properties 
such as mass and the spin.
For this purpose we study  spin zero and spin one-half cases of  dark matter, $D$  which belongs to $SU(2)$ weak doublet and therefore has the charged doublet partner, $D^+$. For the case of scalar dark matter we chose Inert Doublet Model, while for the case of fermion dark matter we suggest the new minimal fermion dark matter  model with only three parameters. 
We choose  two benchmarks for the models under study
which provide the correct amount of observed DM relic density
and consistent with  the current   DM searches.
We focus on the particular process 
$\epe\to  D^+ D^-  \to D D W^+ W^- \to DD(q \bar{q})(\mu^\pm\nu)$ 
at 500 GeV ILC collider which gives rise to the
``{di-jet +$\mu$ +  $\MET$}" signature  and study it at the level of  fast detector simulation, taking into account Bremsstrahlung and ISR effects.
We have found that two kinematical observables -- the energy of the muon, $E_\mu$,
and the angular distribution of $W$-boson, reconstructed from di-jet, $\cos\theta_{jj}$ are very powerful in determination of DM mass and spin, respectively.
In particular we have demonstrated  that in case of fermion DM,
the masses can be measured with a few percent accuracy already at 500 fb$^{-1}$ integrated luminosity. At the same time, the scalar DM model which has about an order of magnitude lower signal, requires   about factor of 40 higher  luminosity to reach the same accuracy in the mass measurement.
We have found that
one can  distinguish fermion and scalar DM scenarios with about 2 ab$^{-1}$ total integrated luminosity or less
without using the information on the  cross sections for benchmarks under study.
The methods of the determination of DM properties which we suggest here
are generic for the models where DM and its partner belong to the 
weak multiplet and can be applied to explore various DM models at future $\epe$ colliders. 
\end{abstract}

\newpage
\tableofcontents
\newpage

%%%%%%%%%%%%%%%%%%%%%%%
%\input{01-intro.tex}
\section{Introduction}
\label{secintro}
%The  search for  Dark Matter (DM) in High Energy Physics experiments (HEP) became one of the primary  goals of the LHC, future colliders as well as non-collider experiments.
The  search for  Dark Matter (DM) in High Energy Physics (HEP) experiments has become one of the primary  goals of the LHC and future colliders, in addition to non-collider experiments. Indeed it is one of the fundamental problems for the HEP community to discover and decode the nature of DM, the existence of which has been confirmed by several independent cosmological observations. 
These include galactic rotation curves, cosmic microwave background fits of the WMAP and PLANCK data, gravitational lensing, the large scale structure of the Universe, and interacting galaxy clusters such as the Bullet Cluster. Despite this multitude of observations strongly suggesting the existence of cold non-baryonic particle DM, many of its properties remain a mystery. 
We do not know what is  the  spin and the mass of DM,
whether it is  involved in non-gravitational interactions, what  symmetry stabilises it and what is the  nature of mediators between the Standard Model (SM) and DM, as well as whether  there is any partner of  DM  particles in the dark sector.

As one of the most active research areas 
in HEP, there are many key papers exploring the vast model landscape of DM and the possibilities to disentangle these models experimentally. DM models under study include SUSY \cite{Jungman:1995df,Ellwanger:2009dp, Giudice:2004tc}, sterile neutrinos \cite{Dodelson:1993je}, general minimal WIMP models \cite{Cirelli:2005uq}, Axions \cite{Kim:2008hd}, Kaluza-Klein DM \cite{Cheng:2002ej}, Universal Extra Dimensions \cite{Hooper:2007qk} and extended Higgs sectors \cite{Boehm:2003hm,Branco:2011iw, Belyaev:2016lok}. Determination of DM properties such as 
spin \cite{Christensen:2013sea, Belyaev:2016pxe, Asano:2011aj} and mass \cite{Christensen:2014yya, Burns:2008va} would be key in the event of a discovery.

%Experiment: Collider \cite{DM LHC REVIEW}, other \cite{}.
 
%
Traditional searches for DM at the LHC via missing energy signatures through mono-jet \cite{Sirunyan:2017jix, Aaboud:2017phn}, mono-$Z$($W$) \cite{Sirunyan:2017hci, Sirunyan:2017onm, Aaboud:2018xdl, Basalaev:2017cpw, Aaboud:2017bja, Aaboud:2017dor}, mono-Higgs \cite{Sirunyan:2018fpy, Sirunyan:2017hnk, ATLAS:2018bvd, Aaboud:2017uak, Aad:2015yga}, DM+ top quarks \cite{Sirunyan:2018dub, Sirunyan:2018gka, Aaboud:2017rzf} and invisible Higgs decays \cite{Sirunyan:2018owy, Aaboud:2019rtt, Aaboud:2017bja} or through potential mediators \cite{Sirunyan:2018xlo, Sirunyan:2018wcm, Sirunyan:2016iap} expand on constraints from LEP on DM charged partner masses \cite{Ellis:1997wva}. DM may also be probed in scenarios where its charged partners are long-lived, providing unique signatures with little background. These scenarios, known as non-prompt searches, include disappearing charged tracks \cite{Aaboud:2017mpt,Sirunyan:2018ldc} and displaced vertices \cite{Aaboud:2017iio,Sirunyan:2018pwn}. 

In  case of generic scenarios where DM is involved in  $SU(2)$ electroweak (EW) interactions and no additional Beyond-the-Standard-Model (BSM) mediators are present, it is quite challenging for the LHC to probe such a DM even in the 100 GeV range. For example,  in~\cite{Barducci:2015ffa} it was shown that even at High Luminosity (HL) LHC, the higgsino-like neutralino DM from the Minimal Supersymmetric Standard Model (MSSM) can be probed only up to about 200 GeV mass with very high transverse momentum mono-jet signature requited to reduce large SM background. There are no current constraints on such a scenario from the LHC and the best limits are set up by LEP on charged DM partner (e.g chargino) mass to be above 100 GeV~\cite{Barducci:2015ffa}.

The most recent global scans of the MSSM \cite{Athron:2017yua} including the neutralino and chargino sector (analogous to split SUSY scenarios) \cite{Athron:2018vxy} reveal a best fit region consistent with a higgsino-bino DM candidate, which is mostly bino (mostly singlet DM will avoid direct detection constraints). These viable points rely on the so-called ``Higgs funnel" annihilation channel (DM mass around half Higgs mass) in order to reproduce the relic density as measured by PLANCK. Such scenarios may be ideally probed by a $\epe$ 500 GeV collider such as the ILC, where the accessible particle spectrum ($<300$ GeV) contains the two lightest neutralinos in addition to the lightest chargino.

In this study we explore the potential of a future $e^+e^-$ collider to probe and distinguish two well motivated minimal  models with DM of spin-0 and 1/2 embedded into an $SU(2)$ weak doublet with  no additional BSM mediator. 
We assume that the DM sector is represented by an EW doublet without loss of generality and that each model includes  DM, $D$,  its charged partner $D^+$ and heavier neutral partner(s), $D_{2}$ with $M_{D_2}> M_D$,   which are odd-particles with respect tothe  $Z_2$ symmetry responsible for  DM stability. We explore the potential of the 500 GeV $e^+e^-$ collider (which can be ILC, FCC-ee, CLIC etc.) to measure $M_D$ and $M_{+}$ masses and distinguish DM spin using semi-leptonic signature from $D^+D^-$ decays. 
The observation of such signature requires a non-vanishing  mass gap $\Delta M=M_{+}-M_D$ ($\gtrsim 10$~GeV) which would provide detectable leptons. 

There are several important advantages of the $\epe$ colliders in comparison to the LHC which motivates our study, including:
a) the background  cross section at $\epe$
collider is lower and the respective signal to background ratio is at least one order of magnitude higher than the one at the LHC;
b) at $\epe$ colliders one can determine not only the missing transverse momentum, but also missing mass, which allows further suppress background without reducing the signal;
c) since the centre-of-mass energy is fixed at $\epe$ colliders, they allow to reconstruct  kinematics of various particles and their distributions in the lab frame, including characteristics of 
$W^\pm$-bosons from $D^\pm$ decays which 
is crucial for determination of DM properties as we demonstrate in our paper.

Our study goes beyond the previous exploration(e.g. \cite{Asano:2011aj} or \cite{Ginzburg:2014ora}) of the  ILC potential to discriminate DM models in several principal aspects:
a)
we explore models with DM of two different spins  and for the first time demonstrate that $W$-boson angular distribution allows to determine the spin of DM even without using the information about the signal cross section;
b) we explore the signature with leptonic final state which has the advantage of keeping background under better  control and more precise determination of the final state energy distributions;
c) we make use of the predicted cross-sections for typical parameter points delivering the correct relic abundance; 
d) we suggest the set of new kinematical observables and cuts which  boost  $e^+e^-$ collider potential  discrimination  of DM models; 
e) we explore both cases for off-shell and on-shell $W$-boson decay; 
f) we use model-independent template based approach to fit kinematic endpoints and determine $D$ and $D^+$ masses using likelihood methods.
In addition  for the case of fermion DM we suggest the new minimal fermion dark matter (MFDM) model with only three parameters.

This paper is organized as following. In Section~2 we discuss models, benchmarks
and analysis setup, in Section~3 we study the signal properties, in Section~4 we perform signal versus  background analysis and find  the potential of  $e^+e^-$ collider to determine DM properties, such as mass and spin. 
Finally in Section~5 we draw our conclusions.

\section{Models and  Benchmarks}
\label{sec:models}
%%%%%%%%%%%%%%%%%%%%%%%%%%%%%%%%%%%%%%%%
%Arran is editing
\subsection{Inert doublet model (I2HDM)}\label{secinert}

The spin-0 or scalar DM (SDM) model which we use as a first case study is  the inert Two Higgs Doublet Model \cite{inert-1,inert-2,inert-3,inert-4,Lundstrom:2008ai-1,Lundstrom:2008ai-2,Lundstrom:2008ai-3} which in addition to SM Higgs doublet contains inert scalar $Z_2$-odd doublet, $\phi_D$, that does not acquire a Vacuum Expectation Value (VEV).  In our paper we call all particles odd under $Z_2$ symmetry $D$-particles, and refer to the $Z_2$ symmetry as $D$-parity.
The scalar sector of the model  is given by
  \begin{equation}
  \mathcal{L} =
  |D_{\mu}\Phi|^2 + |D_{\mu}\phi_D|^2 -V(\Phi,\phi_D)
  \textrm{,}
  \label{eq:lagrangian}
  \end{equation}
where   $V$ is the potential with all scalar interactions compatible with the ${Z}_2$ symmetry:
\begin{eqnarray}
  V &=& -m_1^2 (\Phi^{\dagger}\Phi) - m_2^2 (\phi_D^{\dagger}\phi_D) + \lambda_1 (\Phi^{\dagger}\Phi)^2 + \lambda_2 (\phi_D^{\dagger}\phi_D)^2    \nonumber  \\
  &+&  \lambda_3(\Phi^{\dagger}\Phi)(\phi_D^{\dagger}\phi_D)
  + \lambda_4(\phi_D^{\dagger}\Phi)(\Phi^{\dagger}\phi_D)
  + \frac{\lambda_5}{2}\left[(\Phi^{\dagger}\phi_D)^2 + (\phi_D^{\dagger}\Phi)^2 \right]\,.\label{eq:potential}
\end{eqnarray}
In the unitary gauge, the SM doublet, $\Phi$ and the inert doublet, $\phi_D$  take the form
\begin{equation}
\Phi=\frac{1}{\sqrt{2}}
\begin{pmatrix}
0\\
v+H
\end{pmatrix},
  \qquad
  \phi_D= \frac{1}{\sqrt{2}}
\begin{pmatrix}
 \sqrt{2}{D^+} \\
 D + iD_2
\end{pmatrix},
\end{equation}
where we consider the parameter space in which only the first, SM-like doublet,  acquires a VEV, $v$.
After EW Symmetry Breaking (EWSB), the $D$-parity is still conserved by the vacuum state, which forbids direct coupling of any single inert field to the SM fields
and protects the lightest inert boson from decaying, hence providing the DM candidate in this scenario.
In addition to the SM-like scalar $H$, the model contains one inert charged $D^+$ and two further inert neutral $D$ and $D_2$ scalars. The two neutral scalars of the I2HDM have opposite $CP$-parities, but it is impossible  to unambiguously determine which of them is $CP$-even and which one  is $CP$-odd since
the model has two $CP$-symmetries, $D \to D, D_2 \to -D_2$ and $D \to -D, D_2 \to D_2$, which get interchanged upon a change of basis $\phi_D \to i \phi_D$. This makes the specification of the $CP$-properties of $D$ and $D_2$ a basis-dependent statement.
Therefore, following Ref.~\cite{Belyaev:2016lok},  we denote the two neutral inert scalar masses as $M_{D} < M_{D_2}$, without specifying which is scalar or pseudoscalar, so that $D$ is the DM candidate.

The model can be conveniently described by a five dimensional parameter space\cite{Belyaev:2016lok}
using the following   phenomenologically relevant variables:
\begin{equation}
\label{eq:model-parameters}
M_{D}\,,\quad M_{D_2} > M_{D}\,,\quad M_{+} > M_{D}\,, \quad \lambda_2 > 0\,,\quad \lambda_{345} > -2\sqrt{\lambda_1\lambda_2},
\end{equation}
where $M_{D},M_{D_2}$ and $M_{+} $ are the masses of the two neutral and  charged inert scalars, respectively, whereas   $\lambda_{345}=\lambda_3+\lambda_4+\lambda_5$
is the coupling which governs the Higgs-DM interaction vertex $H D D$.
There is ($\phi_D \to i\phi_D$, $\lambda_5\to-\lambda_5$)  symmetry of the Lagrangian which allows us to chose $\lambda_5 >0$ as a conversion.
The masses of the three inert  scalars are expressed in terms of the parameters of the Lagrangian in Eqs.~(\ref{eq:lagrangian})--(\ref{eq:potential}) as follows:
\begin{eqnarray}
\label{masses}
\begin{array}{lcl}
M_{D}^2 &=& \frac{1}{2}(\lambda_3 +\lambda_4 - \lambda_5) v^2 - m_2^2, \\[2pt]
M_{D_2}^2 &=& \frac{1}{2}(\lambda_3 +\lambda_4 + \lambda_5) v^2 - m_2^2 \ >M_{D}^2, \\[2pt]
M_{+}^2 &=& \frac{1}{2} \lambda_3 v^2 - m_2^2, 
\end{array}
\end{eqnarray}
which represent the only three parameters relevant to our study,
since we explore production of $D$-particles in the gauge interactions at 
$e^+e^-$ colliders.

Constraints on the Higgs potential from requiring vacuum stability and a global minimum take the following  form\cite{Belyaev:2016lok}:
\begin{eqnarray}
\left\{
\begin{array}{lcr}
M_{D}^2 > 0 \text{ (the trivial one)} & \text{for} & |R|<1, \label{eq:scalar-pot1}\\
M_{D}^2 > (\lambda_{345}/2\sqrt{\lambda_1\lambda_2}-1) \sqrt{\lambda_1\lambda_2} v^2 = (R-1) \sqrt{\lambda_1\lambda_2} v^2 & \text{for} & R>1,~\label{eq:scalar-pot2}
\end{array}
\right.
\end{eqnarray} 
where $R=\lambda_{345}/2\sqrt{\lambda_1\lambda_2}$ and  $\lambda_1 \approx 0.129$ is fixed as in the SM by the Higgs mass in Eq. (\ref{masses}).
The latter condition places an important upper bound on $\lambda_{345}$ for a given DM mass $M_{D}$.
Constraints  on the model's parameter space have already been comprehensively explored in a large variety of previous papers~\cite{Ma:2006km,Barbieri:2006dq,LopezHonorez:2006gr,Arina:2009um,Nezri:2009jd,Miao:2010rg,Gustafsson:2012aj,Arhrib:2012ia,Swiezewska:2012eh,Goudelis:2013uca,Arhrib:2013ela,Krawczyk:2013jta,Krawczyk:2013pea,Ilnicka:2015jba,Diaz:2015pyv,Modak:2015uda,Queiroz:2015utg,Garcia-Cely:2015khw,Hashemi:2016wup,Poulose:2016lvz,Alves:2016bib,Datta:2016nfz,inert-1,inert-2,inert-3,inert-4,Belyaev:2016lok}.

%%%%%%%%%%%%%%%%%%%%%
%%%%%%%%%%%%%%%%%%%%%
\subsection{Minimal Fermion DM (MFDM)}\label{secFDM}

The second model we consider here is a minimal model with an EW fermion DM doublet.
The model should respect direct DM constraints coming from the most restrictive DM Direct Dtection (DD) searches from the XENON1T experiment~\cite{Aprile:2018dbl}, and at the same time provide the correct amount (or at least not an over abundance) of relic density.
Therefore the model  must have a mechanism to suppress DM scattering through intermediate Z-bosons and/or Higgs bosons. Among  several candidates for such a mechanism, the most minimal is to introduce Majorana neutral D-odd particles $\chi^0_1$ and $\chi^0_2$ as a part of an EW doublet and split their masses via interactions with the SM Higgs doublet and additional Majorana singlet fermion $\chi^0_s$:
\begin{equation}
\mathcal{L}_{FDM} = \mathcal{L}_{SM} + 
\bar{\psi}(i\slashed{D}- m_\psi)\psi  
+\frac{1}{2}\bar{\chi^0_s}(i\slashed{\partial}- {m_s})\chi^0_s 
-( Y (\bar{\psi}\Phi\chi^0_s)+ h.c.) \ \ ,
\label{eq:MFDM}
\end{equation}

% for details on majorana spinors see e.g Schwartz pg 192

where fermion fields are in bispinor form and $\Phi$ is the SM Higgs doublet. A DM $SU(2)$ vector-like doublet with hypercharge $Y=1/2$, may be defined in terms of majorana states $\chi^0$,  ${\chi'^0}$ as:
\begin{equation}
\psi = \begin{pmatrix} \chi^+ \cr \tfrac{1}{\sqrt{2}}\left( \chi^0+i\chi'^0 \right) \,\end{pmatrix}.
\end{equation}

The model which we suggest and use in our paper has only three new parameters: $m_\psi, Y$ and  $m_s$. This model is more minimal in comparison to the previously studied doublet-singlet model~\cite{Enberg:2007rp,Cohen:2011ec} which has four parameters because of $two$ Yukawa couplings, distinguishing left- and right-handed interactions of Higgs and DM doublets with  $\chi^0_s$. Our choice of the  parity conserving $\psi-\Phi-\chi^0_s$ Yukawa interactions adds just one parameter to the model -- the Yukawa coupling, which is the same for left and right interactions. We have checked that this scenario is radiatively stable. This parity would be spoiled if the DM sector would directly couple to SM fermions, which is eventually not the case. Therefore  our model with just three new parameters is consistent and truly the minimal one.

The Yukawa interaction mixes  $\chi^0$ and $\chi^0_s$ while $\chi^+$ and $\chi'^0$ have the same  mass $m_\psi$ and remain degenerate at tree-level.
This degeneracy is not essential, since $\chi'^0$ decay is driven by the $\chi'^0 \to \chi^0 Z^{(*)}$ process.

We trade  $m_\psi, Y$ and  $m_s$ parameters  for
three physical masses:
\begin{equation}
M_{D}, m_\psi\equiv M_{+}=M_{D'}, \ \ \mbox{and} \ \ M_{D_2},
\end{equation}
corresponding to $(D, D_2, D')$ mass bases of the neutral DM sector, 
which one obtains from  the diagonalization of the  mass matrix in the $(\chi^0,\chi_s,  \chi'^0)$ basis:	
\begin{equation}
\mathcal{M} = 
\begin{bmatrix}
	{m_\psi} & {Yv}& 0 \\[3pt]
	{Yv}& {m_s} & 0 \\[3pt]
	0 & 0 & {m_\psi}
\end{bmatrix} \ \ .
\label{eq:fdm-mass-matrix}
\end{equation}

Since  the mass and gauge  eigenstate of $\chi'^0$ coincide and  have  the mass $m_\psi \equiv M_{+} = m_{D_2}$,
diagonalization should be done only for 
$2\times 2$  upper-left block of the matrix~(\ref{eq:fdm-mass-matrix}).
This diagonalization describes the
rotation of the gauge eigenstates $(\chi^0,\chi^0_s)$  into the  mass  eigenstates $(D,D_2)$ by angle $\theta$ and given  by:
\begin{eqnarray}
\chi^0&=&D \cos\theta - D_2 \sin\theta\nonumber\\
\chi^0_s&=&D \sin\theta + D_2 \cos\theta . 
\label{eq:fdm-mix}
\end{eqnarray}
One can find the following useful relations
from this diagonalization:
\begin{eqnarray}
	m_s &=& M_{D} + M_{D_2} - M_{+} \\
	\tan{2\theta} & = &
	\frac{2Y v}{m_\psi-m_s}
	=
	\frac{2Y v}{2M_+-M_{D_2}-M_D} \\
	\sin{2\theta} & = &
	-\frac{2Y v}{M_{D_2}-M_D}\label{eq:sin2t} \ .
\end{eqnarray}
The relation between Yukawa coupling, $Y$ and physical mass parameters is given by:
\begin{equation}
Y = \pm \frac{\sqrt{(M_{D_2}-M_{+})(M_{+}-M_{D})}}{v} \ .\label{eq:Y}
\end{equation}
The  mass order
\begin{equation}
	M_{D_2}>M_{+}=M_{D'}>M_{D}
\end{equation}
follows from the condition for $Y$ to be real.
The phase of $\chi_s^0$ may be chosen such that $Y$ is positive. 
This MFDM model, with singlet-doublet dark sector content can be mapped into the bino-higgsino MSSM model with all other SUSY particles decoupled, including winos.

In this  model, DM interaction with $Z$-boson is absent  at tree-level.
One should also note that inelastic up-scatterings arising from the $ZDD'$ vertex do not take place for the mass split between $D$ and $D'$  above the recoil energy threshold of direct detection experiments (typically around 1-10 keV) --  the case of our  study.
The spin-independent DM -- nucleon scattering cross-section therefore depends only on the Higgs coupling to DM, $g_{DDh}$,
given by
 \begin{equation}
 	g_{DDh} = -2Y \cos\theta \sin\theta = 2\frac{(M_+-M_D)(M_{D_2}-M_+)}{v(M_{D_2}-M_D)} \, \label{eq:gddh}
 \end{equation}
which follows from Eq.\eqref{eq:sin2t} and~\eqref{eq:Y}.
This coupling can  be small, which provides viable parameter space from DM direct detection point of view. 
In this study we consider scenarios with relatively large $M_+-M_D$ mass gap -- of the order of $W$-boson mass, so
$D^+$ decays to  on-shell (or slightly virtual) $W$ bosons. In this scenario, the 	small value of $g_{DDh}$ coupling is driven by the small  $M_{D_2}-M_+$ value
as one can see from Eq.~\eqref{eq:gddh}. At the same time the correct relic abundance can be provided through the effective  resonant annihilation of DM through the Higgs boson, when   $M_D \simeq M_h/2$ even if  
the value of $g_{DDh}$ is small.
One should note that in this case  $D-D_2$ and $D-D^+$ co-annihilation channels are sub-dominant  due to the relatively large mass split between DM and its partners.
{The implementation of this model in CalcHEP~\cite{CALCHEP} format using LanHEP package~\cite{Semenov:2014rea} is publicly available in High Energy Model Physics Data Base (HEPMDB)~\cite{HEPMDB}
	under this link
	\url{https://hepmdb.soton.ac.uk/hepmdb:0420.0327}.}

\subsection{Benchmark Points}\label{sec:BPs}

In our study we chose two  benchmarks  with different   $D^+ - D$ mass gaps: one -- providing $D^+ \to DW$ decay with
on-mass-shell $W$-boson in the final state and another -- with the  off-mass-shell $W^*$-boson.
We choose model parameters providing the right amount of relic density and satisfying the latest DM DD constraint from XENON1T searches to make sure that chosen benchmarks are  the realistic ones.
The benchmarks are presented in Table~\ref{benchmarks}
together with DM observables, where
abbreviations $SDM$ and $FDM$ denote scalar and fermion DM respectively.
One should note that I2HDM model
	has two more parameters (five versus three) in comparison to MFDM model --
	$\lambda_{345}$ and $\lambda_2$.
First we chose  $M_D,M_D^+,M_{D_2}$
to make the relic density consistent with the results from PLANCK for MFDM model, then use additional parameter 
$\lambda_{345}$ from I2HDM 
to make the relic density from this model to be consistent with PLANCK.
The other parameter -- $\lambda_2$ , which controls the self-interaction of DM, is not relevant to collider phenomenology. We keep 
$\lambda_2=1$ without loss of generality\footnote{The large value of $\lambda_2$ could potentially affect the DM density profile and loop-induced DM annhillation into SM particles. These effects are outside the scope of this paper.}  since it does not affect  any conclusion in this paper.
We chose the same $D$, $D^+$ and $D_2$ masses  for  both models with the aim to explore the ILC potential in distinguishing theories with same mass but different spin of the DM sector.

The relic density, $\Omega h^2$, and spin-independent proton scattering cross-section,  $\sigma_{SI}^p$, were calculated using the micrOMEGAs package \cite{micromegas}.  
\begin{table}[h]
\centering
\label{benchmarks}
\begin{tabular}{|l|l|l|l|}
\hline
\hline
\multicolumn{2}{|l|}{\diagbox[]{Parameters}{Benchmarks}}& BP1 & BP2\\
\hline\hline
\multicolumn{2}{|c|}{$M_{D}$}   & 60  & 60  \\
\multicolumn{2}{|c|}{$M_{+}$} & 160  & 120  \\
\multicolumn{2}{|c|}{$M_{D_2}$} & 160.85  & 120.85  \\
\hline
\hline
\multicolumn{2}{|l|}{I2HDM parameters}& & \\
\hline
\multicolumn{2}{|c|}{$\lambda_{345}$} & $6.5\times 10^{-4}$ & $7.0\times 10^{-4}$ \\
\multicolumn{2}{|c|}{$\lambda_2$} & 1.0 & 1.0 \\
\hline
\hline
\multicolumn{2}{|l|}{DM observables}& & \\
\hline
\multirow{2}{*}{$\Omega h^2$}&
	\multicolumn{1}{|l|}{$SDM$}
	& 0.111 & 0.112\\
	\cline{2-4}
	&\multicolumn{1}{|l|}{$FDM$} 
	& 0.108 & 0.109 \\
\hline
\multirow{2}{*}{$\sigma^p_{SI}$[pb]}&
	\multicolumn{1}{|l|}{$SDM$}
	& $6.17\times 10^{-13}$ & $6.17\times 10^{-13}$\\
	\cline{2-4}
	&\multicolumn{1}{|l|}{$FDM$} 
	& $1.67\times 10^{-11}$& $1.65\times 10^{-11}$ \\
\hline
\hline
\end{tabular}
\caption{Benchmark points for I2HDM and MFDM with DM observables. All masses are given in GeV. Abbreviations $SDM$ and $FDM$ denote scalar and fermion DM respectively.
\label{tab:BP}}
\end{table}

We have recast an existing SUSY analysis using CheckMATE2 \cite{checkmate} and found that the most sensitive search for BP1 and BP2 is the CMS 13TeV search for electroweak production of charginos and neutralinos in multilepton final states~\cite{CMS:2017fdz}. {CheckMATE  evaluates the $r$-number defined as a ratio of the signal cross section and the cross section excluded at 95\%CL, so $r=1$ means that the model is excluded at 95\%CL.
The respective values for BP1 and BP2 are found to be 0.325 and 0.664 respectively, which means that these points are still allowed by the LHC data.}\footnote{{The latest CheckMATE2 sensitivity could be bit higher since the time our analysis have been done.}}
Many of the most stringent LHC constraints on electroweak scale WIMP masses arise from decays mediated by sleptons and sneutrinos of mass $\lessapprox 500$ GeV. In the scenarios explored here we  assume that  all additional SUSY particles (or analogous  particles which could appear in the I2HDM extension) are decoupled. It is worth noting that the global scan of electroweakino DM by the GAMBIT collaboration shows favoured parameter points around our benchmarks \cite{Athron:2018vxy}.

For the models under study one should also respect the  constraints from the  electroweak precision test (EWPT).  EWPT quantities include
 $S$, $T$, and $U$  observables
  that parametrise contributions from beyond standard model physics to electroweak radiative corrections~\cite{Peskin:1991sw}.
For I2HDM the expressions for  $S$ and
$T$ parameters are evaluated in~\cite{inert-2,Arhrib:2012ia} and give
\begin{equation}
	S=-0.016(-0.013), \ \ \ T=-0.00146(-0.00090)
\end{equation}
  values
 for BP1(BP2) respectively.
 The contribution to $U$ parameter for 
 I2HDM can be neglected.
 These  $S$ and $T$ values for our benchmarks are allowed by the current 
 EWPT  fits 
 which (with $U$ fixed to be zero), have the following  central values
 (for SM Higgs
 boson mass 125 GeV)\cite{Workman:2022}:
\begin{equation}
	S=-0.01\pm 0.07, \ \ \ 
	 T = 0.04 \pm 0.06
\end{equation}
 with correlation coefficient $+0.91$.
 For MFDM we have derived the
 expressions  for $S$ and $T$  and present them in the Appendix~\ref{sec:STU}.
For MFDM the values of  $S$
 parameter are
\begin{equation}
	{S=-1.06 \times 10^{-4}(-8.38 \times 10^{-5})}
\end{equation}
for BP1(BP2) respectively,
while $T$ and $U$ parameters are explicitly zero.
This happens because one of the
down parts of the vector-like doublet, corresponding to the 
neutral Majorana  fermion, does not mix and has the same mass as the charged  fermion. 
For details we refer reader to  Appendix~\ref{sec:STU}.
To conclude,  the benchmarks for both models are consistent with the EWPT,
which is expected since the mass split between vector-like fermions is not large and the $S$ an $T$ parameters are  proportional to the mass split squared (in case if they are not explicitly zero like $T$ parameter in MFDM). 
We would like also to note that contrary to the case of  doublet of new chiral fermions, the 
$T$-parameter for which is known for more than 40 years
~\cite{Veltman:1977kh,Chanowitz:1978uj}, the $T$-parameter for  vector-like fermions is model-dependent. Therefore, we would like to stress that the general statement in the  Review of Particle Physics (PDG)\cite{Workman:2022} that vector-like fermion
doublets contribute to $T$ with extra factor of two is not correct in general.

%%%%%%%%%%%%%%%%%%%%%%%%%%%%%%%%%%%%%%%%%%%%%%%%%%%%
\subsection{Analysis setup}\label{sec:setup}

%%%%%%%%%%%%%%%%%%%%%%%%%%%%%%%%%%%%%%%%%%%%%%%%%%%%
In our study we use the following tools to  evaluate the ILC potential to probe properties of DM. We use CalcHEP\cite{CALCHEP} to perform the parton-level signal analysis in section~\ref{sec:signal}, including the study of the finite  width effects from $W$-boson decay, and the effects  from  the  Initial State Radiation (ISR) and Beamstrahlung 
Radiation (B).\footnote{In our setup we have used the ISR scale equal to the $\sqrt{s}$, $\sigma_x+\sigma_y=500$~nm for bunch $x+y$ size,  0.3mm for bunch length and $2\times 10^{10}$ particles in the bunch, corresponding to the standard setup for ILC simulation~\cite{ILD}.}
We use Pythia8 \cite{Sjostrand:2014zea} to simulate  final state radiation and hadronisation effects. Events from Pythia are then passed to Delphes \cite{deFavereau:2013fsa} fast-detector simulator using the ILC card based on the proposed ILD detector \cite{ILD}. 
We have used  the anti-$K_T$ jet clustering algorithm~\cite{Cacciari:2008gp} with the value of radius parameter, R set to 0.8 (instead of its default 0.5) which allows to add the additional soft jets  and  improve dijet reconstruction. 
At this level of simulation we have performed signal and background analysis of the various kinematic distributions to extract $D$ and $D^+$ masses
 to distinguish scalar and fermion DM models
as discussed in detail in  \ref{secmain}.

%%%%%%%%%%%%%%%%%%%%%%%%%

%\input{03-signal.tex}
%%%%%%%%%%%%%%%%%%%%%%%%%%%%%
\section{The properties of DM signal at $\epe$ collider}
\label{sec:signal}
%%%%%%%%%%%%%%%%%%%%%%%%%%%%
%%%%%%%%%%%%%%%%%%%%%%%%%%
In this section, we discuss and develop the generic strategies which we 
suggest to employ in order to discern the masses and spin of DM in the classes of models discussed in the previous section. In order to evaluate the applicability of such strategies we examine the cross-section of the relevant production channels and the dominant background processes.
Besides, we also discuss various kinematical observables and the respective distributions which will be later used 
to discriminate DM properties and optimise signal versus background.

%%%%%%%%%%%%%%%%%%%%%%%%%%%%
\subsection{The signatures}\label{sec:sign}

The
$e^+e^-	\to D^+ D^-$ process, followed by 
subsequent $D^\pm$ decay to $D$ and $W$ and then by $W$ decay to leptons or quarks lead to the following signatures:
\begin{itemize}
	\item {\bf Two di-jets
	+ $\MET$}
 from 

\be 
\epe \to D^+D^- \to  DDW^+W^- \to DD(W^+\to q\bar{q}')
(W^-\to q\bar{q}') \label{eq:sign-WW-jjjj}
\ee

process. This signature 
has large missing transverse mass, $\MET$, and large
$M_{miss}$ while
each di-jet cluster has energy 
$<\frac{\sqrt{s}}{2}$.
\item
 {\bf Di-jet $+$ ($ e$ or $\mu$)
+  $\MET$},
with energy of each di-jet or lepton \
$<\frac{\sqrt{s}}{2}$, with large  \ $\MET$ and large $M_{miss}$, which originates from
the
\be
\epe
\to D^+D^- \to 
DDW^+W^-
\to DD(W\to \ell\nu)(W\to q\bar{q})
\label{eq:sign-WW-enjj}
\ee
process.
\end{itemize}
Here we consider  ${M_+}<M_{D_2}$ case only, for which the 
branching fraction of $D^\pm \to D W^\pm$ decay is 100\%.
At $M_W^*>5$~GeV, the branching ratios for different
channels of $W^*$ decay are roughly identical to those for on-shell $W$
\cite{PDG}. In particular, the fraction
of events with signature \eqref{eq:sign-WW-jjjj} \ is
$0.676^2\approx 0.45$. The fraction of events with signature \eqref{eq:sign-WW-enjj} is $2\cdot 0.676\cdot (2+0.17)\cdot
0.108\approx 0.32$ (here 0.17 is a fraction of $\mu$ or $e$ from
$\tau$ decay). At $M_W^*<5$~GeV the branchings 
$BR(e\nu)$ and $BR(\mu\nu) $ increase, while the di-jet becomes
a set of a few hadrons.

\vskip 0.5cm

One should also mention, that in  case when ${M_+}>M_{D_2}$ (which we do not study here),
 $D_2$ contributes to
 $D^\pm$ decay such as
 \be
 D^\pm \to D_2 W^\pm \to DZ W^\pm
 \ee
 which  gives rise to the additional signatures such as 
 \begin{itemize}
 	\item 
{\bf from (4 di-jets, 0 charged leptons) to (1 di-jet, 5 charged leptons)	},
 \end{itemize}
originating from 
\be
  e^+e^-\to D^+D^-\to DW^+D_2W^-\to DD W^+W^-Z
  \label{eq:sign-WWZ}
\ee
and
\be
   e^+e^-\to D^+D^-\to D_2W^+D_2W^-\to DD W^+W^-ZZ
   \label{eq:sign-WWZZ}
\ee
cascade processes.
Note that the processes with invisible decay $Z\to \nu\bar{\nu}$ (with branching fraction $BR=20\%$) have the same signature as processes \eqref{eq:sign-WW-jjjj} and \eqref{eq:sign-WW-enjj}.
{There is also an additional process leading to $D$-odd particles + leptons in the final state, $e^+e^- \to DD_2 \to DDZ$ which we discuss in  appendix~\ref{secDDA} in details.}

\vskip 1cm

In our study we denote  electron beam energy as
\be
E=\sqrt{s}/2 \ \ . \label{beamE}
\ee

We consider energies $E$ and three-momenta, $\vb{p}$,  of  particles
in different reference systems and  use particle name  in  superscript ($W$ for $W^+$ and $D$ for $D^+$) for energy and  momentum to indicate  their Lorentz frame.  For the lab system (cms for \epe) corresponding quantities are written without superscript. The subscript indicates just the name of the particle to which the physical quantity belongs.
For example, $\vb{p}_W^{D^+}$ is value of three momentum of $W^+$ in the rest frame of $D^+$, and $E_\mu$ is energy of muon in the lab. system.

We supply upper superscript by additional sign $max$, $(+)$, $(-)$  to characterise  the values  for the corresponding energy
which we define in the text.
The energy can depend on some parameters, e.g. 
mass of a virtual $W$-boson ($M^*_W$),
which we indicate in brackets.
For  example, $E^{max}_\mu(M_W^*)$
means the maximum energy of the muon from the $W^*$
with the  $M_W^*$ invariant mass.

%%%%%%%%%%%%%%%%%%%%%%%%%
\subsection{Kinematial observables}\label{sec:observables}
We study the signal from DM originating from
$e^+e^-	\to D^+ D^-$ process, followed by 
$D^\pm$ decay to $D$ and  on-shell or off-shell $W^\pm$ which we denote as $W$
from now on   for both (off-shell or on-shell)  cases. In its turn, $W$ decays to a $q\bar{q}$
pair (di-jet) or  $\ell\nu$.

Respectively,  we will use
several characteristic  kinematical observables for the signal and background
processes relevant to this final state.
\begin{itemize}
	\item 
Among them is the  
missing mass, $M_{miss}$ 
which is invariant mass of the invisible particles system, i.e. invisible mass for pair of DM particles, defined as 
\begin{equation}\label{Mmiss}
	M_{miss} = \sqrt{ \left((\sqrt{s},0,0,0)-\sum_{vis} P_{vis}
	\right)^2}  \ ,
\end{equation} 
where $P_{vis}$ are 4-momenta of visible particles.
\end{itemize}
In the absence of the initial state radiation (ISR) and beamstrahlung  (B)
effects (which we  call  ISR+B effects for brevity from now on)  the minimum value of $M_{miss}$ is $2M_D$, while for SM background (BG)  with just one neutrino in the final state $M_{miss}$ is vanishing.
Therefore, in this particular case BG can be perfectly separated from the signal using just one $M_{miss}$ variable.
In reality ISR+B effects  play an important role as we demonstrate  below.
Therefore we will make use of several other useful kinematical variables, such as:
\begin{itemize}
\item missing transverse momentum, $\MET$
\item
charged lepton energy (muon in particular), $E_\mu$
\item
angle of reconstructed $W$-boson in the LAB  system, $cos\theta_W$
\item 
the energy of $W$-boson reconstructed from the di-jet pair, $E_{jj}$ .
\end{itemize}

%%%%%%%%%%%%%%%%%%%%%%
\subsection{The strategy  for the signal exploration} 

It has been shown that the $e^+e^-$ colliders such as ILC and CLIC provide an excellent opportunity for discovery of DM and study its properties for some promising theories, including Supersymmetry (see, e.g., \cite{TESLA-1}, \cite{TESLA-2}). 
 In this paper we also demonstrate that this is the case for
 the $\epe\to D^+D^-$ process under study with  the signatures discussed above in section~\ref{sec:sign}.
  The cross section of this process for both DM theories 
  we study here is  large enough in comparison with the respective SM background, such that not only DM discovery is possible at ILC and CLIC but also determination of DM properties.
  
The masses $M_+$ and $M_D$ can be determined from the edges of the $W$-boson energy distribution which can be measured using di-jets
originating from $D^+\to DW^\pm \to Djj$ decay chain as we discuss below in section ~\ref{secWdistr}, %\ref{secleptlight+}
(see \cite{WILC-1,WILC-2} for MSSM and \cite{Lundstrom:2008ai-1,Lundstrom:2008ai-2,Lundstrom:2008ai-3} for I2HDM model cases). 
However, this method  provides an accuracy  worse than that which can be achieved by using the lepton energy distribution which we exploit in our study, section ~\ref{secleptlight+}, together with usage of di-jet energy. Indeed, the accuracy of the jet energy resolution  is typically  one order of magnitude worse than muon momentum resolution~\cite{ILDConceptGroup:2020sfq} for ILC, which eventually affects the edges of the respective distributions and  consequently the accuracy of the mass measurement.
The energy distribution of leptons from the $D^\pm \to D W^\pm \to D \nu \ell^\pm$ signal   has two kink points whose positions are determined by the decay chain kinematics
and can be used to determine the $M_+ $ and $M_D$ masses of $D^+$ and $D$ respectively.
The energy distribution of $W$-boson (which can be determined from di-jet decay channel) also has two kinks which can be used as an important complementary way to measure the masses of DM particles. %%%%%%%%%The low and high end kinks can be very close or even overlap either for $E_W$ or $E_\ell$ energy distributions, which would eventually spoil  $M_+ $ and $M_D$ determination. The key point we stress  here is that these kinks  {\it never overlap for both distributions}, 
%as is demonstrated in Fig. \ref{fig:EW_vs_Emu}, meaning that $M_+ $ and $M_D$ masses can be always reconstructed. \tcg{When kinematic edges are not distinguishable in the $E_\mu$ energy distribution, then the $E_W$ distrubition displays a maximal separation of kinematic edges and vice versa. This is a powerful observation which highlights the importance of the two observables in tandem, in order to effectively extract masses in the class of DM models we explore.}

%\begin{figure}[htb]
%	\centering
%	\includegraphics[width=0.49\textwidth]{E_W_kinks.pdf}
%	%
%	\includegraphics[width=0.49\textwidth]{E_mu_kinks.pdf}
%	\caption{The dependence of position of kink of the Muon(left) and W(right) energy distributions on the DM mass, $m_D$, and mass split with its charged partner, $\Delta M_+$. Here solid(dashed) lines correspond to lower(upper) kinks of the respective energy distributions.}
%		\label{fig:EW_vs_Emu}
%\end{figure}

%%%%%%%%%%%%%%%%%%%%%%%%%%%%%%%%%%%

{We would also like to note that charge lepton and $W$-boson energy distributions for both -- spin zero and spin-half DM cases which we study here -- are quite similar as we demonstrate in the following section.
However, the angular distributions of $W$-bosons from $D^+$ decay as well as the signal production cross sections are quite different  for  spin zero and spin-half DM. These two observables
allow one to very clearly distinguish the spin of DM from the signal under study at lepton colliders.
}

The strategy for DM discovery
with the respective  signature  we discuss here   is very different from  the case when the lightest charged $D$-odd particle is charged slepton,
the spin-zero superpartner of the SM charged lepton. In this case,  the  important signal  channel  is the  $e^+e^-\to \tilde{\ell}^+\tilde{\ell}^-\to \ell^+\ell^-\chi_0\chi_0$ process.
This process  has di-lepton  signature 
which is quite clean and well-identifiable but  different from the one  we study. Also, the energy of an observable lepton -- decay product of slepton -- is well measurable {\it in each individual event}, contrary to our case, when similar product of decay, $W$, is seen as di-jet or lepton plus neutrino with worse measurable energy in each individual event. Therefore, the approach used in the analysis of slepton production (cf. \cite{selectron-1,selectron-2,selectron-3}) cannot be applied directly to our study.

%%%%%%%%%%%%%%%%%%%%%%%%%%%%%
\subsection{Cross sections for $D^+D^-$ production}\label{secmain}
%%%%%%%%%%%%%%%%%%%%%%%%%%
It is convenient use the cross section for SM process
\be
\sigma_0\equiv \sigma(e^+e^-\to \gamma\to\mu^+\mu^-) =4\pi\alpha^2/3s \ \ , 
\ee
 which allows to express the QED cross section of  $\epe\to D^+D^-$ process with the photon exchange only as:
\begin{equation}\label{QED_Xsec}
	\sigma_{\gamma} = \begin{cases} 
		\sigma_0 \beta_D  \left[ 1 + \dfrac{2M_+^2}{s}\right]  & \mbox{if } s_D=\dfrac{1}{2} \\
		\\
		\sigma_0 \dfrac{\beta_D^3}{4} & \mbox{if } s_D=0 
	\end{cases}\ \ ,
\end{equation}
where $\beta_D= \sqrt{1-4M_+^2/s}$.
Now we can express the total cross section of $\epe\to D^+D^-$ process as:
\begin{equation}\label{full_Xsec}
	\sigma = \sigma_{\gamma} + \sigma_{\gamma/Z}  + \sigma_{Z} = \sigma_{\gamma} \left[ 1 + \dfrac{\kappa_{\gamma/Z}}{1-\dfrac{M_Z^2}{s}} + \frac{\kappa_{Z}}{\left(1-\dfrac{M_Z^2}{s}\right)^2} \right] \ \ , 
\end{equation}
where  $\sigma_{\gamma/Z}$ and $\sigma_{Z}$ represent the contribution from $\gamma/Z$ interference and
squared diagram with $Z$-boson exchange respectively.
The corresponding $\kappa_{\gamma/Z}$ and $\kappa_{Z}$ coefficients are given by
\begin{align}
	\kappa_{\gamma/Z} &= \frac{ \cos{2\theta_W}(2\cos{2\theta_W}-1) }{4\cos^2{\theta_W}\sin^2{\theta_W}} \approx {0.0867}
	\\
	\kappa_{Z} &= \frac{ \cos^2{2\theta_W}( \cos^2{2\theta_W} + (\cos{2\theta_W}-1)^2)) }{32\cos^4{\theta_W}\sin^4{\theta_W}} \approx {0.162} \ .
\end{align}

\begin{figure}[htb]
	\includegraphics[width=0.5\textwidth]{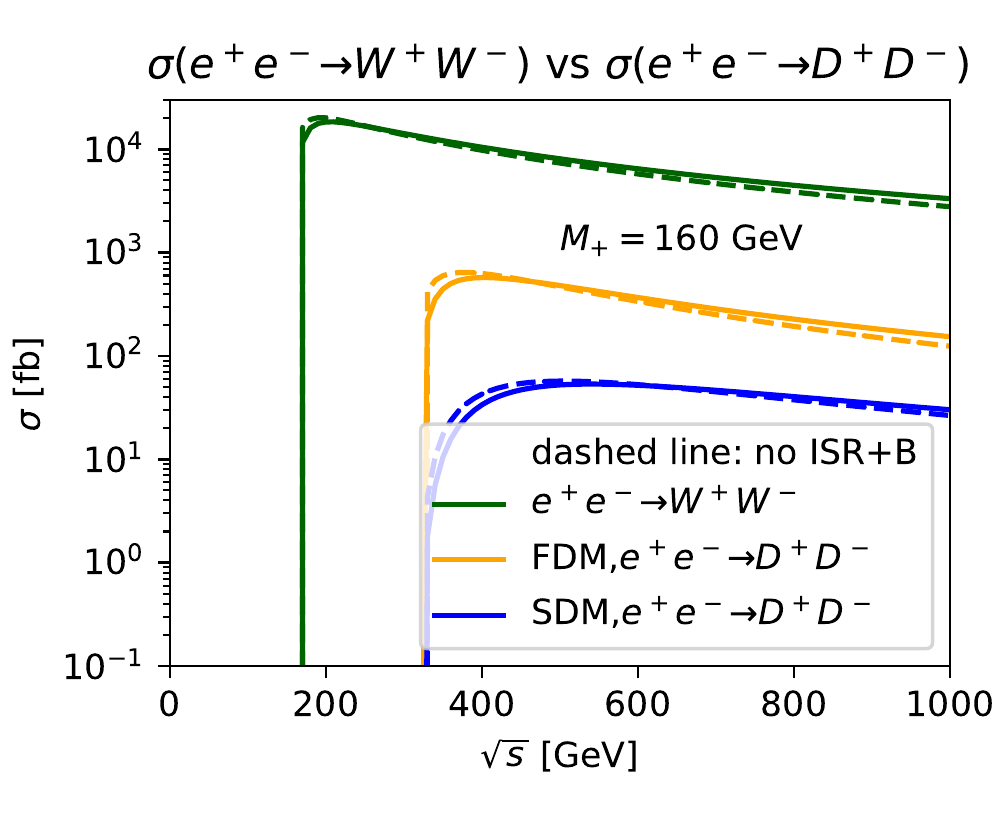}%
	\includegraphics[width=0.5\textwidth]{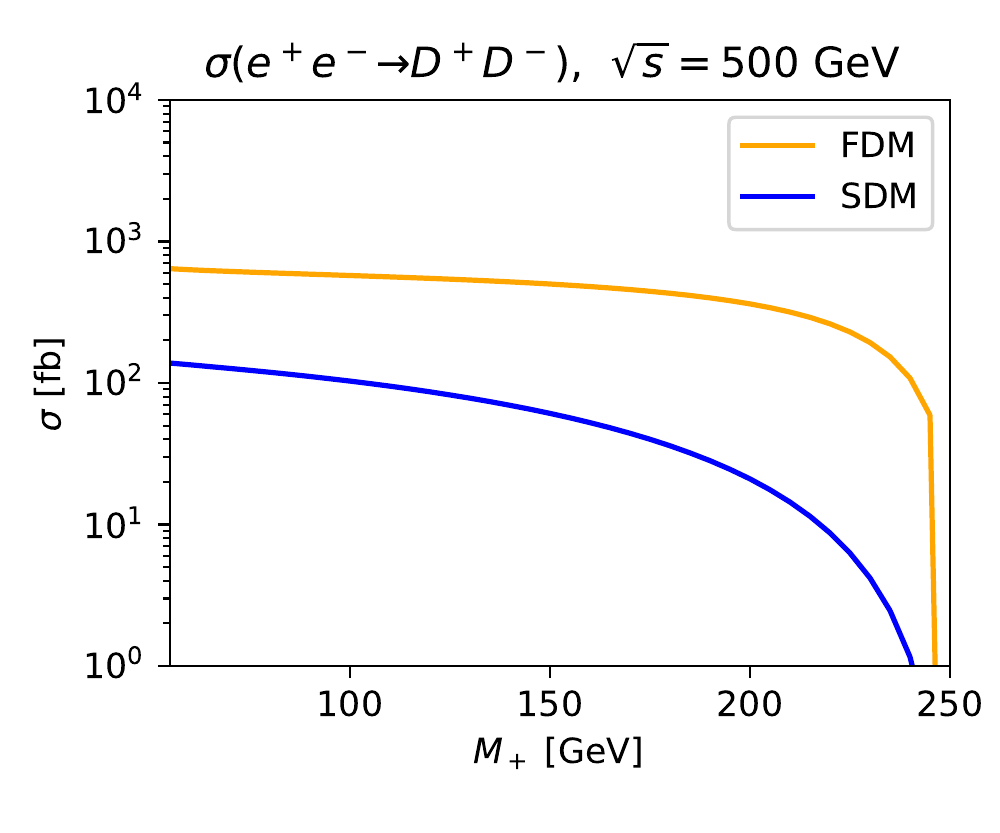}	
	\caption{Left: cross section versus
		$\sqrt{s}$ for background 
		$e^+e^- \to W^+W^-$  process (green) compared to 
	    the cross section of $e^+e^- \to D^+D^-$
	    signal processes for fermion (orange) and scalar(blue) dark matter for BP1 ($M_+=160$~GeV). Solid (dash) lines present results for ISR+B effects
	    switched on (off) 	 respectively.   Right:
	    $e^+e^- \to D^+D^-$ cross section versus $M_+$.
	    \label{fig:signal-xsec}
	}
\end{figure}

The respective signal  cross sections for both models and both benchmarks as well as for  the leading 
SM  background $\epe \to W^+W^-$ are presented in Fig.~\ref{fig:signal-xsec}.

%
%%%%%%%%%%%%%%%%%
%
For our benchmarks, the $\epe \to D^+ D^-$
cross section for the fermion DM (FDM) case is close to $\sigma_0$,
(which makes its use convenient) while for scalar DM (SDM)  case the pross section is about one order of magnitude lower.
 The annual integrated
luminosity $\cal L$ for the ILC
project \cite{TESLA-2} is expected to be  500~fb$^{-1}$ which provides of the order $10^5$ and $10^4$ events for 
FDM and SDM cases respectively. 
The initial ratios of the FDM and SDM signals to the 
$\epe \to W^+W^-$ background are about $\dfrac{1}{10}$ and
 $\dfrac{1}{100}$ respectively.

To measure  the $e^+e^-\to D^+D^-$ cross section at the experimental level one should measure the sum over all processes with signatures \eqref{eq:sign-WW-jjjj} and \eqref{eq:sign-WW-enjj} (that is about 7/9 of the total cross section of $D^+D^-$ production, since our signature does not include dilepton final state which is $2\times(3/9\times3/9)=2/9$). 
When masses $M_+$ is measured, the cross section of
$\sigma(e^+e^-\to D^+D^-)$ process can be calculated and compared
with the measured one. Since the difference between 
the FDM and SDM signal is about one order of magnitude,
the knowledge of the cross section would allow to distinguish DM spin  for these two models. One should note, that in case of Supersymmetry the FDM cross section can be modified by $t$-channel diagrams with the sleptons, which could reduce the cross section by about factor of two, which however would 
still allow to discriminate FDM from SDM case.

%%%%%%%%%%%%%%%%%%%%%%%%%%%%
%%%%%%%%%%%%%%%%%%%%%%%%%%%%
\subsection{$\pmb W$ and charged lepton energy distribution and Dark Matter mass reconstruction}

We proceed  with a discussion of the features of the $W$ and charged lepton energy distributions for processes (\ref{eq:sign-WW-jjjj}) and (\ref{eq:sign-WW-enjj}), comprising the positions of discontinuities and end-points, expressions for which are derived using simple kinematics.

%%%%%%%%%%%%%%%%%%%%%%%%%%%%%%%%%%%%%%%%%%
\subsubsection{$\pmb W$ Energy Distributions}
\label{secWdistr}
First we consider the energy distribution of $W$ (which may be virtual) with mass $M_W^*$. In the regime where $W$ may be produced on-shell (i.e $M_+-M_D>M_W$), then $M_W^*=M_W$. However, when $W$ is produced off-shell its maximum effective mass is $M_W^*=M_+-M_D$ 
for $W^*$ at zero momentum.
In the rest frame of $D^\pm$  we have a two-particle decay $D^\pm\to DW^\pm$. The energy and three-momentum of the $W$ boson in the $D^\pm$ rest frame (labelled by superscript $D$) are given by:

\bear{c}
E^D_W(M_W^*)=\fr{M_+^2 +M_W^{*2}- M_{D}^2}{2M_+},\;\;
p^D_W(M_W^*)=\fr{\sqrt{(M_+^2-M_W^{*2}-M_D^2)^2-4M_D^2M_W^{*2}}}{2M_+}.\\[2mm]
\eear{rkinW}

Denoting $\theta$ as the $W^+$ escape angle in the $D^+$ rest frame with respect to the direction of $D^+$ motion in the laboratory frame, and using $c\equiv\cos\theta$, we find the energy of \ $W^+$ \ in
the laboratory frame to be:
\be
E_W=\gamma_D(E^D_W+c\beta_D p^D_W) \Rightarrow \\\;
E_W^{(-)}(M_W^*)<E_W<E_W^{(+)}(M_W^*), \;\;
\ee
where
\be
E_W^{(\pm)}(M_W^*)=
\gamma_D(E^D_W\pm\beta_D p^D_W) \label{EPWoff} \ \ ,
\ee
with $\gamma_D=\dfrac{\sqrt{s}}{2M_+}$. 

For the on-shell $W$ ($M_W^*=M_W$) case, the kinematical edges of the $W$ energy distribution are

\begin{eqnarray}
E^{(\pm)}_W(M_W)=\fr{E}{2}\left[1+\fr{M_W^2-M_D^2}{M_+^2}\pm
\fr{\sqrt{(M_+^2-M_W^2-M_D^2)^2-4M_D^2M_W^2}}{M_+^2}\sqrt{1-\fr{M_+^2}{E^2}}\right]
\label{EPW}
\end{eqnarray}

where $E$ is the $D^\pm$  energy, which is quite different from delta-function shape of the background distribution, peaking at $E$ in the absence of {ISR+B}. {We show the   $W$-boson energy distribution  in Fig.\ref{fig:EW_and_Emu} for both SDM and FDM cases.}
In reality ISR+B, as we show later,
introduces an important smearing which makes the background non-negligible.

For the off-shell case ($M_+-M_D<M_W$) although these equations hold for events with both a virtual $W$ and real $D$ produced at rest in the $D^\pm$ frame, the kinematic edges are smeared as a result of variation the final state momenta (and consequently the four-momentum of the virtual $W$) over the phase-space. This is  demonstrated below in Fig.\ref{fig:SDM_FDM_EW_OFF} where the kinematic edges are not clearly visible.

In a well known approach, one measures edges in the energy distributions of dijets, representing $W$ coming from $D^\pm\to DW^\pm$ decay~\cite{WILC-1,WILC-2}. However, the
individual jet energies and, consequently, effective masses of dijets cannot be measured with a high precision. The observed lower edge of the $W$ energy distribution in the dijet mode is smeared because of this. One can only hope for a sufficiently accurate measurement of the upper edge of the $W$ energy distribution, $E_W^{+}$ given by Eq. \eqref{EPW}.
Therefore
we suggest to extract the second quantity for derivation of masses from the lepton energy spectra. The lepton energy is measurable with a higher accuracy in comparison to the di-jet one. We will show that the singular points of the energy distribution
of the leptons in the final state with signature \eqref{eq:sign-WW-enjj} 
are kinematically determined, and therefore can be used for a mass measurement. 
%%%%%%%%%%%%%%%%%%%%%%%%%%%%%%%%%%%%%%%%%%%%%%%%%%%%%%%%%%%%%%%%%%%%%%%%%%%%%%%%%%%%%%%%%%%%%%%%%%%%%%%%%%%%%%%%%%%%%%%%%%%%%%%
\subsubsection{Charged lepton energy distributions in
	$\pmb{\epe\to D^+D^-\to DD W^+W^-\to DD q\bar{q}\,\ell\nu}$ %(\ref{sign+WA})
}\label{secleptlight+}
%%%%%%%%%%%%%%%%%%%%%%%%%%%%%%%%%%%%
%\tcg{ The corresponding text in [Ginold] contains inaccuracies. Presented here eq-s cover App. A1 and A2 }

We next study the distribution of events over the muon energy, $E_\mu$. The fraction of such events for each separate lepton, $e^+$, $e^-$, $\mu^+$ or $\mu^-$, is about $1/9 \times 2/3 =2/27 \simeq 0.074$, while their sum is about $4\times 2/27 \simeq  0.30$ of the total cross section of the process.

In the following sections we consider only muons, so that in the $W$ rest frame and the laboratory system with $W$ energy $E_W$ respectively, we have
\be
E_\mu^W=|\vb{p}|_\mu^W=M_W^{(*)}/2\,, \qquad \gamma_W=E_W/M_W^{(*)}, \;\; \beta_W= \sqrt{1-\gamma_{W}^{-2}}.\label{muonrest}
\ee

Just as before, we denote $\theta_1$ as the escape angle of $\mu$ relative to the direction of the $W$ in the laboratory frame and use
$c_1=\cos\theta_1$. The muon energy in the laboratory frame is

\beg
E_\mu=\frac{E_W}{2}\left(1+c_1\beta_W\right).
\label{Emugen}
\eeg 

Muon energies lie between energies $\tfrac{1}{2}\left( E_W \pm \sqrt{E_W^2-M_W^{(*)2}} \right)$. The maximum muon energy, $E_\mu^{max}$, may be determined from the highest value of $W$ energy , i.e $E_W=E_{W}^{(+)}$ from Eq.~\eqref{EPW} (see appendix \ref{D2}):

\be
E_\mu^{max} = \frac{E}{2} \left( 1+\beta_D\right)\left(1-\frac{M_D}{M_+}\right).
\label{EmuMax}
\ee

With a shift of $E_W$ from these boundaries inwards, the density of
states in the $E_\mu$ distribution grows monotonically due to contributions of smaller $E_W$ values up to $E_\mu^{(\pm)}$ values, corresponding
to the lowest value of $W$ energy $E^{(-)}_{W}$ from Eq.~\eqref{EPW}. At these points the energy distributions of muons have kinks, located at $E_{\mu}^{(\pm)}$. Between these kinks, the $E_\mu$-distribution is approximately flat.
The following equation (derived in appendix \ref{D1}):
\be
E_{\mu}^{(\pm)}=
\fr{E_{W}^{(-)}\pm \sqrt{(E_{W}^{(-)})^2-M_W^2}}{2}\label{Emuin}
\ee
gives the upper and lower bounds in the muon energy distributions.

At $M_+-M_D>M_W$, the positions of upper edge in the dijet energy distribution $E^{(+)}_{W}$  \eqref{EPW} and
the lower kink in the muon energy distribution $E_\mu^{(-)}$ \eqref{Emuin} give us two equations necessary for
determination of $M_D$ and $M_+$ (derived in appendix~\ref{D3}):
\be
M_D^2 = M_W^2-M_+^2\left[ \fr{1}{E} \left( \alpha+\beta\right)-1\right],
\ee
\label{eq:md}
\be
M_+^2= 2 \left[\fr{E^2\left(\alpha \beta + M_W^2\right)-\sqrt{E^4(\alpha^2-M_W^2)(\beta^2-M_W^2)}}{(\alpha+\beta)^2}\right],
\label{eq:mplus}
\ee
where $\alpha$ and $\beta$ are defined as:
\be
\alpha=\fr{4E_\mu^{(+)2}+M_W^2}{4E_\mu^{(+)}}, \quad\quad \beta=\fr{4E_\mu^{(-)2}+M_W^2}{4E_\mu^{(-)}}.
\ee

The position of the upper edge in the dijet energy distribution $E^{(+)}_{W}$ should be extracted from all events with signatures \ref{eq:sign-WW-jjjj}, \ref{eq:sign-WW-enjj}, while the position of the lower kink in the muon energy distribution $E_\mu^{(-)}$ can be extracted from events with signature \ref{eq:sign-WW-enjj} only. 

\begin{figure}[H]
	\centering
	\includegraphics[width=0.49\textwidth]{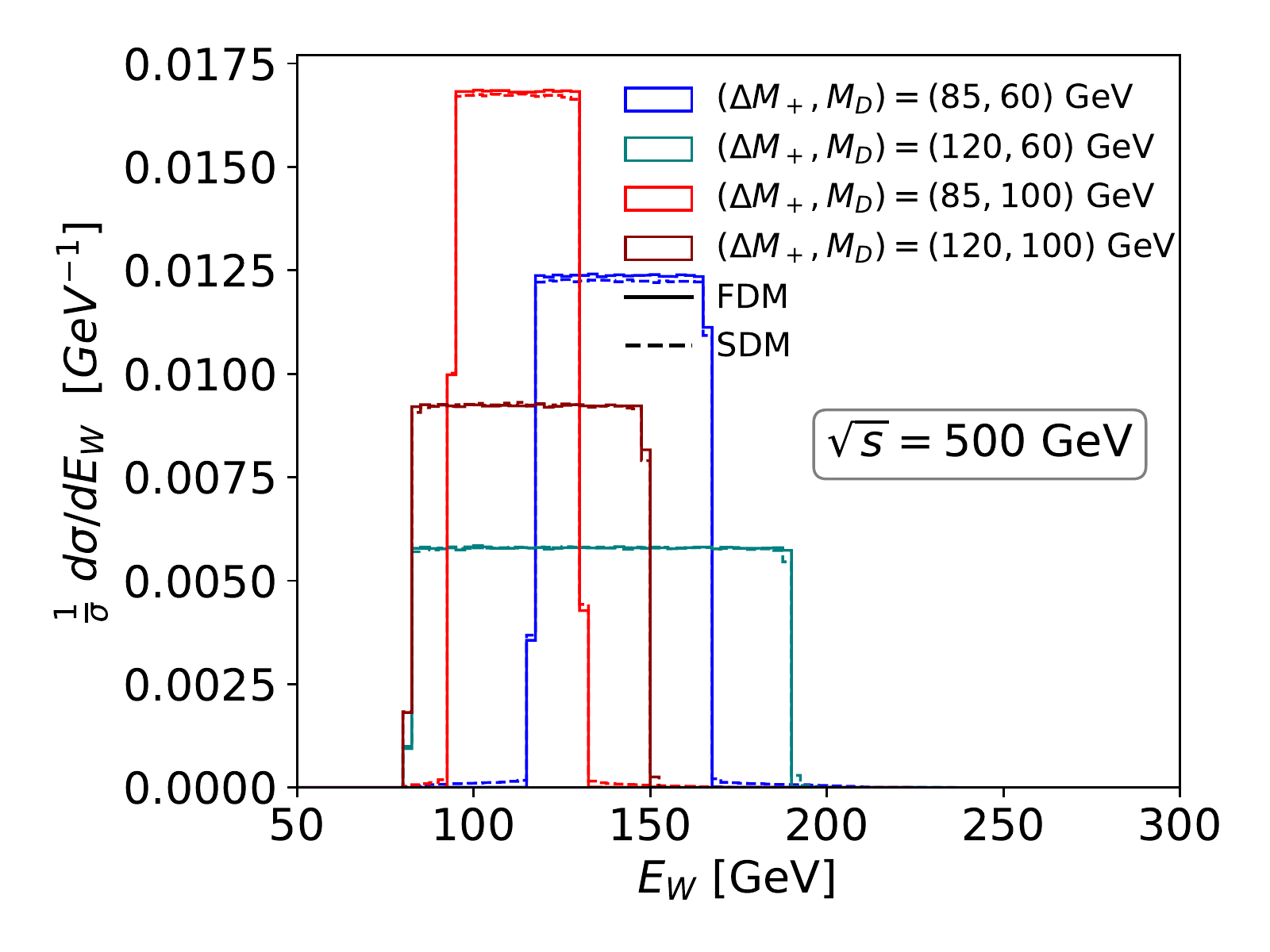}
	\includegraphics[width=0.49\textwidth]{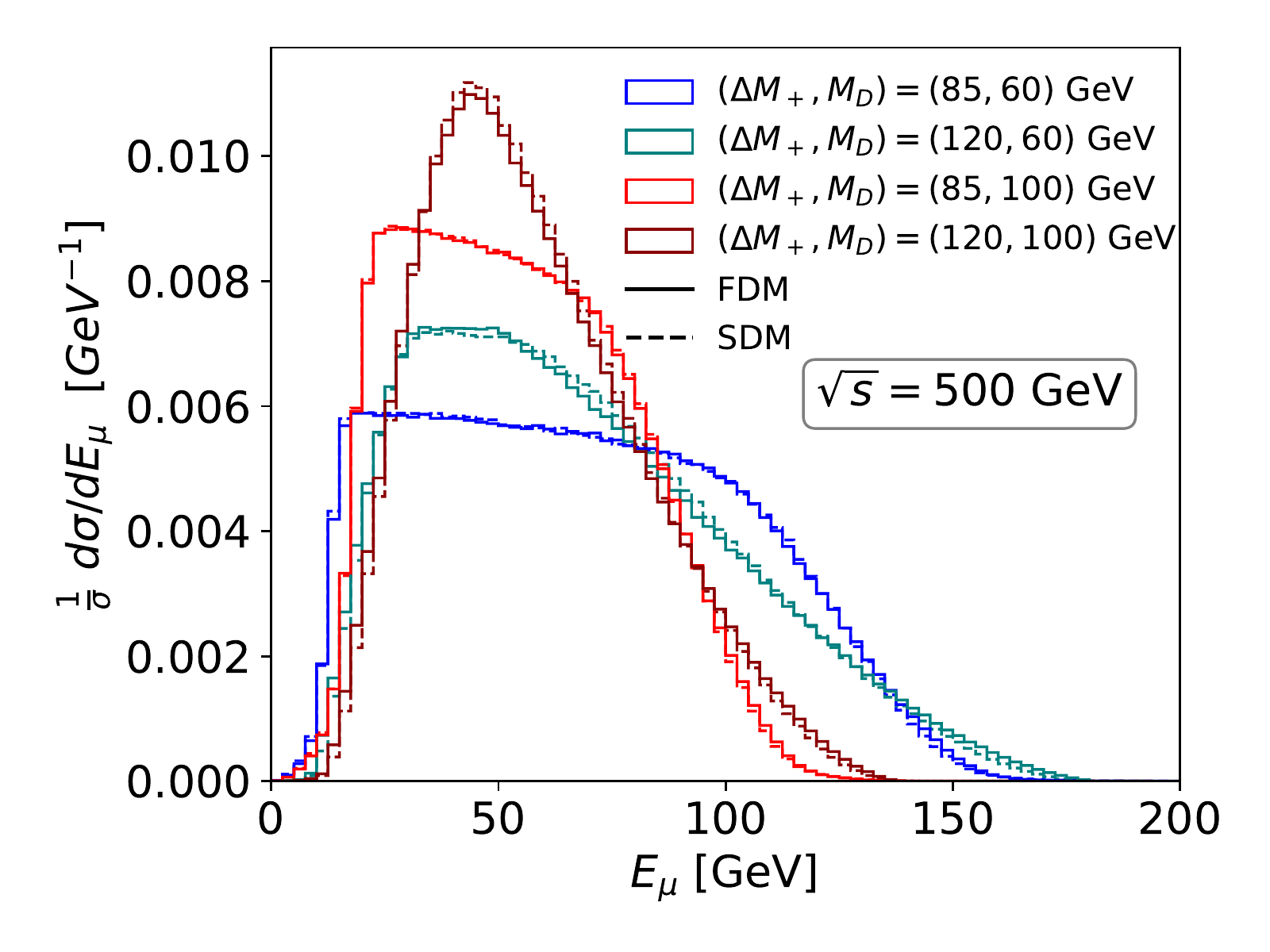}
	\caption{The energy spectra (without ISR+B) of the $W$ (left) and muon (right) for different DM mass, $m_D$, and mass split with its charged partner, $\Delta M_+$. Solid and dashed lines correspond to FDM and SDM respectively.}
	\label{fig:EW_and_Emu}
\end{figure}

If a $D_2$ particle is absent or $M_{D_2}>M_+$, the results \eqref{Emugen}-\eqref{Emuin}  are valid since
one can neglect the  interference between
the signal and SM diagrams as we discuss below.
The shape of the energy distribution of leptons (with one peak or two kinks) allows to determine which case is realized, $M_+-M_D>M_W$ or $M_+-M_D<M_W$. The energy distributions (without ISR+B effects) of muons $E_\mu$ alongside $E_W$ are presented in Fig.\ref{fig:EW_and_Emu} for both SDM and FDM cases.

While spin does not affect the shape of distributions in Fig.~\ref{fig:EW_and_Emu}, the different DM masses and mass split scenarios can  be easily distinguished by using both energy spectra  $E_W$ and $E_\mu$. For the $W$ energy distributions in the left plot, increasing $\Delta M_+$ spreads the energy distribution across a larger range. The $W$ is produced nearly at rest when near the $D$-$D^+$ mass split but can have a larger range when given more energy and is more boosted. 

This is opposite in the case of muon energy distributions in the right plot, where increasing $\Delta M_+$ narrows the distributions. 
When the width of $E_W$ is minimal, this maximises the width of $E_\mu$ because the muon can go exactly along $W$ or opposite this direction, making the muon distribution as wide as possible.
If $W$ has larger phase space then it is not aligned along the direction of $D^+$. 
In this case the $W$ energy is not fixed and can be varied which leaves less phase space to muon for its energy variation.

In the $W$ energy distributions, increasing DM mass shifts the distributions to lower energies. This is also the case for muon energy distributions, as the tails extend to higher energies for smaller DM mass. Since a heavier DM is produced, this takes a larger fraction of the system's energy, giving less energy to the $W$ and muons for the same input energy.

\begin{figure}[H]
	\centering
	\includegraphics[width=0.49\textwidth]{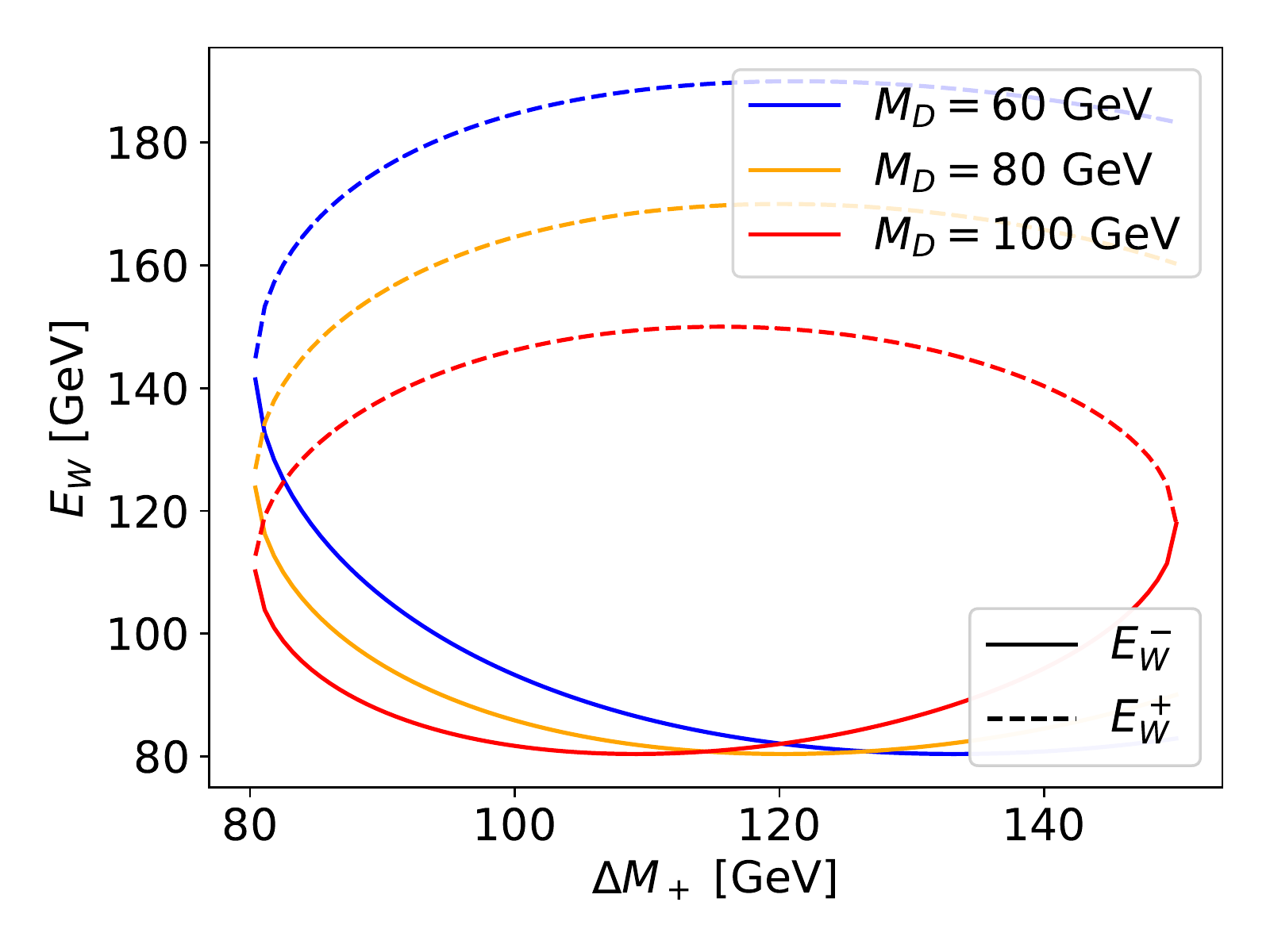}
	\includegraphics[width=0.49\textwidth]{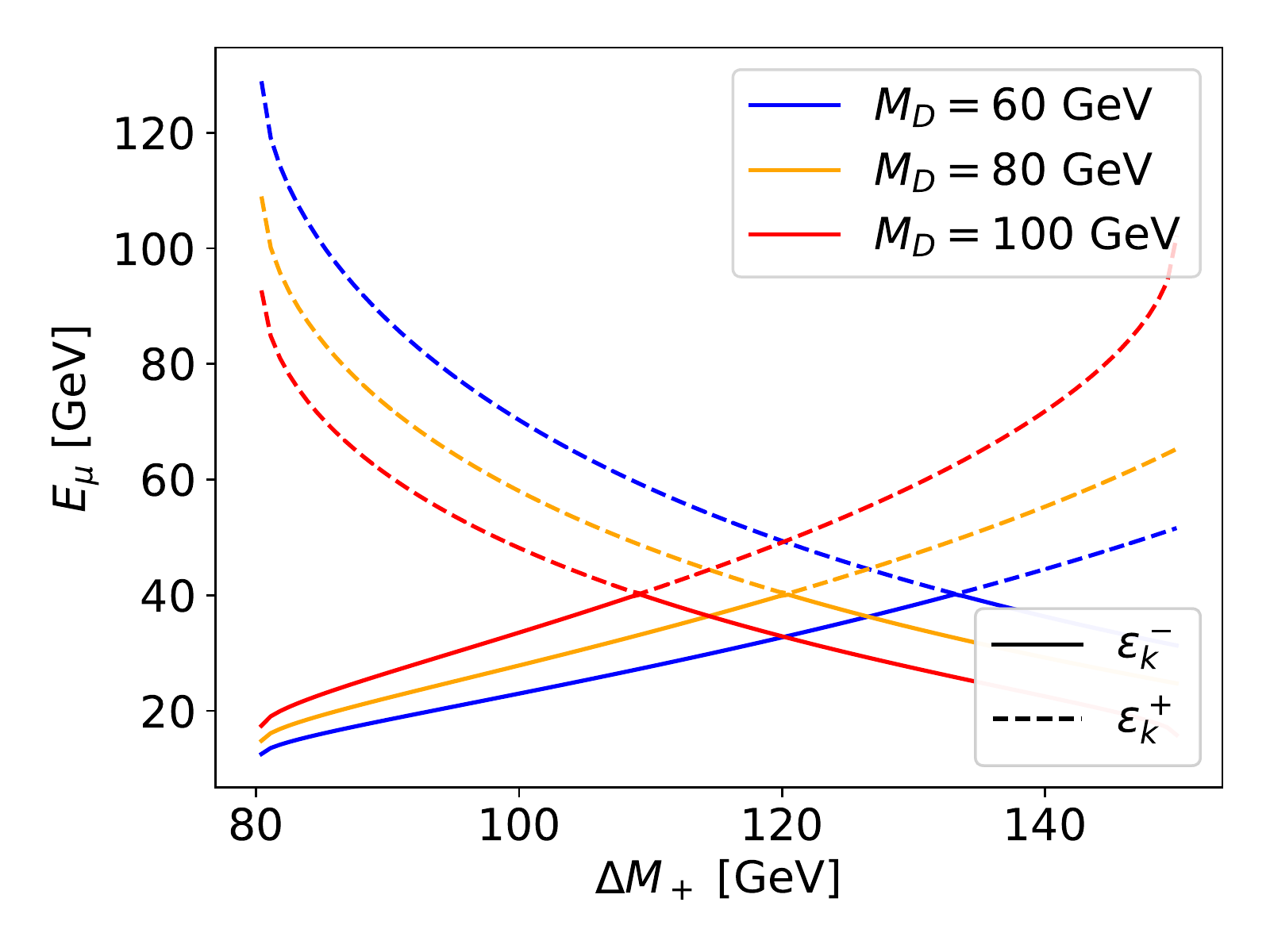}
	\caption{The dependence of position of kink of the $W$(left) and muon(right) energy distributions on the DM mass, $m_D$, and mass split with its charged partner, $\Delta M_+$. Here solid(dashed) lines correspond to lower(upper) kinks of the respective energy distributions.}
	\label{fig:EW_vs_Emu}
\end{figure}

The low and high end kinks can be very close or even overlap either for $E_W$ or $E_\mu$ energy distributions, which would eventually spoil  $M_+ $ and $M_D$ determination. The key point we stress  here is that these kinks  {\it never overlap for both distributions simultaneously}, 
as is demonstrated in Fig. \ref{fig:EW_vs_Emu}, meaning that $M_+ $ and $M_D$ masses can be always reconstructed. {When kinematic edges are not distinguishable in the $E_\mu$ energy distribution,  the $E_W$ distribution displays a maximal separation of kinematic edges and vice versa. This is an important feature of the signal  which highlights the complementary power of the two observables which allows to effectively extract DM masses in the whole parameter space relevant to the ILC signal under study.}

Observation of events with signature \ref{eq:sign-WW-jjjj}, \ref{eq:sign-WW-enjj} will be a clear {\it signal} for DM particle candidates. The non-observation of such events will allow to find lower limits for masses $M_+$, like \cite{Lundstrom:2008ai-1,Lundstrom:2008ai-2,Lundstrom:2008ai-3}.  At $M_+<\frac{\sqrt{s}}{2}$, the cross section $\epe\to D^+D^-$ is a large fraction of the total cross section of \epe annihilation, making this observation a very realistic task.

%%%%%%%%%%%%%%%%%%%%%%%%%%%%%%
\subsubsection{Distortion of the energy distributions from width effects, ISR+B and intermediate $\tau$s}\label{secdist}
%%%%%%%%%%%%%%%%%%%%%%%%
A more detailed analysis reveals two main sources of distortion of the energy distributions (we neglected them in our preliminary analysis).

\begin{enumerate}
	\item The final width of $W$ and $D^\pm$  leads to a
	blurring of the singularities derived. This effect increases with
	the growth of $M_+-M_D$.
	\item The energy spectra under discussion will be smoothed due to  QED
	initial state radiation (ISR) and beamstrahlung (B). %The ISR and FSR spectra are machine
	\item Smearing from intermediate $\tau$ leptons in the cascade {\bf  $\pmb{D^-\to DW^-\to D\tau^-\nu\to D\mu^-\nu\nu\nu}$}
\end{enumerate} 

\begin{figure}[htb]
	\centering
	\includegraphics[width=0.49\textwidth]{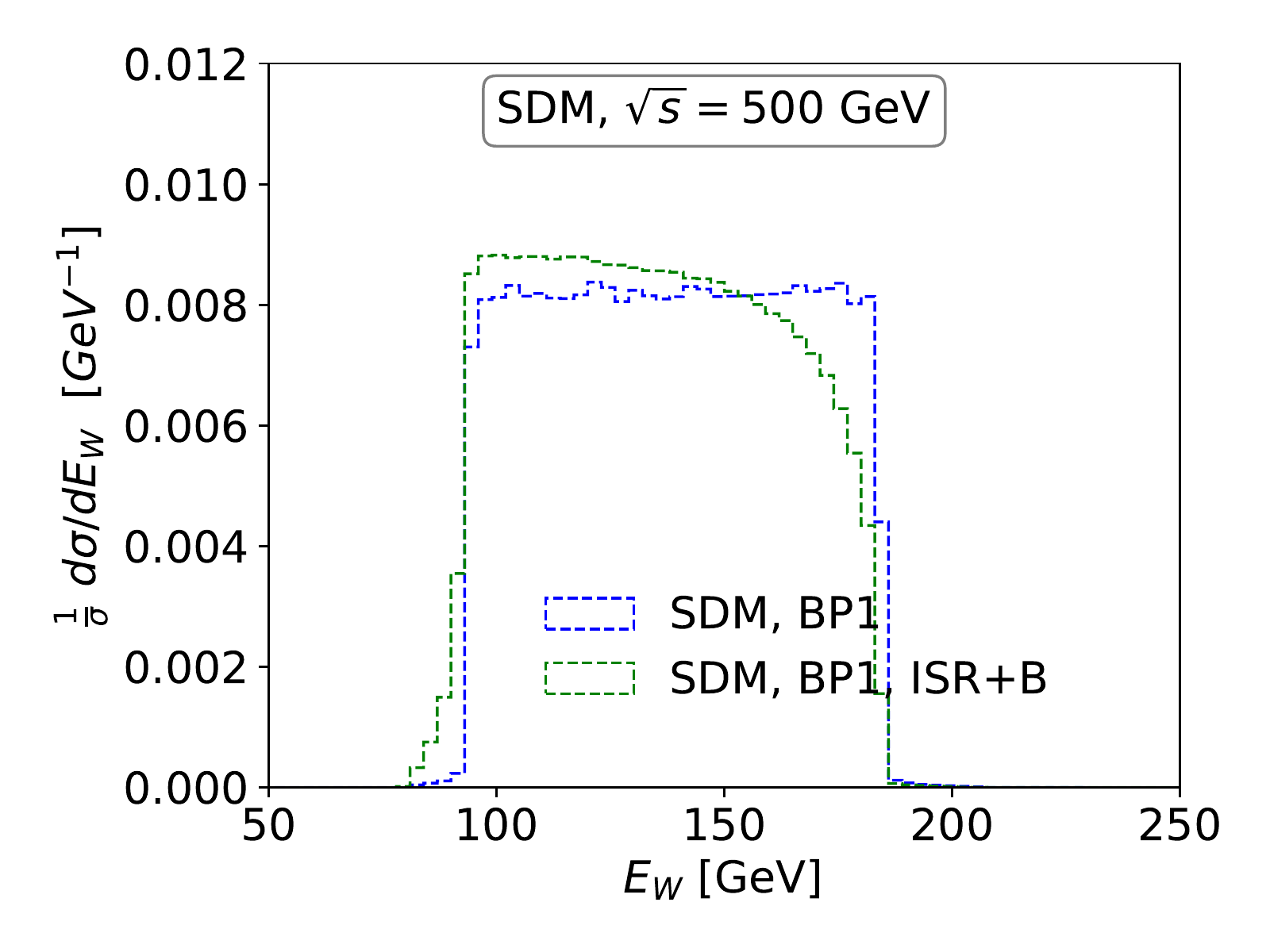}
	\includegraphics[width=0.49\textwidth]{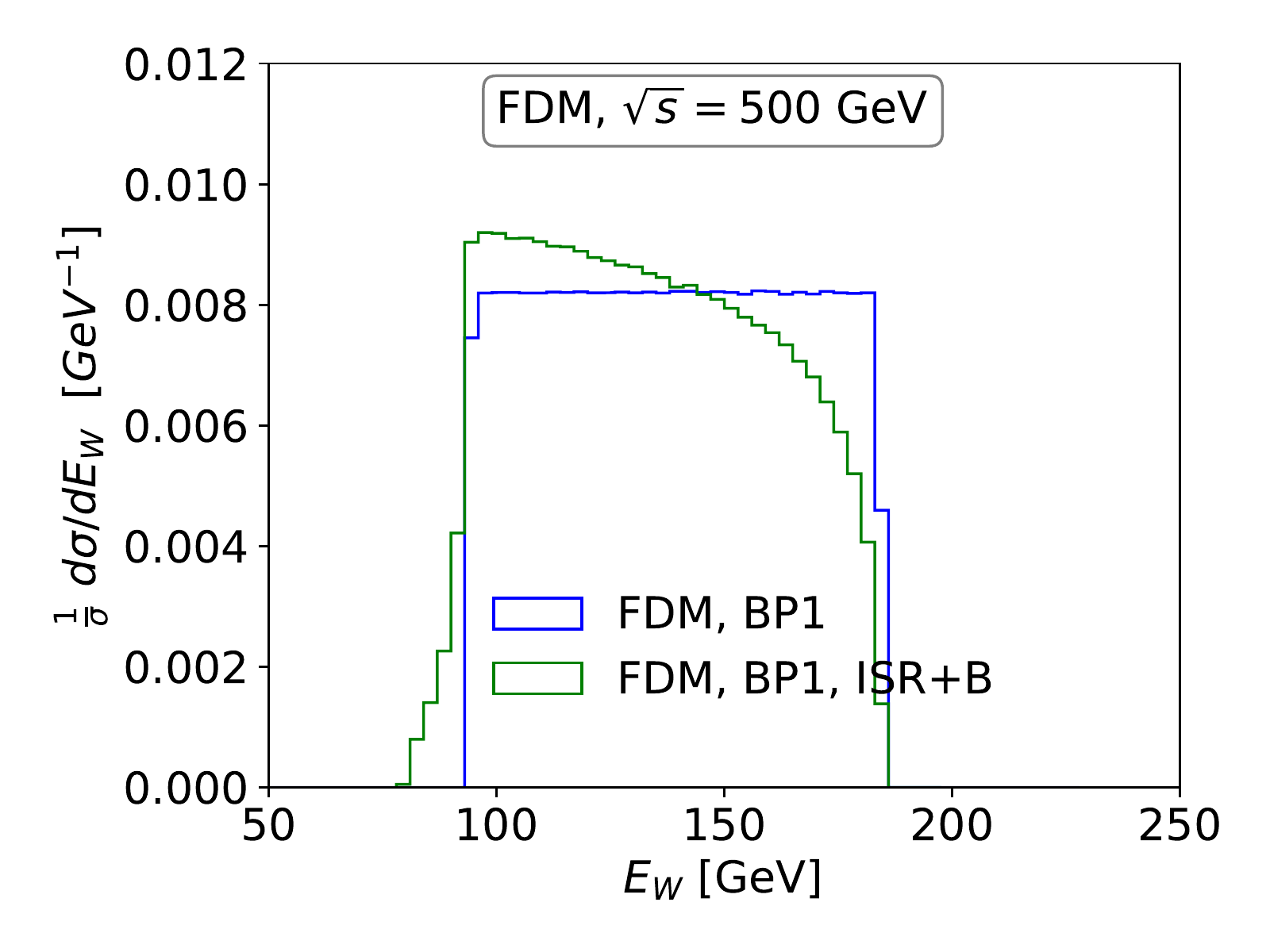}
	\caption{The $W$ energy distribution for BP1, the on-shell $W$ case, for SDM (left) and FDM (right).
		\label{fig:SDM_FDM_EW}}
\end{figure}

\begin{figure}[htb]
	\centering
	\includegraphics[width=0.49\textwidth]{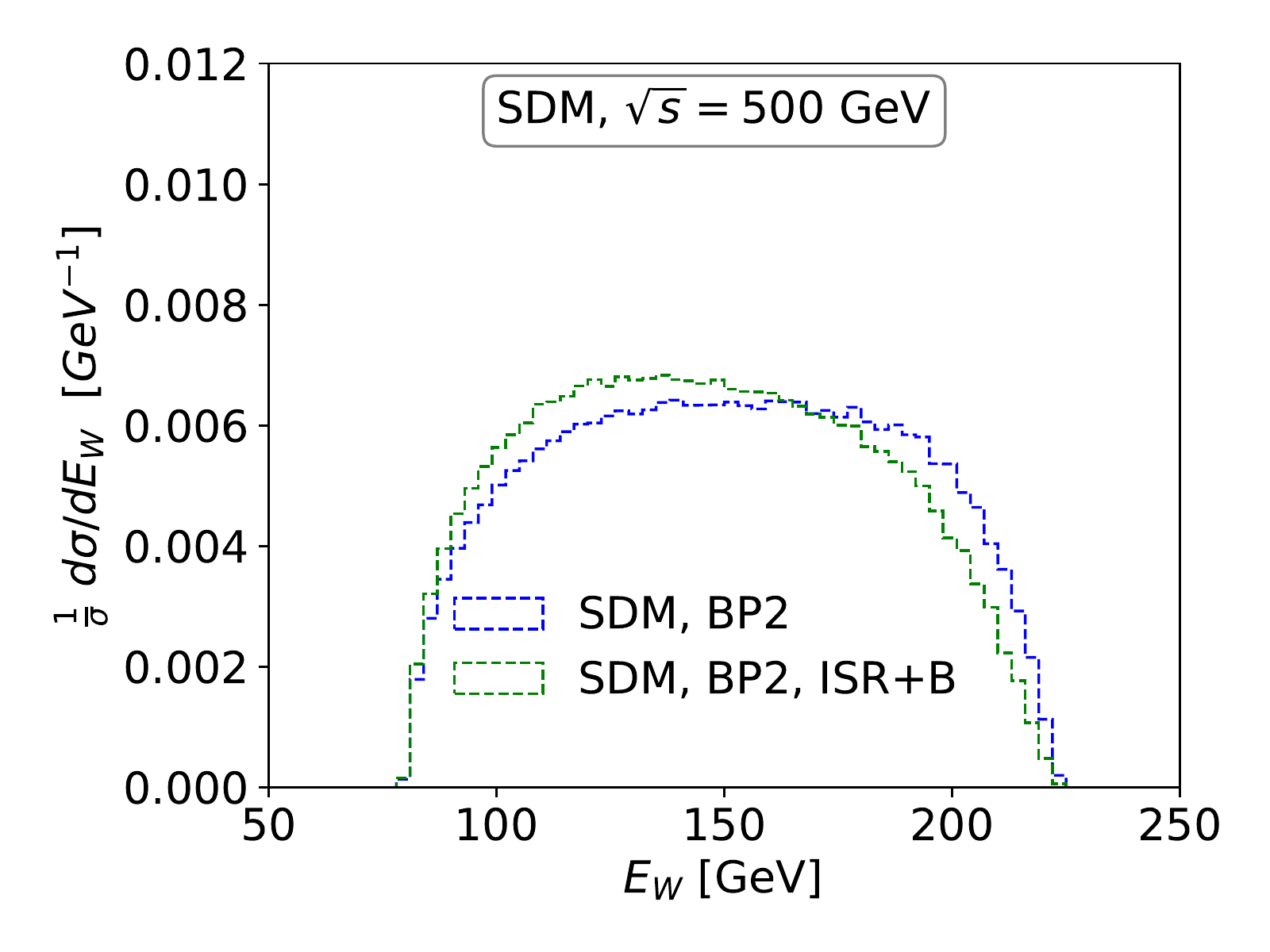}
	\includegraphics[width=0.49\textwidth]{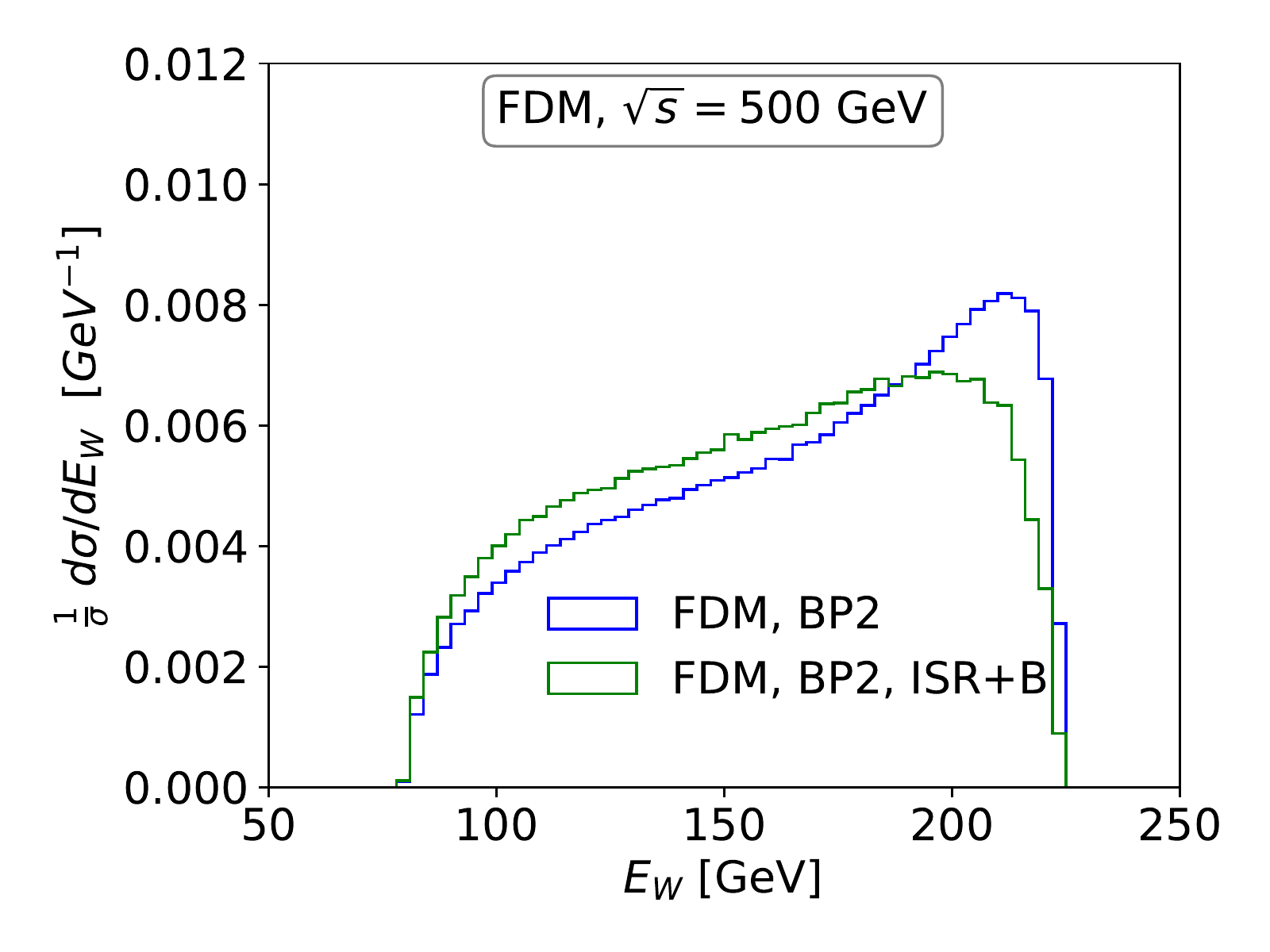}
	\caption{The $W$ energy distribution for BP2, the off-shell $W$ case, for SDM (left) and FDM (right).
		\label{fig:SDM_FDM_EW_OFF}}
\end{figure}

For the on-shell $W$ energy distributions with scalar and fermion DM shown in Fig.~\ref{fig:SDM_FDM_EW} the upper and lower edges in $E_W$ are clearly visible. However, the ISR+B  smearing effect increases the uncertainty in edge identification, especially for the upper edge in $E_W$.
For the off-shell $W$ case, its energy distributions in Fig.~\ref{fig:SDM_FDM_EW_OFF} show no visible kinks or edges, making it impossible to determine DM masses, regardless of ISR+B effects.
\begin{figure}[htb]
	\centering
	\includegraphics[width=0.49\textwidth]{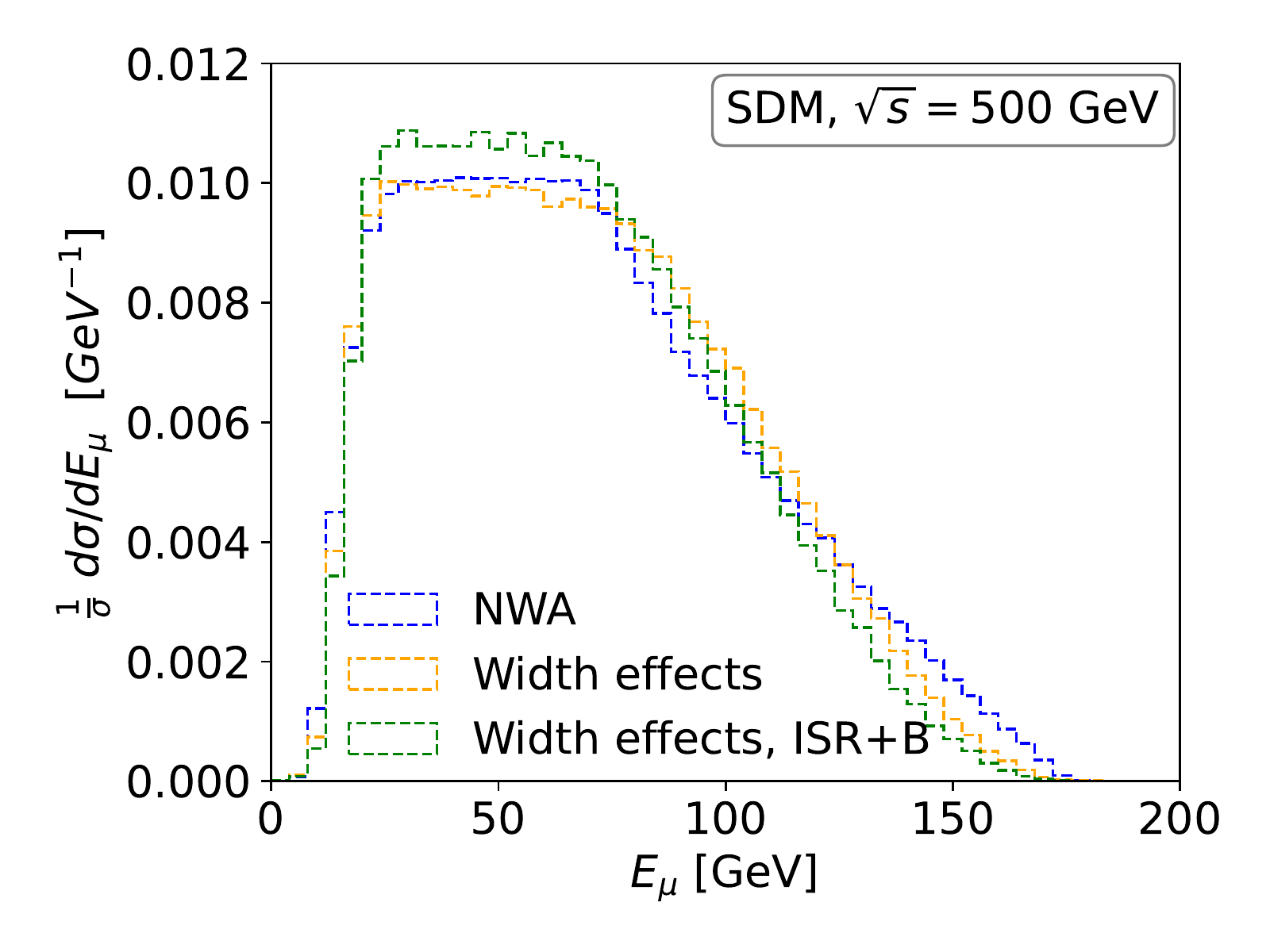}
	\includegraphics[width=0.49\textwidth]{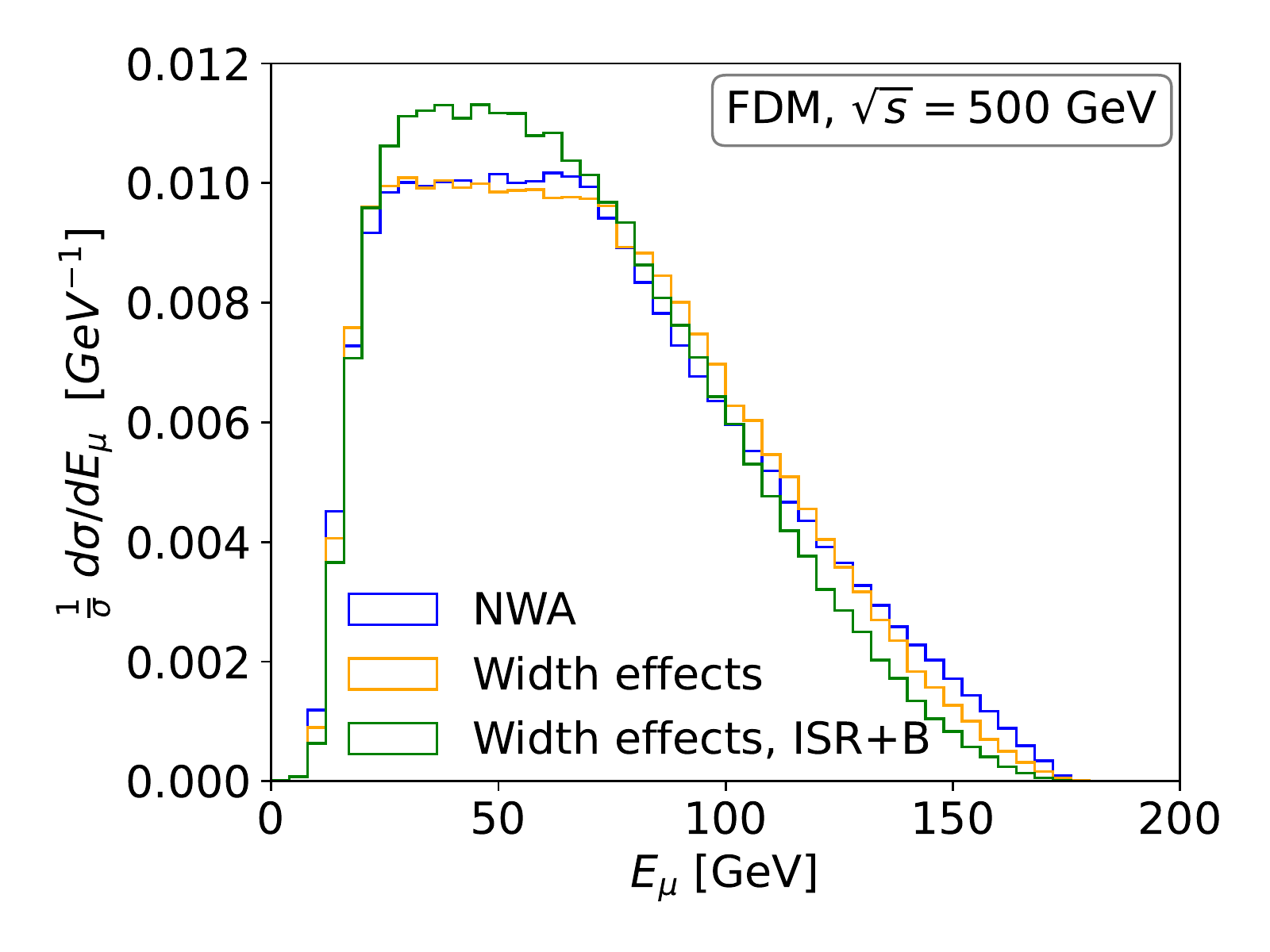}
	%	\caption{Muon energy distributions for the fermion dark matter model using BP1 values, for 3 cases: the simplified case where the distribution is independent of $\theta_1$, the full $2 \rightarrow 4$ process, and the full $2 \rightarrow 4$ process including radiation effects.}
	\caption{Muon energy distribution at BP1 for SDM (left) and FDM (right).
		\label{fig:SDM_FDM_Emu}}
\end{figure}

The effect of ISR+B which distorts  the parton level muon energy distributions is presented in Fig.~\ref{fig:SDM_FDM_Emu} left and right for SDM and FDM respectively (for BP1). 
The blue line corresponds to production of $D^\pm D W^\mp$ and subsequent decay of $W$ boson, i.e $W$-width effects are not included. The yellow line corresponds to simulation of the full production cross-section, taking into account all widths. This effect smooths $E_\mu^{(+)}$ considerably, but the dominant distortion comes from the effects of ISR+B, as shown by the green line. In CalcHEP, ISR is modelled using equation by Jadach, Skrzypek, and Ward \cite{Skrzypek:1991}, and Bremsstrahlung by that of P. Chen \cite{Chen:1991wd}. The key observation here is that the left hand kink, $E_\mu^{(-)}$, remains visible.

The process  {\bf  $\pmb{D^-\to DW^-\to D\tau^-\nu\to D\mu^-\nu\nu\nu}$} also modifies the spectra just discussed. The energy
distribution of $\tau$ produced in the decay $W\to \tau\nu$ is the same as that for $\mu$ or $e$ (within the accuracy of $\sim (M_\tau/M_W^*)^2$). Once produced, $\tau$ decays to $\mu\nu\nu$ in 17 \% of cases (the same for decay to $e\nu\nu$). These muons are added to those discussed above. In the $\tau$ rest frame, the energy of muon is $E_\mu^\tau=y\,M_\tau/2$ with $y\leqslant 1$. The energy spectrum of muons is $dN/dy=2(3-2y)y^2$. %(see textbooks). 
The signal evaluation is presented as energy distributions of muons in the Lab frame. It is clear that
this contribution is strongly shifted towards the soft end of the entire muon energy spectrum.

\begin{figure}[htb]
	\begin{center}
		\includegraphics[width=0.49\textwidth]{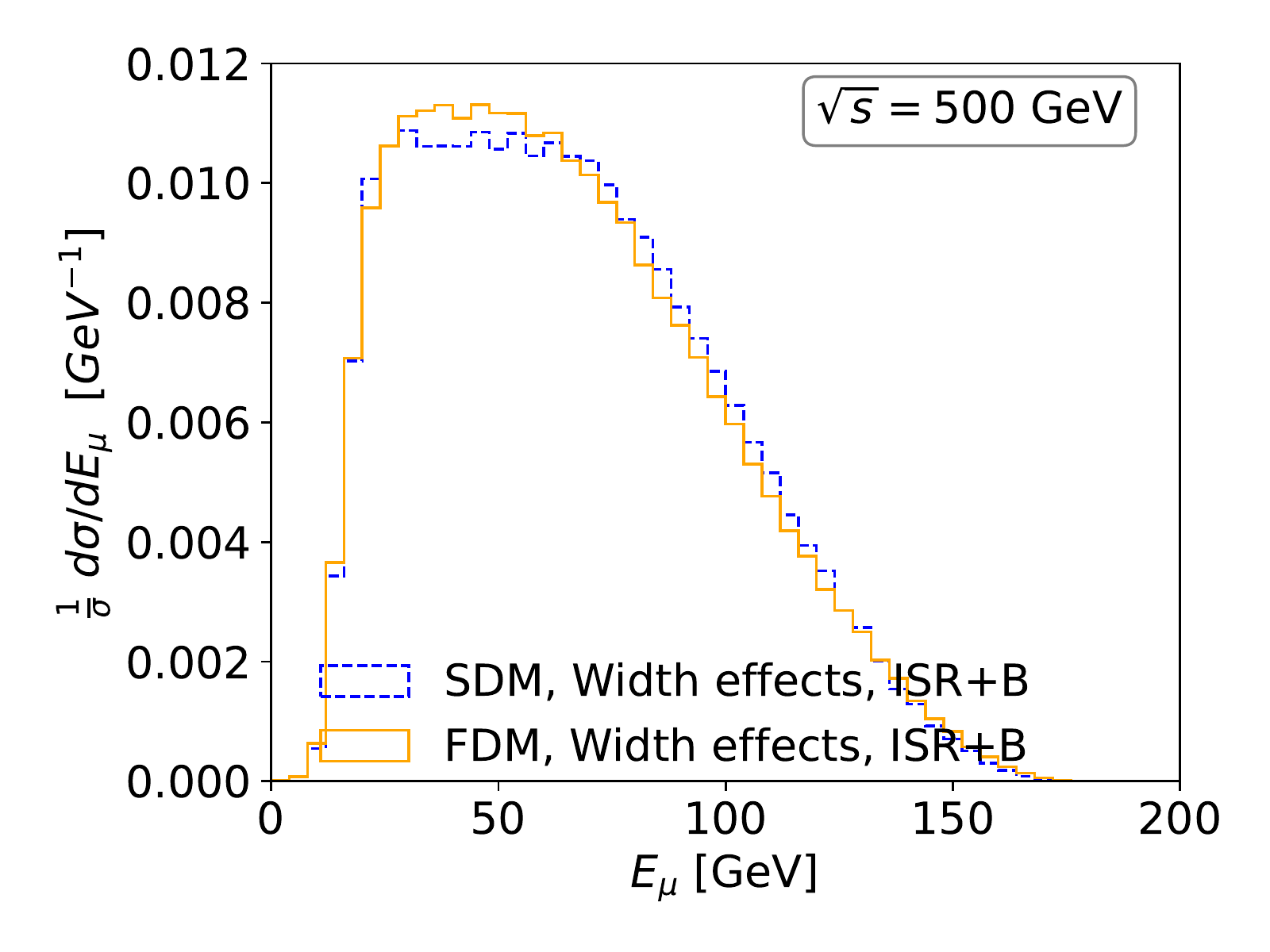}
		\includegraphics[width=0.49\textwidth]{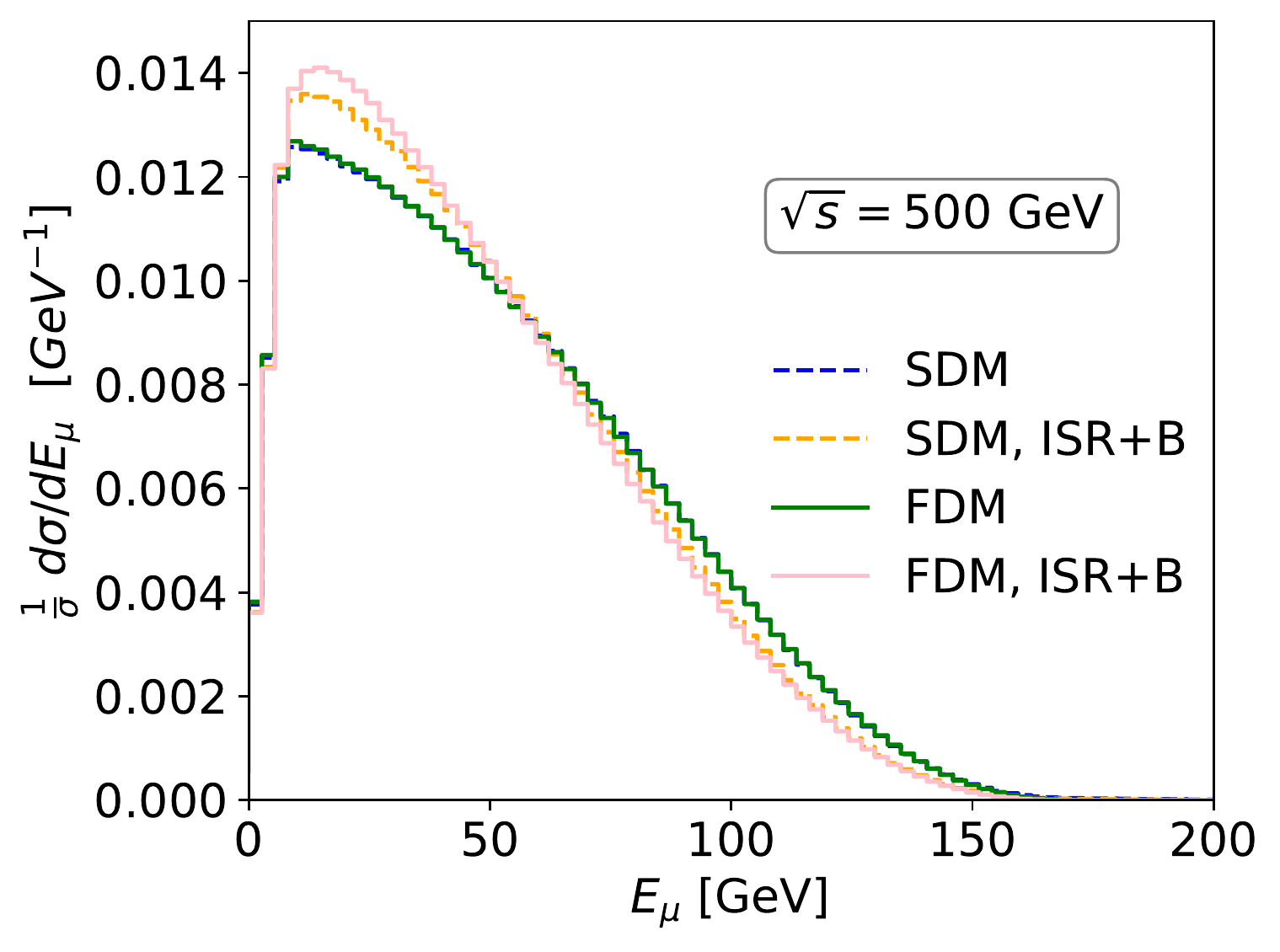}
		\caption{Comparing SDM and FDM normalised muon energy distributions for BP1 (left) and BP2 (right) values, including width and ISR+B effects.\label{fig:SDMvsFDM_Emu}} 
	\end{center}
\end{figure}

In Fig.~\ref{fig:SDMvsFDM_Emu}, we compare the normalised muon energy distributions for SDM and FDM, including all width and ISR+B effects.
Since positions of kinks are kinematically determined, it is not surprising that calculations for distinct models (containing different angular dependence) demonstrate variations in shapes, but do not perturb the position of kinks. We see that for FDM, $E_\mu^{(+)}$ is less well preserved than for SDM. Also, the higher energy tail demonstrates small difference in behaviour, however does not change the endpoint, $E_\mu^{(+)}$, required for measurement of mass. This small difference between overall shapes suggests that muon energy is not a good observable to differentiate between spins of DM, but conversely that it is a good observable for spin-independent measurements of mass. Note that spin correlations were taken into account from $2\to 4$ process.

The right plot in Fig.~\ref{fig:SDMvsFDM_Emu} shows the muon energy distribution for the off-shell $W$ decay case. 
In this case, $E_\mu^{(-)}$ is easily distinguishable between DM spins, including some differences in the shapes of the distribution tails. However, $E_\mu^{(+)}$ does not exist in this case, so one must rely solely on $E_\mu^{(-)}$ for mass and spin determination for the off-shell case. {In the next section we discuss the observable which can be used for DM spin determination.}

%\clearpage
%%%%%%%%%%%%%%%%%%%%%%%%%%%%
%%%%%%%%%%%%%%%%%%%%%%%%%%%%
\subsection{Angular distributions for Dark Matter spin discrimination}
\label{subsec:angular}
We have found that the remarkable observable, which can distinguish the spin of DM, is the angular distribution of $W$-boson with respect to the electron beam in the lab frame.
We found that $D^+$ angular distribution is determined by the spin of DM (and in case of $s$-channel SM vector mediators -- photon and $Z$-boson). For spin zero and spin one-half of DM:
\begin{equation} 
\frac{d\sigma}{d\cos{\theta_{D^\pm}}} \propto
\left\{
\begin{array}{lr}
1-\cos^2{\theta_{D^\pm}},& \text{for scalar } D^\pm\\
&\\
1 + \dfrac{s-4M_+^2}{s+4M_+^2}\cos^2{\theta_{D^\pm}},& \text{for fermion }D^\pm .
\end{array}	
\right.
\label{eq:dm-angle}
\end{equation}

\begin{figure}[htbp]\label{fig:AngDists}
	\begin{center}
		\includegraphics[width=0.5\textwidth]{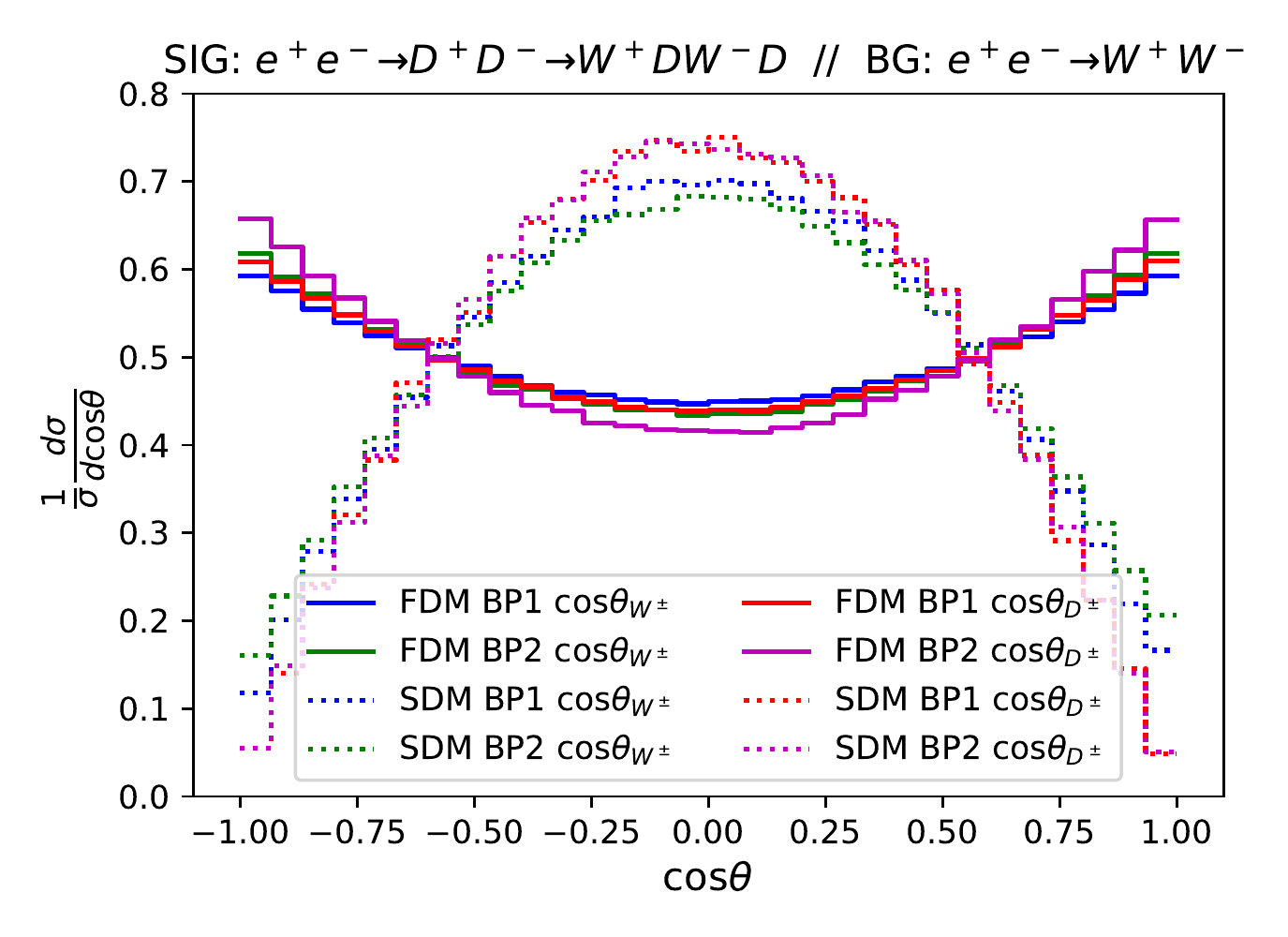}%
		\includegraphics[width=0.5\textwidth]{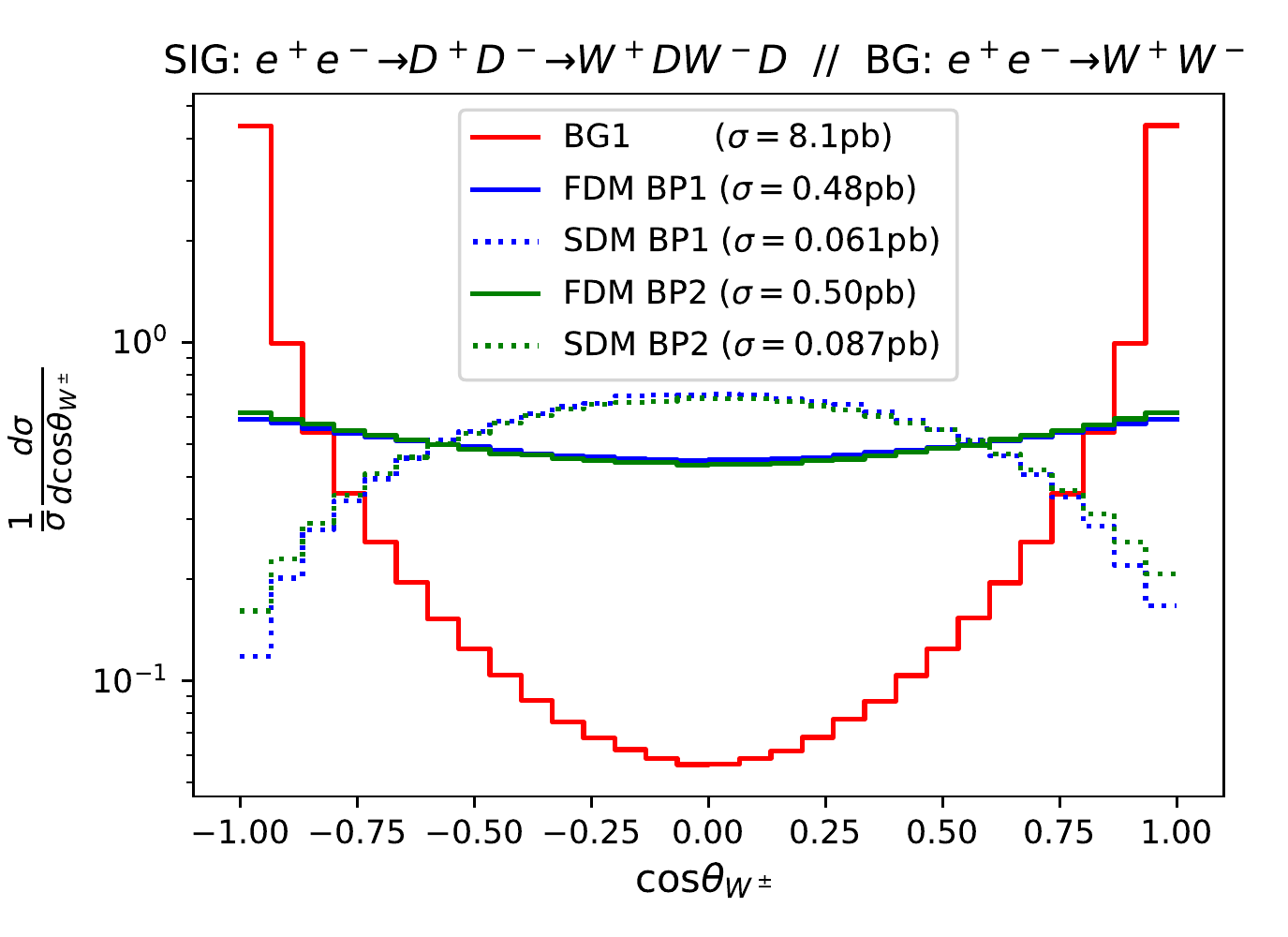}%
		\caption{Left:
			comparison of the  $W^\pm$ and $D^\pm$ angular distributions respect to beam direction in the lab frame.
			Right: the angular distribution of $W^\pm$ with respect to beam direction in the lab frame for signal and background processes.
			\label{fig:dm-angle} }
	\end{center}
\end{figure}

On the other hand, the angular distribution of $W$-boson from  $D^{\pm}$ decay is strongly correlated with $D^{\pm}$  one. 
This can be observed from Fig.~\ref{fig:dm-angle}(left), where we  present normalised angular distributions for $D^\pm$ and  $W^\pm$ for two benchmarks of the fermion and scalar DM cases.  One can see that the shapes of the $D^\pm$ distributions, given by Eq.\eqref{eq:dm-angle} determines the angular distribution of its decay product, $W^\pm$, whose angular distribution  is very close to its parent,  $D^\pm$.
While the (inverted parabolic) shape of $D^\pm$ angular  distribution is the same for different masses of DM in case of scalar DM,
the shape of angular distribution for the fermion DM case
has mild dependence on DM mass and in the extreme case of the $D^\pm$ production at the threshold it becomes flat, but still clearly distinguishable from the inverted parabolic shape of the angular distribution for the scalar DM case.

The different shapes of the distributions for DM with different spins have a very simple physical explanation. Since the mediator for $D^+D^-$ production is the spin-one SM vector bosons -- photons and $Z$-boson, only left-left(LL) or right-right(RR)  spin configuration for the initial $e^+e^-$ state is allowed. In case of scalar DM, the forward-backward  scattering of $D^+D^-$ pair is forbidden, since forward-backward $D^+D^-$ pair can not form orbital momentum equal to one to match the spin of the mediator.
This angular momentum conservation is reflected in $(1-\cos^2\theta)$ dependence of the angular distribution of the scalar $D^\pm$ particles.
At the same time, in case of fermion DM
the forward-backward scattering of $D^+D^-$ with their LL or RR final state
spin configuration (matching  spin one mediator) is naturally allowed, which is reflected in 
$\left(1+\dfrac{s-4M_+^2}{s+4M_+^2}\cos^2\theta\right)$ functional form  of the angular distribution of the fermion $D^\pm$ particles.

It is important  to stress that angular distributions
of $W^\pm$ are very close to $D^\pm$ ones for both  on-shell and off-shell $W$-boson cases. This makes the approach of distinguishing DM  with different spins  applicable to the whole model parameter space,
once $W$-boson (on-shell or off-shell) is reconstructed from the di-jet.
\begin{figure}[htbp]
	\begin{center}
		\includegraphics[width=0.55\textwidth]{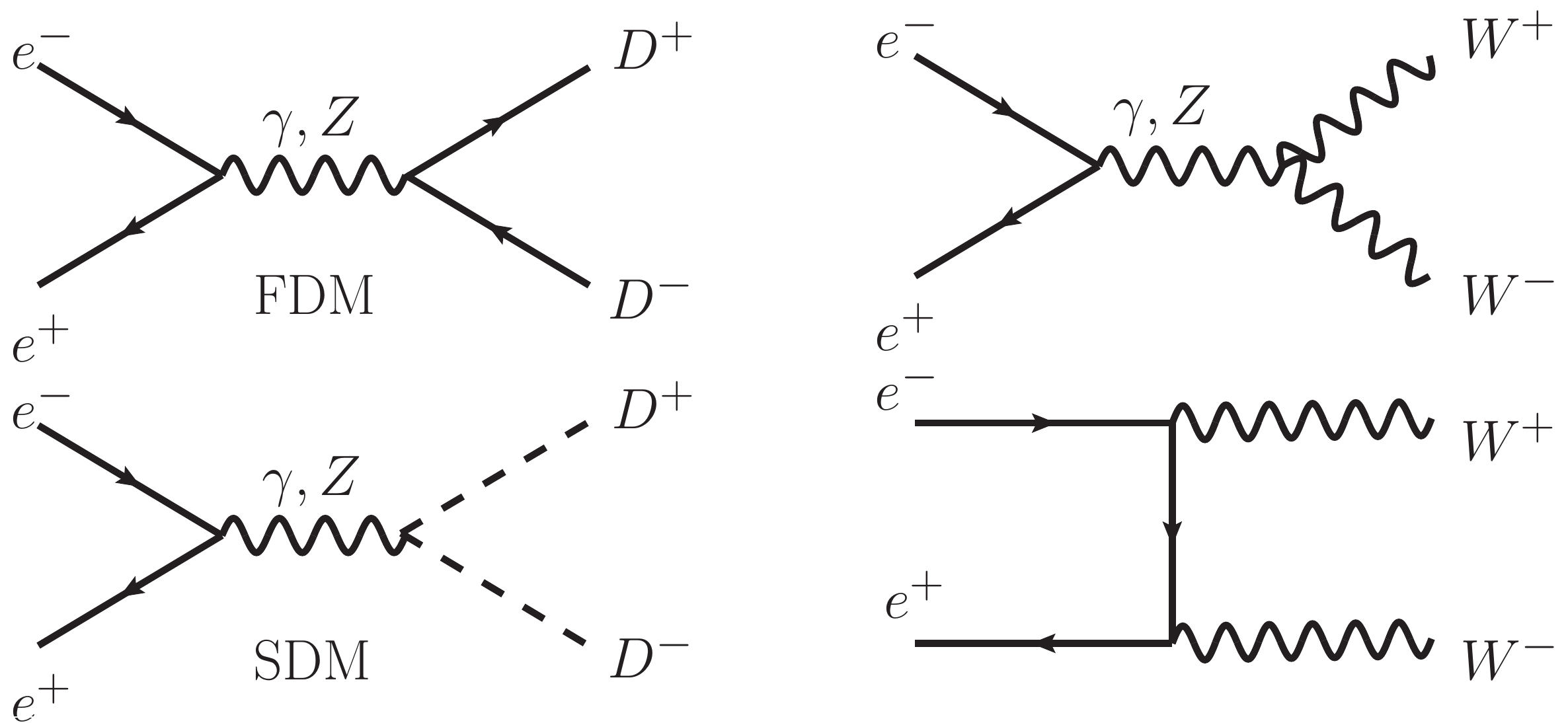}
		\caption{Signal (left) and background (right) diagrams
			\label{fig:diags}
		}
	\end{center}
\end{figure}

Another remarkable property of the signal angular distributions  for both spin zero and spin one-half DM is that they are very different from the the background.
This is demonstrated in Fig.~\ref{fig:dm-angle}(right),
where  we  present normalised angular distributions for $W^\pm$ for both benchmarks from both DM models
as well as  the leading $e^+e^- \to W^+ W^-$ background.
One can see that the background distribution has very pronounced forward-backward peak even  in comparison  with the distribution for  the fermion DM.
The reason for this is the $t$-channel diagram with the electron exchange for the
background, shown in Fig.~\ref{fig:diags}(right)
which plays an important role and provides the gauge invariance together with the s-channel $\gamma/Z$ diagram.
This is contrary to the  signal case, which  has only s-channel $\gamma/Z$ diagram (Fig.~\ref{fig:diags}(left))
for the $D^+D^-$. In case of an additional $t$-channel diagram
for sleptons production which could take place in case of Supersymmetry, the angular $D^\pm$ and respectively $W^\pm$
distributions will be still quite different from $W^+W^-$, as shown in~\cite{Christensen:2014yya}.

Therefore, the angular distribution of $W^\pm$ from $D^\pm$
is very powerful observable to discriminate the spin of DM. Moreover, as we have found in our study, this variable is the generic one for the whole parameter space of a given model and therefore it allows one to successfully distinguish signal models as well as background between each other as we demonstrate below.

%

%%%%%%%%%%%%%%%%%%%%%%%%%%%%
%%%%%%%%%%%%%%%%%%%%%%%%%%%%

%\input{04-analysis.tex}
\section{Signal versus background analysis and determination of Dark Matter mass and spin}

In this section, we study  various  background processes and suggest strategies to optimise the statistical significance of the signal as well as the signal to background ratio (which controls the impact of systematic errors). This analysis uses  the kinematical observables and distributions discussed in section~\ref{sec:signal}. The  "backgrounds'' discussed here are of two distinct types:

\begin{enumerate}[I]
\item SM backgrounds - reducible model-independent backgrounds, which influence observation of the model.
\item Model specific irreducible backgrounds - those which can potentially obscure precise measurement of the shape and kinematic features (as presented in section~\ref{sec:signal}) of signature (\ref{eq:sign-WW-enjj}).
\end{enumerate}

Backgrounds of type-II 
have the same final state particles as the signal process $\epe \to D^+D^- \to
W^+ W^- DD$. Since the aim of the analysis is to accurately measure the masses of the models and determine the spin of DM, our strategy is to minimise both types of background without significantly distorting the shape of the signal. In this section we perform signal versus background optimisation at the  fast detector simulation level and present results demonstrating the possibility of mass and spin determination of  DM which is the main aim of this paper.

\subsection{Background processes}\label{sec:backgrounds}
%%%%%%%%%%%%%%%%%%%%%%%%%%%%%%%%%%%%%%

In this study we focus on the signal process\eqref{eq:sign-WW-enjj} 
with  the respective 
``{\bf di-jet $+$ ($ e$ or $\mu$)
	+  $\MET$}" signature.
The total cross section for 
background processes  providing this signature
is $\sim 10-100$ times more  than that  of the signal, therefore
we explore kinematical distributions and optimise kinematical cuts to maximise signal significance and improve signal to background ratio.
\\
\\
\noindent There are several backgrounds contributing to this signature which  we include in our analysis, such as:
\begin{itemize}
	\item[\textbf{BG1}]
The process $\epe\to W^+W^-$. This is  by far the dominant background {and is of type-I}. The cross section of this process itself is one (two) orders of magnitude higher than the 
signal from fermion(scalar) DM, as was discussed earlier and illustrated  in Fig.~\ref{fig:signal-xsec}.
There are several kinematical observables which allow to suppress BG1, for example:
\begin{itemize}
	\item[(a)] the energy of each dijet for BG1 is $E_{jj}=\frac{\sqrt{s}}{2}$, while for the signal 
   $E_{jj}$ will be below $\frac{\sqrt{s}}{2}-M_D$, 
   so suitable cut on $E_{jj}$ should strongly suppress BG1.
   The main obstacles for this cut to work perfectly are the 
   ISR+B effects as well as the effect of the detector energy smearing as we demonstrate below.	
   \item[(b)] the missing mass, $M_{miss}$,
   is zero for BG1 for the ideal detector and no ISR+B effects.
   If this would be the case, then $M_{miss} > 2M_D$ would remove BG1 completely since, for the signal, the minimal value of  $M_{miss}$ is $2 M_D$.
   This is not the case, as we know, 
   therefore taking into account ISR+B effects as well as realistic detector resolution is crucial for this study.
 \end{itemize}
   One should note, however, that in spite of  ISR+B and detector  effects, both kinematical variables are very 
efficient for BG1 suppression.

%%%%%%%%%%%%%%%%%%%%%
\item[\textbf{BG2}]
A type-II background. The process
\begin{equation}
	\epe\to W^- D D^+ \to W^- W^+ D D
\end{equation}
which has just one $D^\pm$ in the intermediate sate leads to the same final state as the signal process.
To simplify the discussion, we detail the case where $M_{D_2}>M_+$ only(as is the case with the benchmarks analysed here).
The contribution of this process is at least  $\alpha_W$ times less then that of the signal process since it is genuine
$2\to 3$ process.
The interference of the BG2 process  with the signal  is also relatively small since this interference is proportional to the small $D^\pm$ width.
The overall  contribution of BG2 to the total BG is below  1\%.

\item[\textbf{BG3}]
A type-II background. The process
\begin{equation}
\epe\to DD_2\to DD^+W^-\to DDW^+W^-
\end{equation}
could  also contribute to the signature under study.
 This background is absent if $M_{D_2}<M_+$ or $M_{D_2}+M_D>\sqrt{s}$.
If the rate of this process is large enough, it will also be observable via $\epe\to DDZ$ final state~\eqref{DDAZ}. The cross section $\sigma(\epe\to DD_2)$ is of the same order as $\sigma(\epe\to D^+D^-)$ in general but can be suppressed in case of FDM by the small value of singlet-doublet mixing when $M_{D_2}-M_+$ mass split is small (which is the case of our benchmarks dictated by the DM DD constraints). The kinematics of this background is quite different from the signal: all visible particles follow  $D_2$ direction and therefore will be mostly  in just one hemisphere in
contrast to the signal process. Therefore, the contribution of this
background process may be suppressed  by
application of suitable respective cuts.
Moreover, in case $M_{D_2} < M_+ + M_W$ (the case of our benchmarks) this background comes from genuine $2\to 3 $ process
($\epe \to DD^+W^-$) with the intermediate $D_2$ and therefore 
it is suppressed by $\alpha_W$ in comparison to the signal under study.

\item[\textbf{BG4}]
A type-I background. There could be an additional pure SM BG process
\begin{equation}
	\epe\to W^+ W^- Z
\end{equation}
contributing to the signature under study,
with  large $\MET$ and $M_{miss}$  carried
away by  neutrinos from $Z$-boson produced in association with 
$W^+ W^-$ pair.
 The corresponding cross section is suppressed  at least  by  $\alpha_W$
 in comparison to BG1.

\end{itemize}

\subsection{Signal versus background analysis and ILC discovery potential}

Here we present the signal versus background analysis and ILC discovery potential for DM  at nominal integrated luminosity of 500$fb^{-1}$ (expected at end of Run 1). For our analysis we use tools and  setup  discussed in Section~\ref{sec:setup}.

\begin{figure}[H]
	\hspace*{-0.0cm}\includegraphics[width=0.50\textwidth]{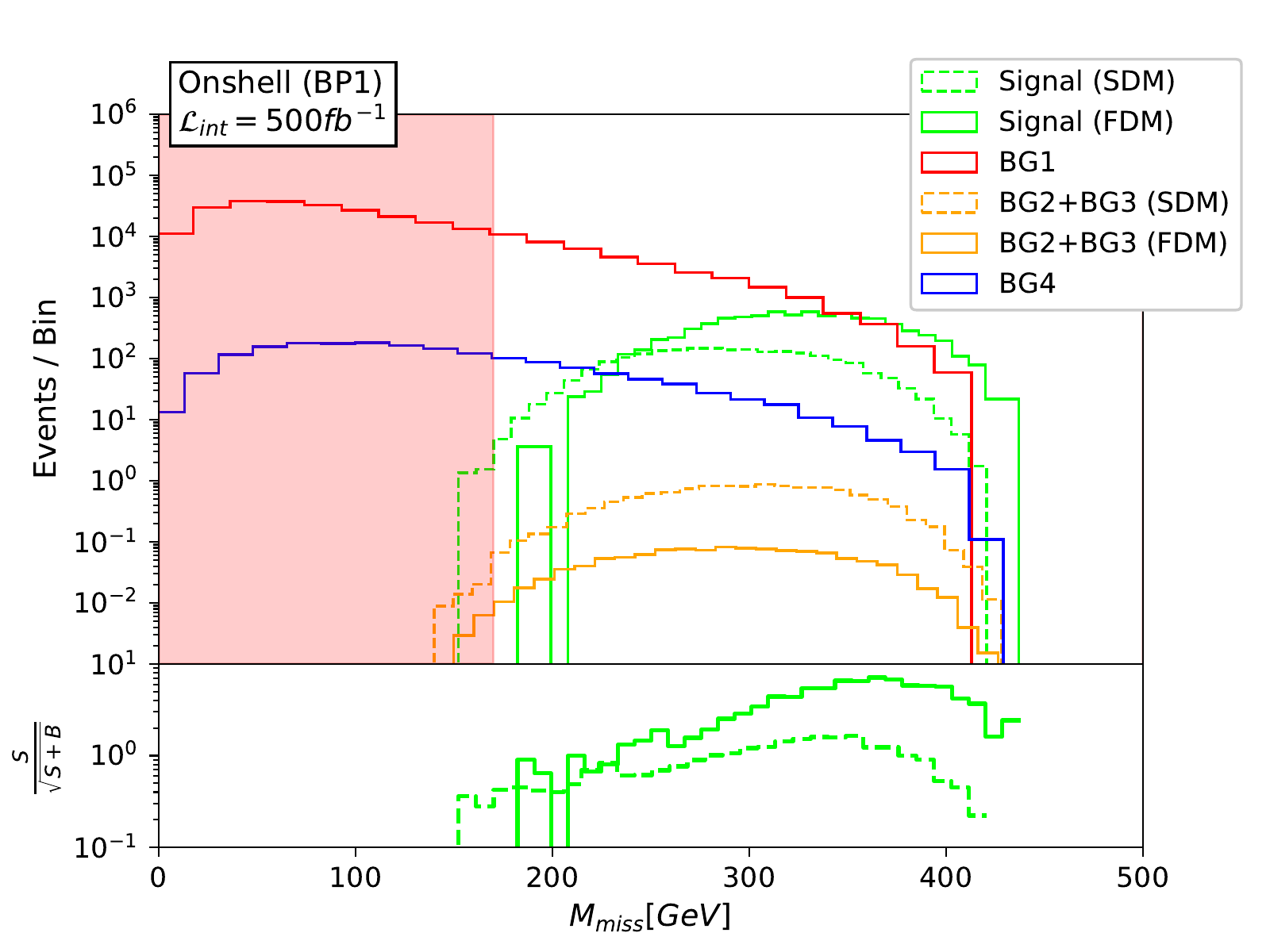}%
	\hspace*{-0.3cm}\includegraphics[width=0.50\textwidth]{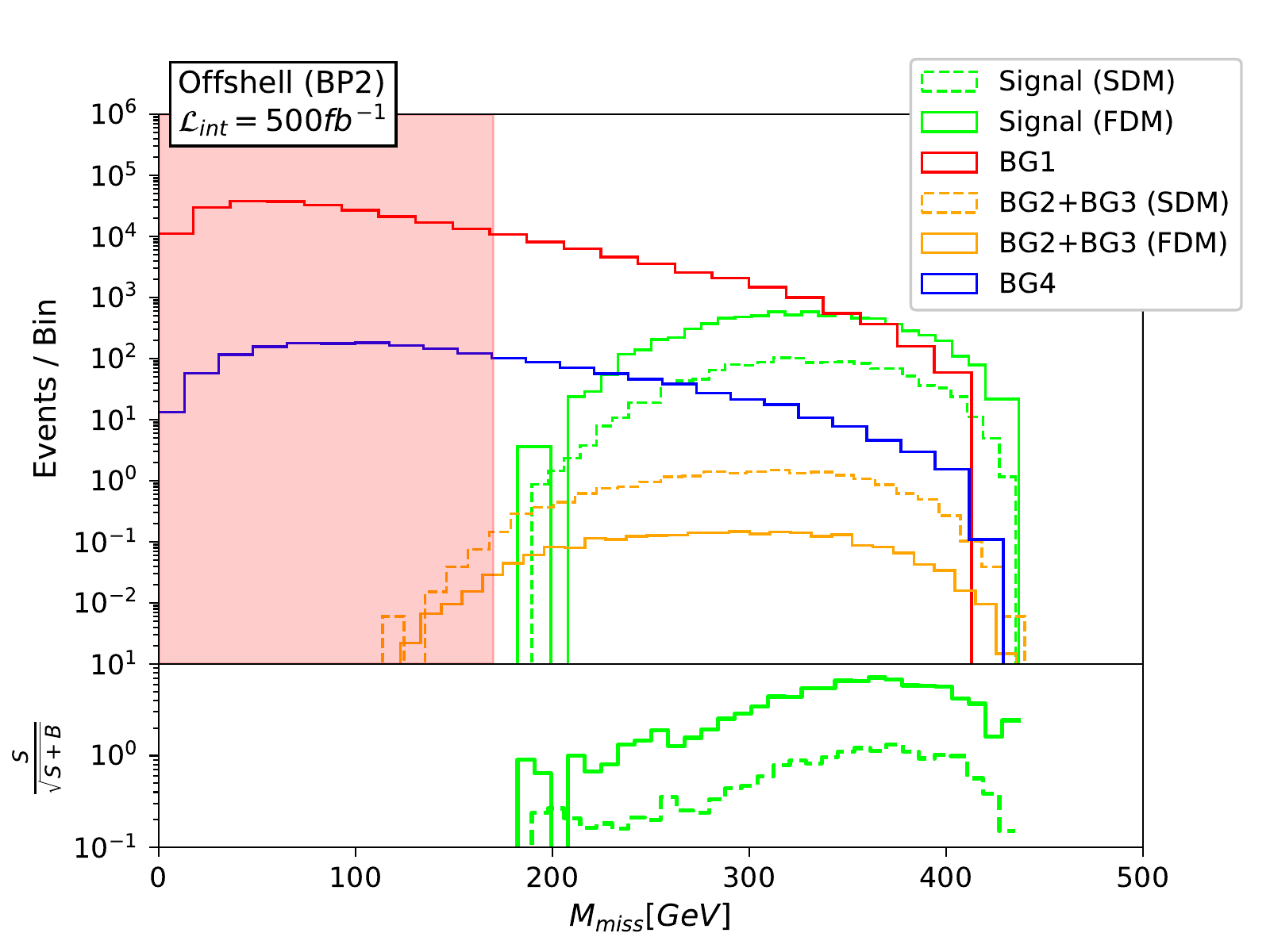}
	\caption{Missing mass at detector level for the signal and background processes in both SDM and FDM, for BP1 (left) and BP2 (right).}
	\label{figMmissCut}
\end{figure}

\begin{figure}[H]
\hspace*{-0.cm}	\includegraphics[width=0.50\textwidth]{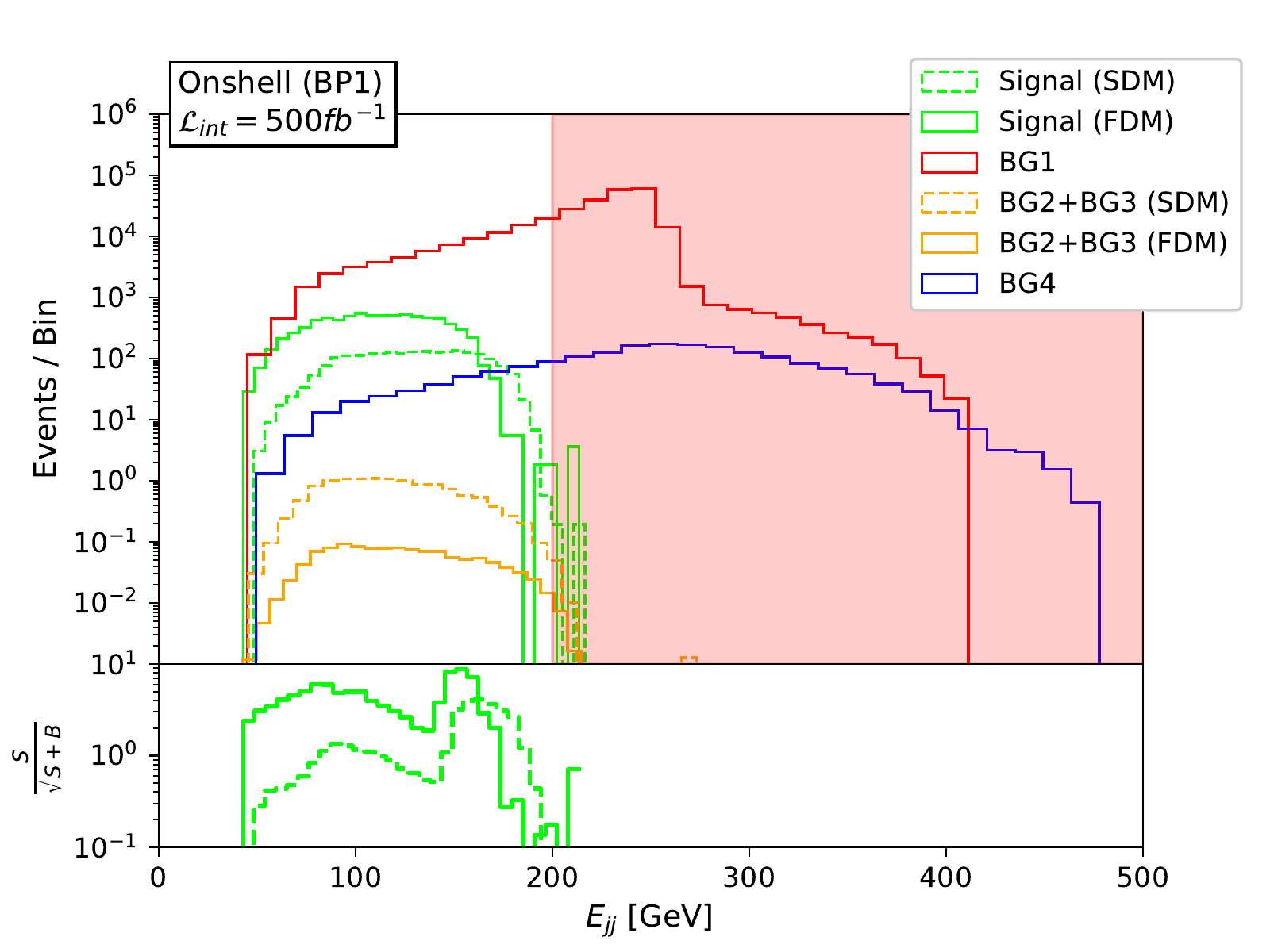}%
\hspace*{-0.3cm}	\includegraphics[width=0.50\textwidth]{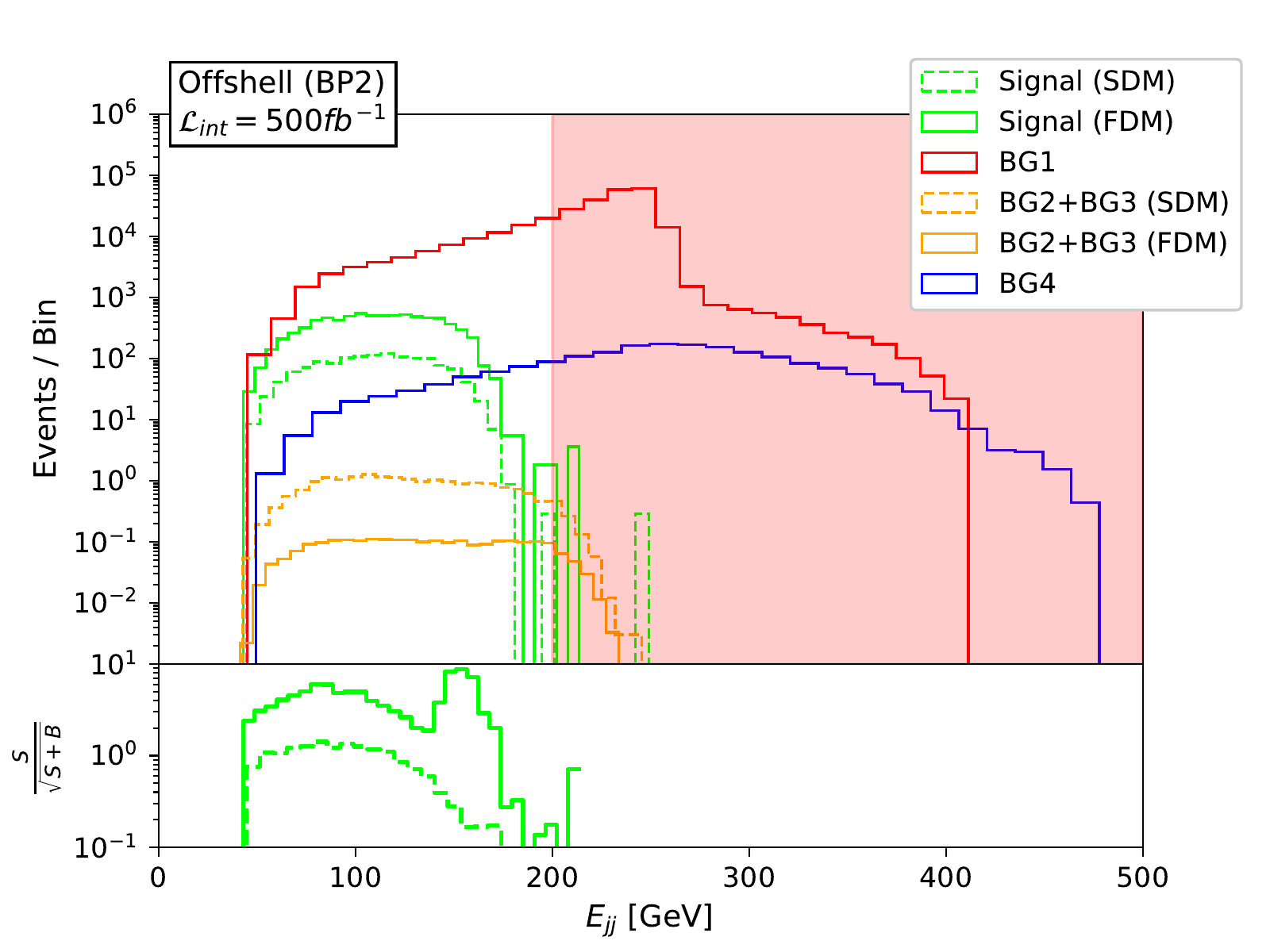}
	\caption{Energy of $W$ boson reconstructed from dijet at detector level.}
	\label{figEjjCut}
\end{figure}

\begin{figure}[H]
\hspace*{-0.0cm}\includegraphics[width=0.50\textwidth]{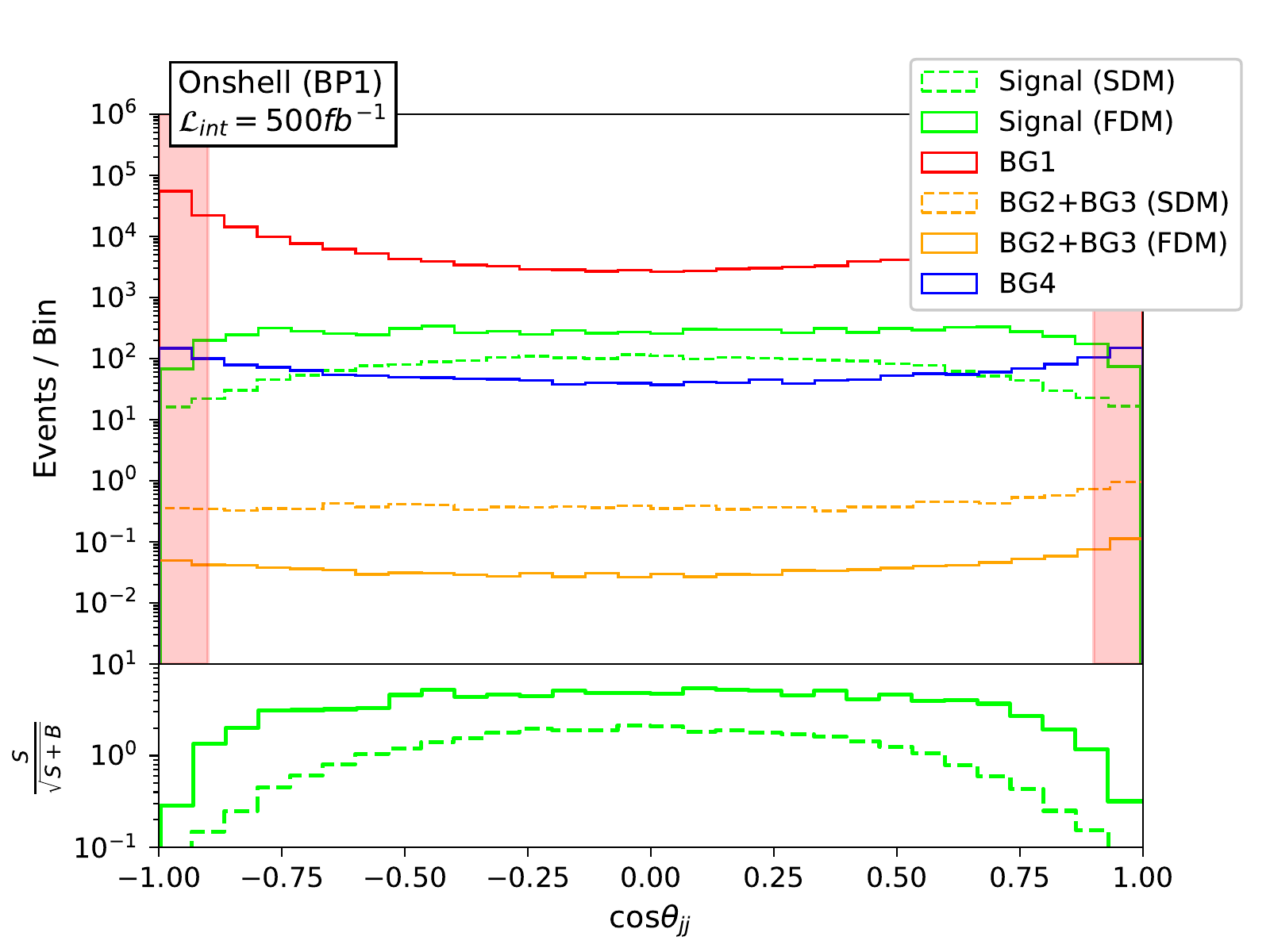}	
\hspace*{-0.3cm}\includegraphics[width=0.50\textwidth]{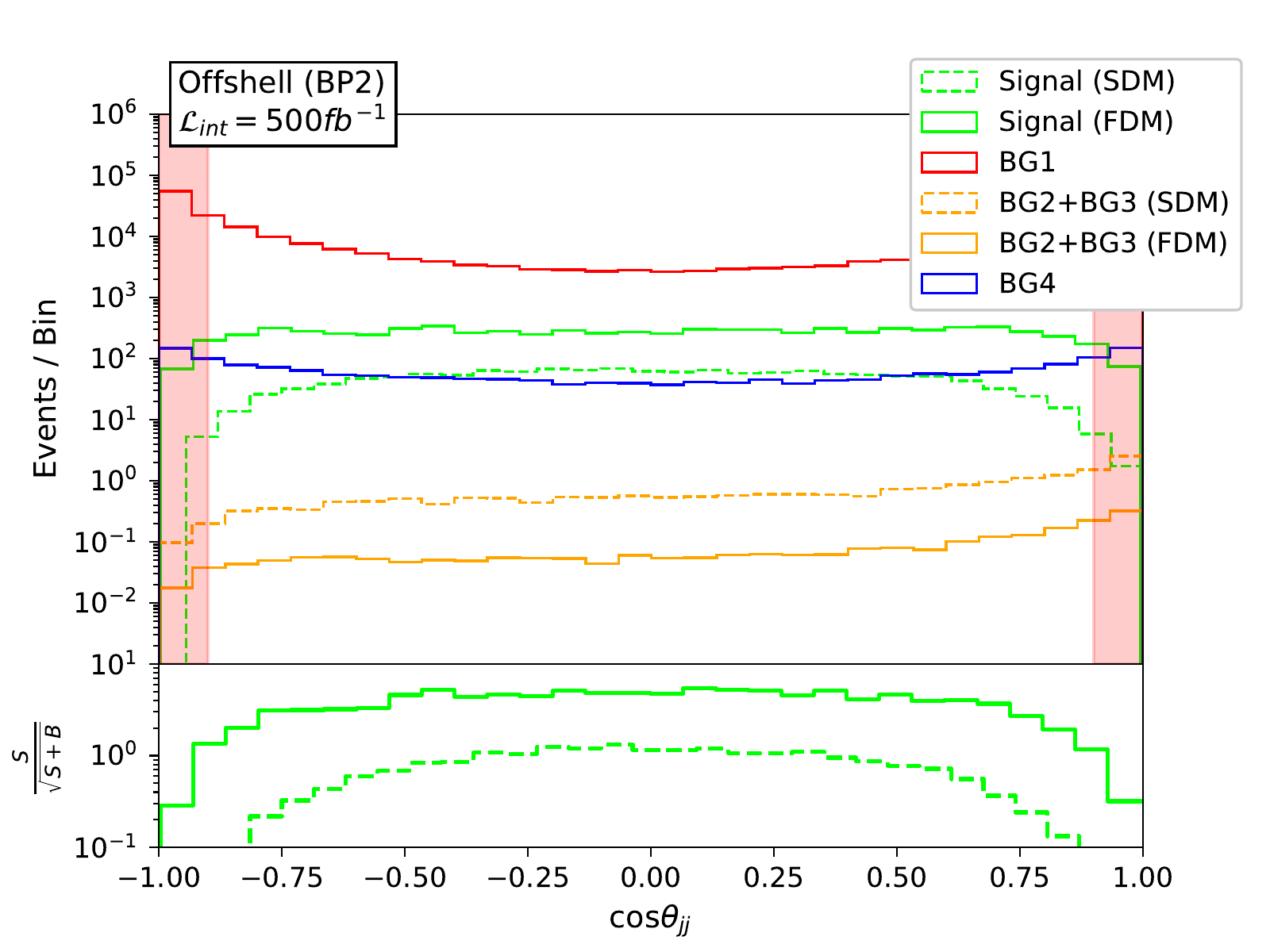}	
\caption{Scattering angle of $W$ as reconstructed from dijet at detector level.}
\label{figCosThetaCut}
\end{figure}

In Figs.\ref{figMmissCut},\ref{figEjjCut} and \ref{figCosThetaCut} we present the key signal and background 
distributions at fast detector simulation (Delphes) level for $M_{miss}$, $E_{jj}$ and $\cos\theta_{jj}$
respectively for both benchmarks -- BP1(left) and BP2(right).

One can see that missing mass, $M_{miss}$, given by Eq.~\eqref{Mmiss} has a an expected peak at low values as well as  a long tail towards the large values (where the signal ``lives") due to the ISR+B and detector energy smearing effects.
These effects are crucial for the correct BG estimation
and obviously should be taken into account.
In the region of large  $M_{miss}$
due to the these effects the dominant BG1 is non-negligible -- it is comparable to FDM signal
and about one order of magnitude above the SDM signal.
As we mentioned above without these effects BG1 would be 
simply a delta function at zero and could be trivially removed.
In the low panel of Figs.\ref{figMmissCut}--\ref{figCosThetaCut} we also present  $S/\sqrt{S+B}$ distribution to give an idea about the statistical signal significance  before cuts application.
The di-jet energy distribution (Fig.~\ref{figEjjCut}) of BG1 and BG4 exhibits a longer tail towards  higher values of $E_{jj}$ than the signal, as a result of a $t$-channel process mediated by an electron neutrino. These are detector level distributions, where we have applied an initial veto requiring at least 2 jets and a single muon. 
Besides this longer tail,  BG1 and BG4 peak at higher values than the signal because these backgrounds do not contain DM pair in the final state. One can see that the difference between the signal and BG1 or BG4 expected at the parton level is preserved also at the detector level.
One can also see that the shape of $\cos\theta_{jj}$ distribution (Fig.~\ref{figCosThetaCut}) discussed 
earlier at the parton level in section~\ref{subsec:angular} is also preserved at the detector level which brings an excellent potential to 
discriminate the DM spin.
One should note that model-dependent BG2+BG3 can be safely neglected even for the case of SDM, for which
these backgrounds are one order of magnitude higher than for FDM (due to the small singlet-doublet mixing effect
in case of FDM).
In the  analysis from now on we study 
``{\bf di-jet + $\mu$	+  $\MET$}" signature
and quote the respective  numbers for the event rates. 
Based on the properties of the signal and background distributions we propose simple kinematical cuts
which are indicated in Figs.\ref{figMmissCut}--\ref{figCosThetaCut}
by red shaded regions.
The corresponding cut-flow analysis
are presented in Tables \ref{tab:cutflowTable2} and \ref{tab:cutflowTable}.
 Table \ref{tab:cutflowTable2}
 presents details of the evaluation of 
SM  $B_I$ backgrounds and efficiency of the cuts,
while Table \ref{tab:cutflowTable}
presents cut-flow analysis of the signal (S) as well as
background $B_{II}$ which actually becomes part of the signal since it comes from the non-SM diagrams.
This table  also presents the   S/B ratio 
which is equal to $(S+B_{II})/B_I$
as well as the  signal
significance
\begin{align}
	\alpha(\delta_{sys}) &= \frac{S}{\sqrt{S+B}+\delta_{sys}(S+B)} \ \ ,
	\label{eq:alpha-sys}
\end{align}
which  includes statistical and systematic uncertainty of the signal.
The numbers are presented 
for  the 500$fb^{-1}$ integrated luminosity
and for two values of systematic uncertainty,
$\delta_{sys}=0$ and $\delta_{sys}=0.01$.
The starting point of Tables ~\ref{tab:cutflowTable2} and \ref{tab:cutflowTable} is the reconstruction level
(Reco Level) after Delphes simulation for which  require at least two jets with $p_T^j>20$~GeV and one muon 
with $p_T^\mu>20$~GeV.
\begin{table}[!htb]
	\begin{center}
			\begin{tabular}{l|llll}
				\hline
				\multicolumn{5}{c}{\bf \large SM BG cut flow} \\ 
				\cline{1-5}
				Cut &BG1 &BG4 &$B_I$ &$\eps_{B_{I}}$\\
				\hline
               Parton Level &   $6.600 \times 10^{5}$ & $1.947 \times 10^{4}$ & $6.795 \times 10^{5}$ &   --- \\
               Reco Level & $2.921 \times 10^{5}$ & $1.842 \times 10^{3}$ & $2.939 \times 10^{5}$ & 0.433 \\
               $M_{miss}>170$ & $4.053 \times 10^{4}$ & $4.881 \times 10^{2}$ & $4.101 \times 10^{4}$ & 0.140 \\
               $E_{jj}<200$ & $3.718 \times 10^{4}$ & $2.993 \times 10^{2}$ & $3.748 \times 10^{4}$ & 0.914 \\
               $|\cos{\theta_{jj}}|<0.9$ & $1.902 \times 10^{4}$ & $2.332 \times 10^{2}$ & $1.925 \times 10^{4}$ & 0.514 \\
               $|\cos{\theta_{\mu}}|<0.9$ & $1.456 \times 10^{4}$ & $1.981 \times 10^{2}$ & $1.476 \times 10^{4}$ & 0.767 \\
               \hline
			\end{tabular}%
		\caption{\label{tab:cutflowTable2}
			Cutflow for the SM BG (BG1 and BG4), which are BP independent.}
	\end{center}
\end{table} 
\begin{table}[!htb] 
	\begin{center}
	\resizebox{\columnwidth}{!}{%
\begin{tabular}{l|ccccccc|ccccccc}
	\multicolumn{15}{c}{\bf \large BP1 cut flow} \\ 
	\hline  
	 & & & &\textbf{SDM}   & & & & & & &\textbf{FDM}  & & &\\ \cline{2-15} 
\multirow{2}{*}{Cut} &\multirow{2}{*}{S}   &\multirow{2}{*}{$\eps_S$}   &\multirow{2}{*}{$B_{II}$}   &\multirow{2}{*}{$\eps_{B_{II}}$} &\multirow{2}{*}{$\dfrac{(S+B_{II})}{B_I}$}   &\multicolumn{2}{c|}{$\alpha(\delta_{sys})$}   &\multirow{2}{*}{S}   &\multirow{2}{*}{$\eps_S$}  &\multirow{2}{*}{$B_{II}$}   &\multirow{2}{*}{$\eps_{B_{II}}$}   &\multirow{2}{*}{$\dfrac{(S+B_{II})}{B_I}$}   &\multicolumn{2}{c}{$\alpha(\delta_{sys})$} \\
&&&&&&$\alpha(0)$&$\alpha(0.01)$&&&&&&$\alpha(0)$&$\alpha(0.01)$\\
\hline
               Parton Level & $4.519 \times 10^{3}$ &         --- &     16.55 &          --- &      0.007 &        5.464 &             0.589 & $3.556 \times 10^{4}$ &         --- &     1.540 &          --- &      0.052 &        42.06 &             4.448 \\
               Reco Level & $2.185 \times 10^{3}$ &       0.484 &     12.56 &        0.759 &      0.007 &        4.016 &             0.623 & $1.848 \times 10^{4}$ &       0.520 &     1.185 &        0.769 &      0.063 &        33.06 &             5.017 \\
               $M_{miss}>170$ & $2.182 \times 10^{3}$ &       0.999 &     12.52 &        0.996 &      0.054 &        10.50 &             3.411 & $1.845 \times 10^{4}$ &       0.999 &     1.174 &        0.991 &      0.450 &        75.67 &             22.01 \\
               $E_{jj}<200$ & $2.182 \times 10^{3}$ &       1.000 &     12.49 &        0.998 &      0.059 &        10.96 &             3.663 & $1.844 \times 10^{4}$ &       1.000 &     1.168 &        0.994 &      0.492 &        78.00 &             23.18 \\
               $|\cos{\theta_{jj}}|<0.9$ & $2.132 \times 10^{3}$ &       0.977 &     10.64 &        0.852 &      0.111 &        14.58 &             5.921 & $1.651 \times 10^{4}$ &       0.895 &     0.946 &        0.810 &      0.858 &        87.30 &             30.20 \\
               $|\cos{\theta_{\mu}}|<0.9$ & $2.027 \times 10^{3}$ &       0.951 &     9.587 &        0.901 &      0.138 &        15.65 &             6.816 & $1.542 \times 10^{4}$ &       0.934 &     0.851 &        0.899 &      1.045 &        88.77 &             32.43 \\

\hline
	\multicolumn{15}{c}{\bf \large BP2 cut flow} \\ 
\hline  
& & & &\textbf{SDM}   & & & & & & &\textbf{FDM}  & & &\\ \cline{2-15} 
\multirow{2}{*}{Cut} &\multirow{2}{*}{S}   &\multirow{2}{*}{$\eps_S$}   &\multirow{2}{*}{$B_{II}$}   &\multirow{2}{*}{$\eps_{B_{II}}$} &\multirow{2}{*}{$\dfrac{(S+B_{II})}{B_I}$}   &\multicolumn{2}{c|}{$\alpha(\delta_{sys})$}   &\multirow{2}{*}{S}   &\multirow{2}{*}{$\eps_S$}  &\multirow{2}{*}{$B_{II}$}   &\multirow{2}{*}{$\eps_{B_{II}}$}   &\multirow{2}{*}{$\dfrac{(S+B_{II})}{B_I}$}   &\multicolumn{2}{c}{$\alpha(\delta_{sys})$} \\
&&&&&&$\alpha(0)$&$\alpha(0.01)$&&&&&&$\alpha(0)$&$\alpha(0.01)$\\
\hline
               Parton Level & $4.519 \times 10^{3}$ &         --- &     16.55 &          --- &      0.007 &        5.464 &             0.589 & $3.556 \times 10^{4}$ &         --- &     1.540 &          --- &      0.052 &        42.06 &             4.448 \\
               Reco Level & $1.352 \times 10^{3}$ &       0.299 &     19.98 &        1.207 &      0.005 &        2.487 &             0.387 & $7.894 \times 10^{3}$ &       0.222 &     2.376 &        1.543 &      0.027 &        14.37 &             2.213 \\
               $M_{miss}>170$ & $1.352 \times 10^{3}$ &       1.000 &     19.81 &        0.992 &      0.033 &        6.566 &             2.147 & $7.894 \times 10^{3}$ &       1.000 &     2.328 &        0.979 &      0.193 &        35.70 &             11.11 \\
               $E_{jj}<200$ & $1.351 \times 10^{3}$ &       1.000 &     19.18 &        0.968 &      0.037 &        6.858 &             2.309 & $7.891 \times 10^{3}$ &       1.000 &     2.194 &        0.943 &      0.211 &        37.05 &             11.84 \\
               $|\cos{\theta_{jj}}|<0.9$ & $1.345 \times 10^{3}$ &       0.995 &     15.85 &        0.826 &      0.071 &        9.369 &             3.848 & $7.616 \times 10^{3}$ &       0.965 &     1.761 &        0.802 &      0.396 &        46.46 &             17.61 \\
               $|\cos{\theta_{\mu}}|<0.9$ & $1.308 \times 10^{3}$ &       0.973 &     14.71 &        0.928 &      0.090 &        10.32 &             4.551 & $7.262 \times 10^{3}$ &       0.954 &     1.614 &        0.917 &      0.492 &        48.94 &             19.70 \\                

\hline
\end{tabular}%
}
    \caption{\label{tab:cutflowTable}
    	Cutflow for BP1 (top) and BP2 (bottom) with efficiency and significances, $\alpha(\delta_{sys})$  for the 500 fb$^{-1}$ integrated luminosity.
    	 See details in the text.
    	    	}
    \end{center}
\end{table}
In this example of the   cut flow analysis 
we apply a simple set of cuts, which demonstrate that
cuts on  $M_{miss}$, $E_{jj}$, $\cos\theta_{jj}$
and $\cos\theta_\mu$ can be chosen such that their  cumulative efficiency  will be as high as about $90\%$ for the signal and only about 5-7$\%$ for the background. This is demonstrated in Tables~\ref{tab:cutflowTable2}  and \ref{tab:cutflowTable} where we present the relative efficiency of each cut after the consequent application one cut after another (see columns $\eps_S$ and $\eps_{B_I}$ for signal and background respectively), so their  cumulative efficiency
is equal to  their product. 

One should also note that the efficiency  of the 
initial (pre-cut) signal selection is about factor of two higher for  BP1 in comparison to BP2.
This is because the jets and muon from the off-shell $W$-boson decay in case of BP2 are eventually softer in comparison to the on-shell $W$-boson decay,
so the respective efficiencies for jet reconstruction
and muon identification with $p_T>20$ GeV are respectively lower for the BP2.

The kinematical cuts on the key kinematical variables listed in Tables~\ref{tab:cutflowTable2} and \ref{tab:cutflowTable} can be further optimised using various techniques, including 
multivariate cuts analysis, boosted decision trees analysis or even neural net analysis at more sophisticated level.
However, as one can see, we can achieve 
high enough significance even  without sophisticated set of cuts and therefore limit ourselves for  this study by  analysis based on simple cuts on key variables we have found.

In Table \ref{tab:DiscoverySignif} we present the luminosity required to observe ``{\bf di-jet + $\mu$	+  $\MET$}" signature  from  SDM and FDM models  at $5\sigma$ level
for zero and 1\% systematic error values,
denoted by $\alpha(0)$ and $\alpha(0.01)$ values.
We use the cut flow from  Tables~\ref{tab:cutflowTable2} and \ref{tab:cutflowTable}. The luminosity is calculated using the asimov data set, and as such represents an ``expected" luminosity for discovery. The number of signal and background events 
for the respective luminosities is large enough to assume the statistical distribution is approximately Gaussian. 

\begin{figure}[htbp]
\centering
	\includegraphics[width=0.66\textwidth]{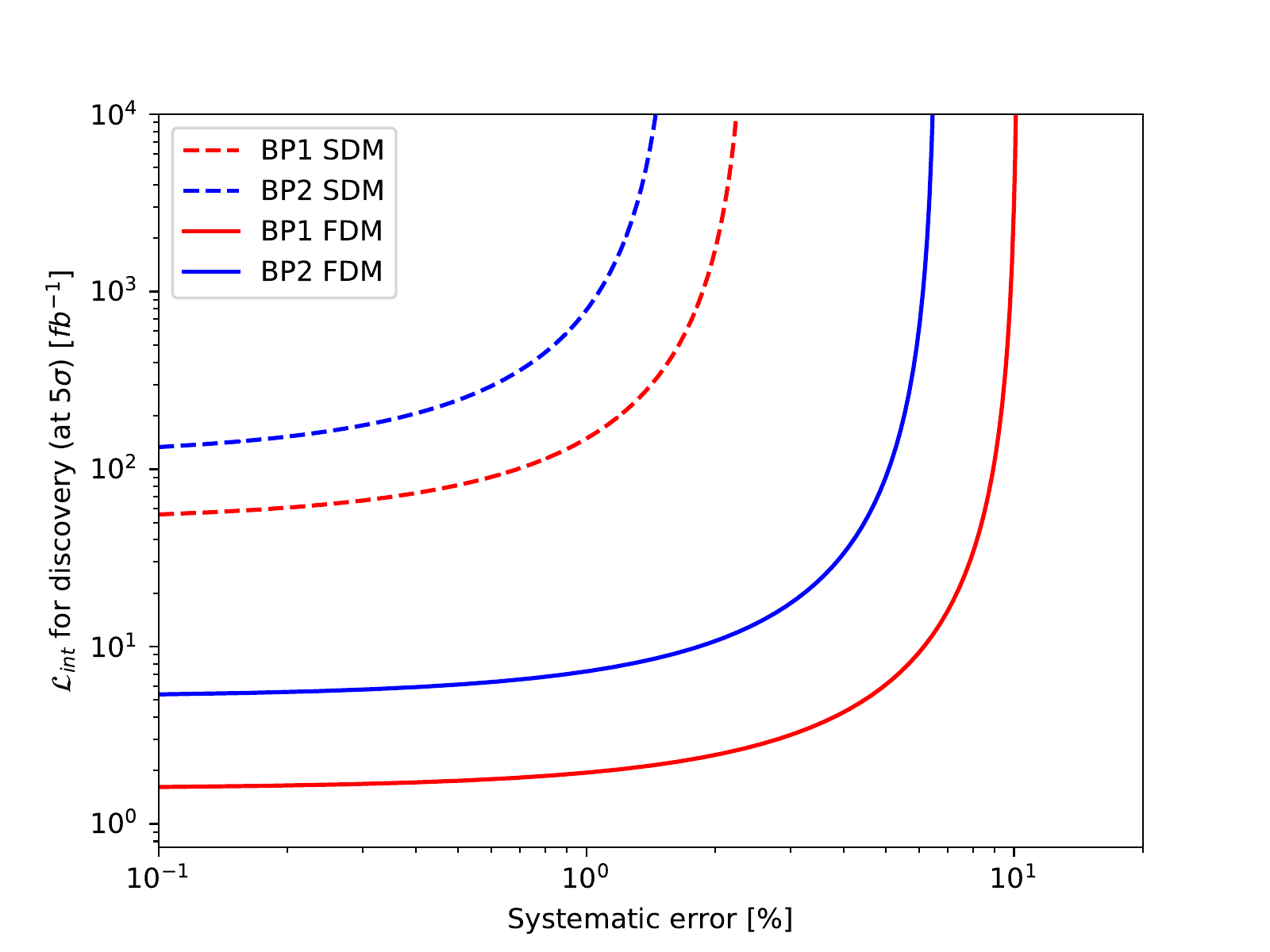}
\caption{Luminosity required for a $5\sigma$ DM signal excess above SM backgrounds 
as a function of  the systematic error percentage based on $\alpha(\delta_{sys})$ given by Eq.~\eqref{eq:alpha-sys}.}   
\label{fig:L_vs_SysErr}
\end{figure}

\begin{table}[htbp]
\centering
\begin{tabular}{|l|l|c|c|}
\hline
\multicolumn{2}{|c}{} &
\multicolumn{2}{|c|}{Luminosity required for discovery (at $5\sigma$)/$fb^{-1}$} \\ 
\cline{3-4}
\multicolumn{2}{|c|}{} &  $\alpha(0)$ & $\alpha(0.01)$                             \\ \hline
\multicolumn{1}{|l|}{\multirow{2}{*}{SDM}}  & BP1 & 51.1  & 149.     \\
                                           & BP2 & 117.   & 789.     \\ \hline
\multicolumn{1}{|l|}{\multirow{2}{*}{FDM}} & BP1 & 1.59   &1.95 \\
\multicolumn{1}{|l|}{}                     & BP2 & 5.21   &7.25  \\ \hline
\end{tabular}
\caption{Table demonstrating the expected luminosity (in fb$^{-1}$) required to observe a $5\sigma$ excess above SM backgrounds.    \label{tab:DiscoverySignif}}
\end{table}
In Figure \ref{fig:L_vs_SysErr} we present the luminosity required for discovery as the function of the value of  the systematic error
using  formula~\eqref{eq:alpha-sys} for $\alpha$.
From this plot, we see that if systematic errors are above  few percent, then discovery of the  SDM benchmarks  via the channel under study alone at ILC is problematic for simple cut-based analysis we use here.
However, various studies at future $\epe$ colliders~\cite{Bian:2015zha},~\cite{DeBlas:2019qco} including ILC
shows that the control of the systematic error for the leading $W^+W^-$ background 
at  1$\%$ level is quite realistic.

One should also add that the background can be further controlled and reduced using electron-positron beam polarisation.
We have checked that, for example, in case of 80\% polarisation of both electron(right) and positron(left)  beams the  background is reduced by about factor of 20 while the  signal from both models drops down by less than factor of two only. 

%%%%%%%%%%%% MASS %%%%%%%%%%%%%%%%%%%%%%%%%%%
%%%%%%%%%%%%%%%%%%%%% MASS 
\subsection{DM mass determination}\label{MassDetermination}
\subsubsection{Kinematic fitting}
In our analysis we  approximate shape of the muon energy distribution using a piecewise function. The functional form has power law dependence for the tail regions, and a constant for the plateau region between the two kinks, described by the following function: 
\begin{equation}\label{funcForms_BP1}
f(E_\mu) = 
\begin{cases} 
b \left(\dfrac{E_\mu}{E_\mu^{(-)}} \right)^a  &\mbox{if } E_\mu \leq E_\mu^{(-)} \\
\\
b & \mbox{if } E_\mu^{(-)} < E_\mu < E_\mu^{(+)}  
\\
\\
b\left(
1-\dfrac{E_\mu-E_\mu^{(+)}}{E_\mu^{max}-E_\mu^{(+)}}
\right)^c & \mbox{if } E_\mu^{(+)} \leq E_\mu < E_\mu^{max} 
\\
\\
0 & \mbox{if } E_\mu \geq E_\mu^{max}  \ \ \ ,
\end{cases}
\end{equation}
where $E_\mu^{(-)}$ and $E_\mu^{(+)}$ are positions of the left and right kinks of the muon energy distribution respectively.
The fit of  the muon energy distribution  for  BP1
using the  function~\eqref{funcForms_BP1} was made  for  the expected number of events, simulated at the Delphes level for each model. 
We have generated the pseudo-experimental data set  neglecting theory errors. This data set corresponds to the expected number of events in each bin from a large MC sample (often called the \textit{Asimov data set}), allowing us to evaluate the expected  resolutions  for $D$ and $D^+$masses. The discrepancy between the piecewise function~\eqref{funcForms_BP1} and the detector level distributions for each model results in a bias on the estimator. To capture this we include a methodological error (conservatively set to be 10\% of signal for each bin). We have found that this value of the methodological error gives  the  result consistent with the input mass and the error from  $\chi^2$ profile.

The number of signal+background events  is evaluated for realistic statistics, corresponding to the cross-section times luminosity for each process. Cuts outlined in Table \ref{tab:cutflowTable} are applied, except the 
cut on $\cos\theta_{jj}$ since  this cut smears $E_\mu^{(-)}$ considerably, whilst other cuts give approximately uniform modulation of the signal distribution.
{The profile $\chi^2$ is calculated by minimising over nuisance parameters $a,b,c$  
	as well as  $M_D$ and $M_+$ masses expressed 
	via $E_\mu^{(-)}$ ,$E_\mu^{max}$
	values  of the fit.}
{The mean collision energy after ISR+B was found to be
	477.78 GeV  from the simulation.
	This value was  used in the fit. As a result, we have found the minimum of this profiled $\chi^2$, corresponding to the global minimum of the fit, where  $M_D, M_+$ are also allowed to vary.}
\begin{figure}[htbp]
	\hspace*{-1.cm}\includegraphics[width=1.1\textwidth]{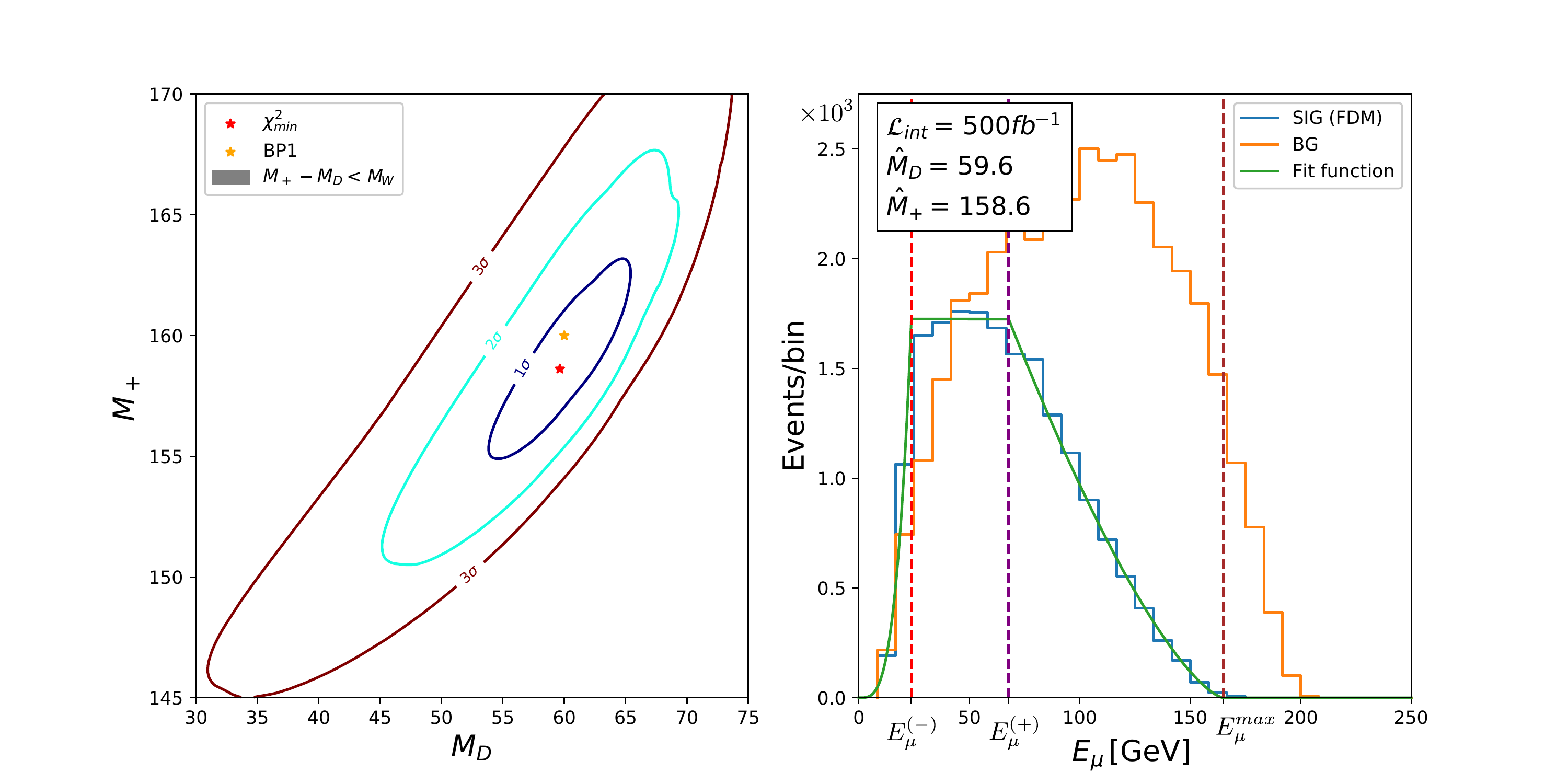}
	\caption{Profile $\chi^2$ value for kinematic fitting of BP1 (left) and muon energy distribution with best fit (right), for FDM.}
	\label{fig:mass-fit-BP1-fdm}
\end{figure}

\begin{figure}[htbp]
	\hspace*{-1.cm}\includegraphics[width=1.1\textwidth]{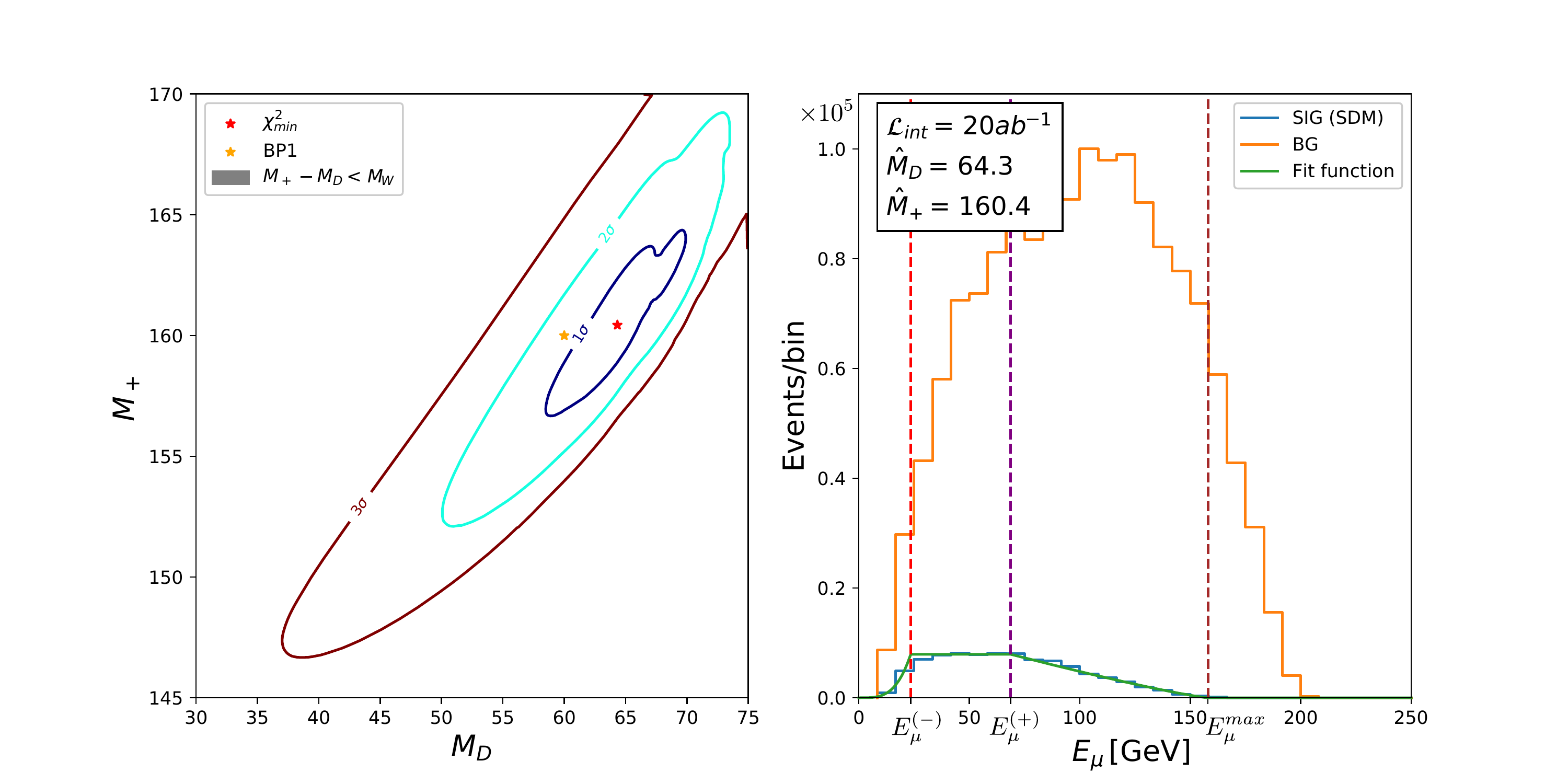}
	\caption{Profile $\chi^2$ value for kinematic fitting of BP1 (left) and muon energy distribution with best fit (right), for SDM.
		\label{fig:mass-fit-BP1-sdm}}
\end{figure}

{The result of this fit is illustrated in Figs.~\ref{fig:mass-fit-BP1-fdm} and \ref{fig:mass-fit-BP1-sdm}
	for FDM (500 fb$^{-1}$) and SDM (20 ab$^{-1}$) cases respectively. The right panels of these figures present $E_\mu$ distributions for signal and background
	as well as piecewise functions~\ref{funcForms_BP1} determined 
	from the fit together with the $E_\mu^{(-)},E_\mu^{(+)}$  parameters which have determined 
	the values of  $M_D$ and $M_+$.
	The left panels of the Figs.~\ref{fig:mass-fit-BP1-fdm} and \ref{fig:mass-fit-BP1-sdm} present the 1-2-3-$\sigma$ contours in $M_D$--$M_+$ plane from  the  likelihood  variation at the respective confidence levels
	which allows us to  determine  the errors on the $M_D$ and $M_+$
	masses.
	Although this shape fitting procedure is less sensitive than template fitting (which we discuss in the next section for BP2 case), it has the key advantage of being model independent. 
	The example of the numerical output of the fit is given in Table~\ref{tab:mass-fit-BP1}, which presents the values
	of the $M_D$ and $M_+$ determined from the fit as well as the 
	accuracy of their measurement.
	From this table one can see that the input values of $M_D$ and $M_+$ -- 60 and 160 GeV respectively -- are consistent with the 
	fit and its accuracy, which is a good cross check of the approach we are using. One can also observe that 
	the accuracy of the FDM mass determination  at  500 fb$^{-1}$ is similar to that of the SDM at  20 ab$^{-1}$.  The determination of SDM masses 
	at 500 fb$^{-1}$ is quite problematic using the simple set of cuts and piecewise function fitting method which requires
	approximately 40 times more luminosity to
	obtain comparable precision for SDM  as for  FDM model. 
	
	\begin{table}[htbp]
		\centering
		\begin{tabular}{llll}
			\hline
			&                            & $500 fb^{-1}$               & $20 ab^{-1}$          \\ \hline
			\textbf{FDM} & \multicolumn{1}{l|}{$M_D$} & $58.4^{+5.7}_{-6.0}$        & $57.6^{+ 1.9}_{-2.2}$ \\
			& \multicolumn{1}{l|}{$M_+$} & $158.1^{+4.0}_{ - 3.7}$     & $157.4^{+2.7}_{-2.4}$   \\ \hline
			\textbf{SDM} & \multicolumn{1}{l|}{$M_D$} & $66.0 ^{+19.2}_{-64.3}$ & $64.3^{+3.2}_{-6.1}$  \\
			& \multicolumn{1}{l|}{$M_+$} & $161.3^{+14.7}_{-52.8}$     & $161.0^{+3.3}_{-3.9}$ \\ \hline
		\end{tabular}
		\caption{Mass resolutions for BP1 kinematic fitting procedure.
			\label{tab:mass-fit-BP1}}
	\end{table}

	\subsubsection{Template fitting}
	For off-shell $W$-bosons in case of 
	``{\bf di-jet + $\mu$	+  $\MET$}" signature
	for BP2, there are no longer clear kinematic edges in $E_\mu$ distribution. Therefore, in this case we have employed  the  Monte Carlo template fitting method. 
	We have produced a 13x13 grid  of   muon energy  distributions 
	for $M_D=168.0-170.0$ GeV  and $M_+ = 118.0-122.0$ GeV ranges. Then, using a 2D linear interpolation in $(M_D, M_+)$ space we have  calculated the $\chi^2$ contour presented in Figs.~\ref{chi2_FDM_BP2} and \ref{chi2_SDM_BP2} 
	together with the result of the fit.
	Although this strategy yields better accuracy, information about the overall normalisation of the distributions -- the cross section -- was used, making this approach more model dependent. 
	Indeed, from Table~\ref{tab:mass-fit-BP2} which presents details of the template fit of BP2 parameters,
	one can see this method provides an order of magnitude better accuracy in case of FDM in comparison to BP1 piecewise function fit. At the same time, the  template fit of BP2 parameters in case of SDM provides similar accuracy to that of  BP1 piecewise function fit.
	The reason for this is quite simple:
	when using the template fit, we do not assume
	10\%  methodological error contrary to  the case of piecewise function fit. In case of FDM this error is dominant, that is why, template fit which does not have such an uncertainty improved the accuracy of the mass measurement of the fermion DM.
	At the same time in case of SDM, because of its lower signal, the statistical uncertainty is comparable to the  methodological error and therefore when we exclude the later one in template fit, we do not achieve a big effect on improving of the accuracy of the mass measurement.

	\begin{figure}[htbp]
		\hspace*{-1.cm}\includegraphics[width=1.1\textwidth]{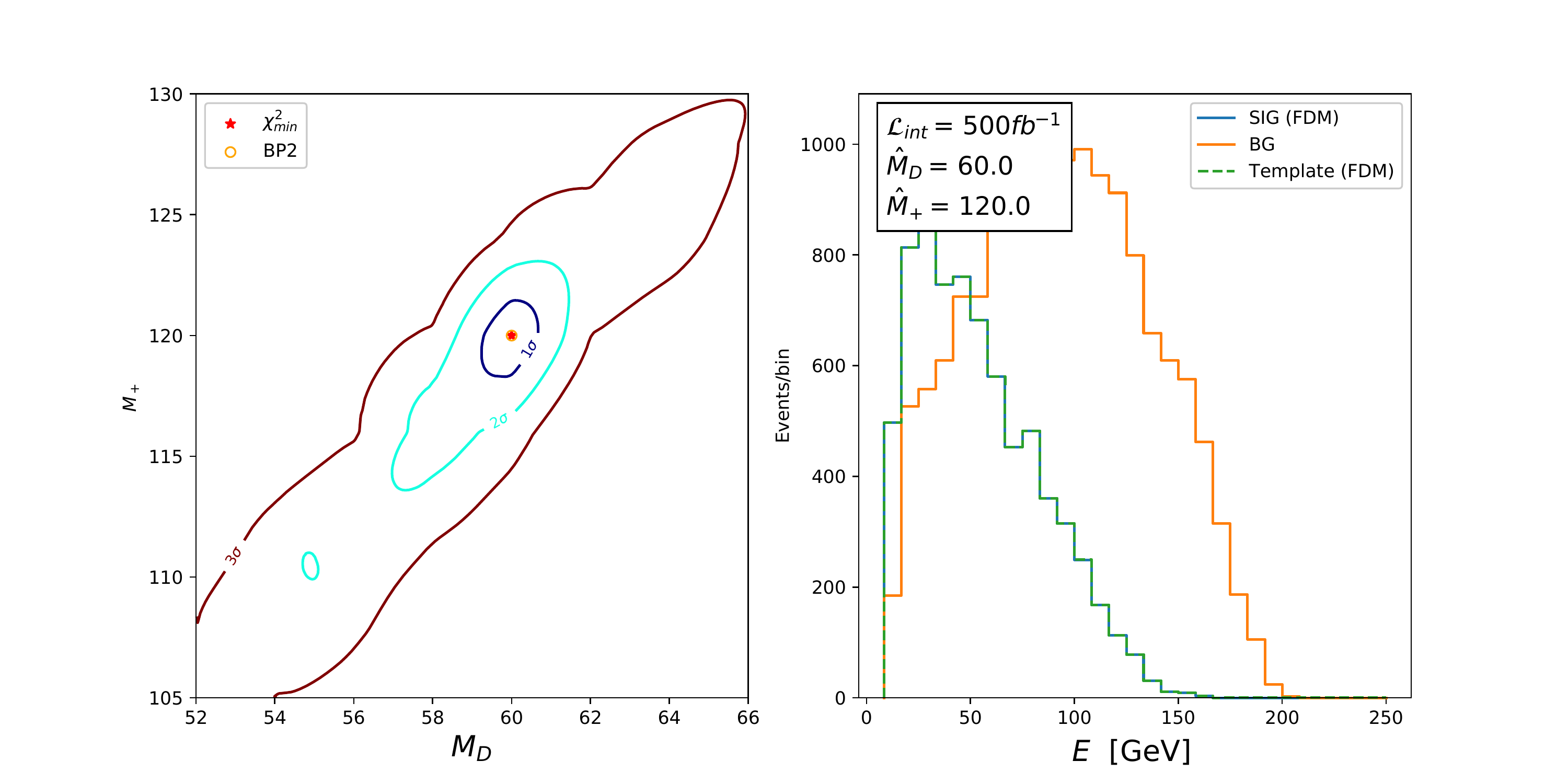}
		\caption{Profile $\chi^2$ value for template fitting of BP2 (left) and muon energy distribution with best fit (right), for FDM.}
		\label{chi2_FDM_BP2}
	\end{figure}
	
	\begin{figure}[htbp]
		\centering
		\hspace*{-1.cm}\includegraphics[width=1.1\textwidth]{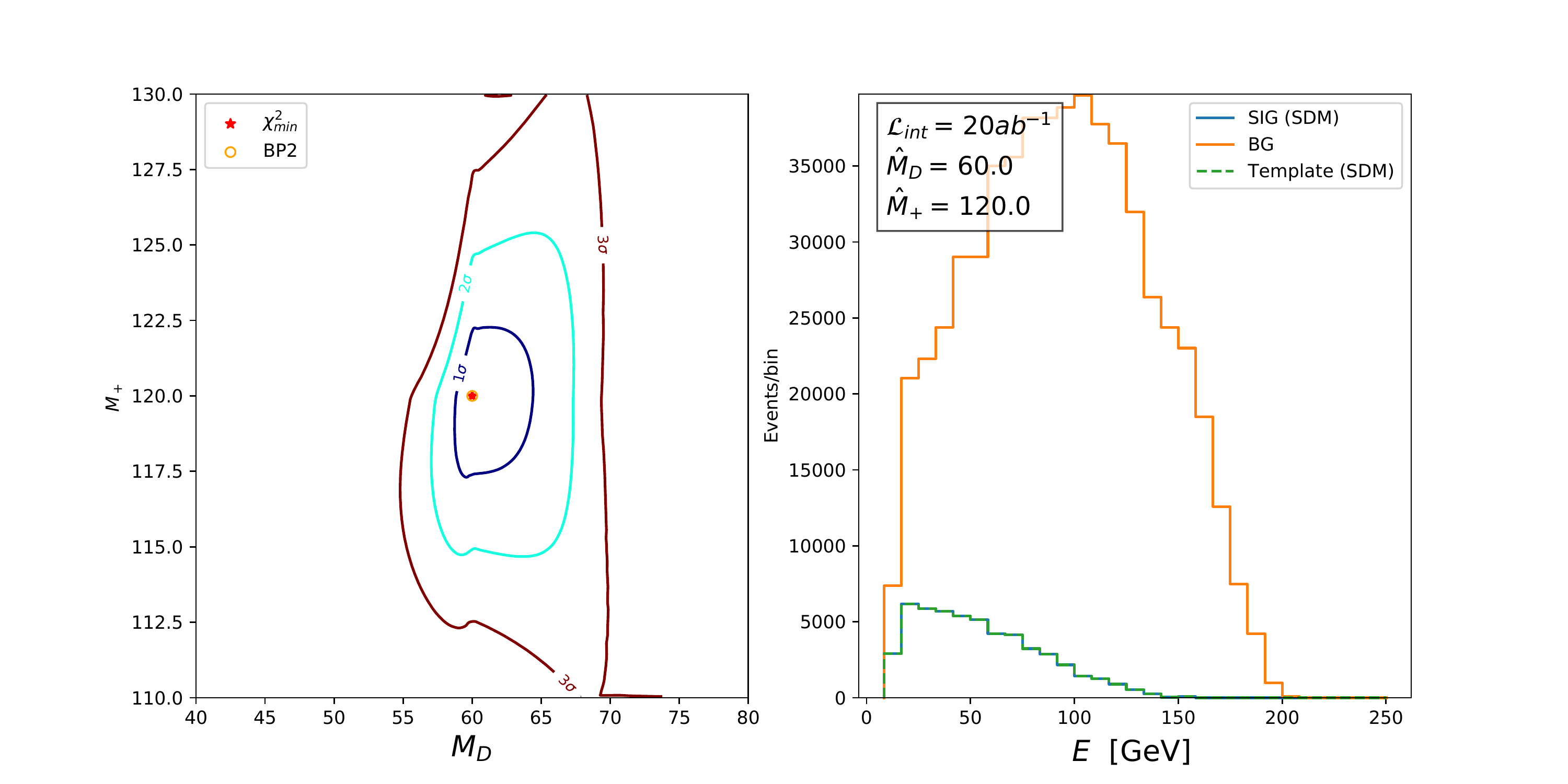}
		\caption{Profile $\chi^2$ value for template fitting of BP2 (left) and muon energy distribution with best fit (right), for SDM.}
		\label{chi2_SDM_BP2}
	\end{figure}

	\begin{table}[htb]
		\centering
		\begin{tabular}{llll}
			\hline
			&                            & $500 fb^{-1}$           & $20 ab^{-1}$          \\ \hline
			FDM & \multicolumn{1}{l|}{$M_D$} & $60.0^{+0.7}_{-0.8}$    & $60.0^{+0.1}_{-0.1}$  \\
			& \multicolumn{1}{l|}{$M_+$} & $120.0^{+1.5}_{ - 1.7}$ & $120.0^{+0.2}_{-0.3}$ \\ \hline
			SDM & \multicolumn{1}{l|}{$M_D$} & $60.0 ^{+24.1}_{-19.7}$ & $60.0^{+4.4}_{-1.3}$  \\
			& \multicolumn{1}{l|}{$M_+$} & $120.0^{+22.3}_{-45.9}$ & $120.0^{+2.3}_{-2.7}$ \\ \hline
		\end{tabular}
		\caption{Mass resolutions for BP2 shape fitting procedure.
			\label{tab:mass-fit-BP2}}
	\end{table}

%%%%%%%%%%%% SPIN %%%%%%%%%%%%%%%%%%%%%%%%%%%
\subsection{Spin discrimination}

As discussed in Section~\ref{subsec:angular},
the angular distribution of $W$-boson reconstructed from di-jet is very important observable to distinguish  scalar and fermionic DM as well as to distinguish the signal from the background.

We have performed  a binned composite likelihood analysis to estimate discriminating power of these distributions from Fig.~\ref{figCosThetaCut}, assuming that a signal of one model is present. We assume that the mass of the DM is precisely known, noting that more complete treatment would involve a simultaneous fit of mass and spin. Events have been  generated for the model assigned to "Assumed nature" in Table \ref{table:shapeAnalysis}, before the statistical comparison with the alternative model is conducted. We perform the analysis for two cases: using only the shape (signal strength becomes a nuisance parameter $\mu$, which may vary to maximise the likelihood) and also using the signal strength predicted by the specific model (in which case $\mu =1$). In Table~\ref{table:shapeAnalysis} we present the luminosity required to exclude a given hypothesis at the expected 95\% confidence level. Distributions used in this analysis were taken after the $M_{miss}$ and $E_{jj}$ cuts of Table~\ref{tab:cutflowTable} only.
From Table~\ref{table:shapeAnalysis} one can see first
that if one would use the information about the signal cross section, one can  discriminate the spin of DM with   integrated luminosity of the order of 10 fb$^{-1}$ only.
However this discrimination is quite model dependent.
On the other hand, even if we do not use the signal cross section
as a discriminator, one can still distinguish FDM from SDM scenarios with about 2 ab$^{-1}$ total integrated luminosity
in the worst case scenario of ``Assumed nature" model  for  BP1 or BP2. This discrimination is purely based on the shape of the
$\cos\theta_{jj}$ distribution, demonstrating its important power.
\begin{table}[H]
	\centering
	\begin{tabular}{|l|l|l|l|l|}
		\hline
		& \multicolumn{4}{l|}{$\mathcal{L}_{int}$ to differentiate  at 95\% CL $/fb^{-1}$}         \\ \cline{2-5} 
		& \multicolumn{2}{l|}{Shape only}           & \multicolumn{2}{l|}{Shape and cross-section} \\ \hline
		Assumed nature & 
		SDM                  & FDM                & SDM                  & FDM                   \\ \hline
		BP1	& $9.8 \times 10^2$  & $30$ 			  & 1.9   & 3.4                   \\
		BP2 & $2.3 \times 10^3$  & $1.2  \times 10^2$ & 9.6   & 13.                  \\ \hline
	\end{tabular}
	\caption{Integrated luminosity required to discriminate between spin of DM within these models using binned composite likelihoods. \label{table:shapeAnalysis}}
\end{table}

\section{Conclusions}\label{secconcl}
 %%%%%%%%%%%%%%%%%%%%

In this study we explore the potential of the $\epe$ colliders
to discover and determine the properties of DM,
such as mass and the spin.
The results of this study are applicable for future $\epe$ colliders such as ILC or FCC-ee.

We study two cases of minimal  models with DM, $D$ of  spin zero and spin one-half, which belongs to $SU(2)$ weak doublet with the hypercharge $1/2$ and therefore has the charged doublet partner, $D^+$.
For the case of scalar DM we chose Inert Doublet Model,
while for the case of fermion DM we suggest the new minimal fermion dark matter (MFDM) model with only three parameters. 
In this MFDM model  the $SU(2)$ DM  weak doublet  interacts
with the singlet  Majorana fermion and the SM Higgs doublet.
In comparison to the previously studied doublet-singlet model with different left- and right-handed interactions of Higgs,  DM doublet and the singlet, MFDM has these couplings equal to each other and this 
structure is  preserved against the quantum corrections.

We suggest two benchmarks for the models under study
which provide the correct amount of observed DM relic density
and satisfy the current   DM direct (and indirect) detection as well as LHC  constraints.
{We also provide the values of oblique $S,T,U$ parameters which are consistent with the current  electroweak precision tests.  For the  case of MFDM this result is new -- 
	we have evaluated formulas for  $S,T,U$ parameters and found that for this model the values of  $T$ and $U$
	are zero because of the model's structure, while the values of $S$ parameter is very close to SM one. }

We chose the particular process
$\epe\to  D^+ D^-  \to D D W^+ W^- \to DD(q \bar{q})(\mu^\pm\nu)$ 
providing the signature 
``{\bf di-jet +$\mu$ +  $\MET$}" at 500 GeV ILC collider and study it at the level of the fast detector simulation, taking into account Bremsstrahlung and ISR effects.
As a result, we have found several key kinematical characteristics
which allow to optimise signal to background ratio, discover and identify properties of  DM properties.
Among them are: the missing mass, $M_{miss}$,
the muon energy, $E_\mu$, and  the angular distribution of $W$-boson, reconstructed from di-jet, $\cos\theta_{jj}$. 

In particular, we have shown that  $E_\mu$ distribution
in case of on-shell $W$-boson from $D^+ \to D W^+$ decay (BP1),
has characteristic  points (kinks) , whose positions are kinematically
determined by masses of $D^+$ and $D$. The successful determination of these points allows to precisely determine these masses as we have demonstrated using the piecewise function fitting procedure which we suggest in this paper. For the model parameter space with off-shell $W$-boson (BP2)
we have demonstrated the success of the template fit of  $E_\mu$, distribution, since it does not have kinks in this case.
In particular we have demonstrated  that in case of fermion DM,
the masses can be measured with few percent accuracy already at 500 fb$^{-1}$ integrated luminosity. At the same time, the scalar DM model which has about order of magnitude lower signal, requires about factor of 40 higher  luminosity to reach the same accuracy in the mass measurement.
One should also add that the background can be further controlled and reduced using electron-positron beam polarisation.
We have also found 
the background can be further strongly suppressed by using the electron-positron beam (right-left) polarisation which  only mildly affects the signal rate.

We have also found that  $\cos\theta_{jj}$ distribution
is crucial for the determination of the DM spin.
To the best of our knowledge, we have shown for the first time that
it allows to
 distinguish fermion and scalar  DM scenarios with about 2 ab$^{-1}$ total integrated luminosity or less
 without using the information on BP1 or BP2   cross sections. This discrimination is purely based  on the shape of the
 $\cos\theta_{jj}$ distribution, demonstrating its important role.

The methods of the identification of DM properties we suggest here
are generic for the models where DM and its partner belong to the 
weak multiplet and can be applied to explore various DM models at future $\epe$ colliders {such as ILC, CLIC, FCC-ee.}

\section*{Acknowledgments}

This work was supported in part by grants RFBR and NSh-3802.2012.2, Program of Dept. of Phys. Sc.
RAS and SB RAS ``Studies of Higgs boson and exotic particles at LHC"  and Polish Ministry of
Science and Higher Education Grant N202 230337. IG thankful to A. Bondar, E. Boos, A.~Gladyshev,
A. Grozin, S. Eidelman, I.~Ivanov, D.~Ivanov, D.~Kazakov, J.~Kalinowski, K.~Kanishev, P.~Krachkov
and V.~Serbo for discussions. We would like to thank Tristan Hosken for his contribution on the
very early stage of this study. Authors acknowledge the use of the IRIDIS High Performance
Computing Facility, and associated support services at the University of Southampton to complete
this work.  AB and DL acknowledge  support from the STFC grant ST/L000296/1 and Soton-FAPESP
grant.
The work of AP was  carried out within the scientific program 
``Particle Physics and Cosmology" of Russian  National Center for Physics and
Mathematics.
\newpage
\appendix
%%%%%%%%%%%%%%%%%%%%%%
\section{Evaluation of the $S$ and $T$ parameters for MFDM}
\label{sec:STU}

{
	To find the contribution of new physics to the oblique parameters  one should evaluate  quantum corrections to the masses of   vector bosons and their mixings, $\Pi_{ZZ}(p^2)$, $\Pi_{Z \gamma}(p^2)$,
	$\Pi_{\gamma \gamma}(p^2)$, $\Pi_{WW}(p^2)$ {(or in unbroken gauge basis, 3,0 labels refer to $W^3$, $B$ gauge bosons respectively)}, defined by the effective
	Lagrangian
	\begin{equation}
	{\mathcal L}_{\rm oblique} =  \frac{1}{2} Z_\mu \Pi_{ZZ} (p^2) Z^{\mu}
	+ \frac{1}{2} \gamma_\mu \Pi_{\gamma \gamma} (p^2) \gamma^\mu
	+ Z_\mu \Pi_{Z \gamma} (p^2) \gamma^\mu 	
	+ W_\mu^+ \Pi_{WW} (p^2) W^{-\mu} \ \ ,
	\end{equation}
	where vacuum polarisation functions $\Pi$'s can be expanded in powers of $p^2$:
	\begin{equation}
	\Pi (p^2) = \Pi(0) + p^2\, \Pi'(0) + \frac{(p^2)^2}{2}\, \Pi''(0) + \dots
	\ \ ,
	\end{equation}	
since the new physics scale is expected to be high.
As for I2HDM, for MFDM we are evaluating here $S$, $T,$ and  $U$ observables~\cite{Peskin:1991sw} among the complete set of seven oblique
parameters~\cite{Barbieri:2004qk}
which are related to  $\Pi$ functions as follows:
\begin{align}
	S &\equiv \frac{4 c_W^2 s_W^2}{\alpha} \left[ \Pi'_{ZZ}(0) - \frac{c_W^2-s_W^2}{c_W s_W} \Pi'_{Z \gamma}(0) - \Pi'_{\gamma \gamma}(0) \right] = \frac{4 s_W^2}{\alpha} \frac{g}{g'} \Pi'_{30}(0) \\
	T &\equiv \frac{1}{\alpha} \left[ \frac{\Pi_{WW}(0)}{M_W^2} - \frac{\Pi_{ZZ}(0)}{M_Z^2} \right] = \frac{1}{\alpha} \frac{\Pi_{33}(0)-\Pi_{WW}(0)}{M_W^2} \\
	U &\equiv \frac{4 s_W^2}{\alpha} \left[\Pi'_{WW}(0) - \frac{c_W}{s_W} \Pi'_{Z \gamma}(0) - \Pi'_{\gamma \gamma}(0) \right] - S = \frac{4 s_W^2}{\alpha} \left[ \Pi'_{WW}(0)-\Pi'_{33}(0) \right]
\end{align}
where $\alpha=\alpha_{em}(M_Z)$.
In the following expressions 
we adopt 
$\Pi^{(')}(0)\equiv \Pi^{(')}$ notation, i.e. omit $(0)$ argument of  $\Pi^{(')}$ functions,

To find $S$, $T$, and  $U$ values
we have calculated   $\Pi$  functions 
for the interaction Lagrangian of the general form:
\begin{equation} 
	{\mathcal L}_{V\psi\psi}
	=
	\bar{\psi_1}\left( 
	g_V\gamma^\mu -
	g_A\gamma^\mu\gamma^5
	\right)
	\psi_2 V_\mu  + h.c.
\end{equation}
using dimensional regularisation.
We have found that 
\begin{equation} 
	\Pi^{(')}=
	\frac{1}{4\pi^{2}}
	\left(\left(g_{V}^{2}+g_{A}^{2}\right)
	{\Pi}_{V\!+\!A}^{(')}
	+\left(g_{V}^{2}-g_{A}^{2}\right) {\Pi}_{V\!-\!A}^{(')}\right)
	\label{eq:pi} \  \ ,
\end{equation}
with   ${\Pi}_{V\pm A}^{(')}$ 
given by
\begin{eqnarray}
	{\Pi}_{V\!+\!A} & = & 
	-\frac{1}{2}\left(m_1^2+m_2^{2}\right)
	\left (div + L \right )
	- \frac{1}{4} (m_1^2+m_2^{2})
	-\frac{\left(m_1^{4}+m_2^{4}\right)}{4\left(m_1^2-m_2^{2}\right)}
	\ln\left(\frac{m_2^{2}}{m_1^2}\right)\label{eq:pivpa}
	\\
	\Pi_{V\!-\!A} & = & m_1 m_2\left(div+L +1+\frac{\left(m_1^2+m_2^{2}\right)}{2\left(m_1^2-m_2^{2}\right)}\ln\left(\frac{m_2^{2}}{m_1^2}\right)\right)\label{eq:pivma}
	\\
	{\Pi}'_{V\!+\!A}& \!\!=\!\!\! & 
	\left ( \frac{1}{3}div + \frac{1}{3}L \right ) \!+\!\frac{m_1 ^{4}-8m_1^2m_2^2+m_2  ^{4}}{9\left(m_1^2-m_2^2\right)^{2}} +\frac{\left(m_1^2+m_2^2\right)\left(m_1 ^{4}-4m_1^2m_2^2+m_2  ^{4}\right)}{6\left(m_1^2-m_2^2\right)^{3}}\ln\left(\frac{m_2^2}{m_1^2}\right) \label{eq:pipvpa}
	\\
	{\Pi}'_{V\!-\!A} & = & 
	m_1 m_2  
	\left(
	\frac{\left(m_1^2+m_2^2\right)}{2\left(m_1^2-m_2^2\right)^2}+\frac{m_1^2m_2^2}{\left(m_1^2-m_2^2\right)^{3}}
	\ln\left(\frac{m_2^2}{m_1^2}\right)\right),
	\label{eq:pipvma}
\end{eqnarray}
where $div=\dfrac{1}{\epsilon}+ln(4\pi)-\gamma_\epsilon$, \ \ $L=\ln\left(\dfrac{\mu^{2}}{m_1 m_2}\right)$ and $m_1$, $m_2$
are the fermion masses in the loop.
For $m_1=m_2\equiv m$ the expressions for  ${\Pi}_{V\pm A}^{(')}$ 
are given by
\begin{eqnarray}
	&
	\ \ \ \ 
	\Pi_{V\!+\!A}=-m^{2} div-m^{2} \ln\left(\frac{\mu^{2}}{m^2}\right) \ \ , 
	\ \ \
	&\Pi_{V\!-\!A}=m^{2} div +m^{2} \ln\left(\frac{\mu^{2}}{m^2}\right)
	\label{eq:pi1}
	\\
	&
	\Pi'_{V\!+\!A}=\frac{1}{3} div
	+\frac{1}{3}\ln\left(\frac{\mu^{2}}{m^2}\right) 
	-\frac{1}{6} \ \ , \ \ \
	&\Pi'_{V\!-\!A}=\frac{1}{6}
	\label{eq:pip1}
\end{eqnarray}
One should note that in~\cite{Cynolter:2008ea} the identical expressions have been found in the context of generic model with vector-like fermions, with the exception of
two errors/typos in the expressions for  $\Pi'_{V+A}$ given by Eqs.~(\ref{eq:pipvpa}) and (\ref{eq:pip1}). The modification of $\Pi'_{V+A}$ presented in ~\cite{Cynolter:2008ea}  such that $L \to L/3$ in addition to squaring the denominator of the third term gives results identical to our independent calculation. We assume these mistakes are typos as taking the limit of their complete expression for $\Pi'_{V+A}$ gives rise to a further divergence, and the equal mass limit presented in~\cite{Cynolter:2008ea} has inconsistent dimensionality.

For the model with vector-like fermions ($g_V=1, g_A=0$) such as MFDM, the expressions for $S$ and $T$ observables
are given in terms of
\begin{equation}
	\Pi^{(')}_V=\Pi^{(')}_{V\!+\!A}+\Pi^{(')}_{V\!-\!A}
	\label{eq:piv}
\end{equation}
which follows from Eq.~\ref{eq:pi}.
Using  Eqs.~(\ref{eq:pivpa}-\ref{eq:pip1}) and Eq.~\ref{eq:piv}.
one finds the following expressions for $S$ and $T$ observables for MFDM:
\begin{eqnarray}
	S&=& \frac{1}{\pi}
	\left[
	\cos^2\theta \Pi_V'(M_{D'},M_D) 
	+ \sin^2\theta \Pi_V'(M_{D'},M_{D_2})
	- \Pi_V'(M_+,M_+) 
	\right]\label{eq:S}
	\\
	T&=& \frac{1}{4 \pi M_W^2 s_W^2}
	\left[ 
	\Pi_V(M_+,M_{D'})  + 
	\cos^2\theta \Pi_V(M_+,M_D)   + \sin^2\theta \Pi_V(M_+,M_{D_2})  \right.
	\nonumber\\
	&&
	\left.
	\quad\quad\quad\quad\ \ 
	-\Pi_V(M_+,M_+) 
	-\cos^2\theta \Pi_V(M_{D'},M_D)   -\sin^2\theta \Pi_V(M_{D'},M_{D_2}) 
	\right]
	\label{eq:T}\ \ ,
\end{eqnarray}
where $\theta$ is the $\chi^0$-$\chi^0_s$
mixing angle defined by Eq.~\ref{eq:fdm-mix}.
The coefficients infront of $\Pi$ and $\Pi'$ functions
in Eqs.~(\ref{eq:S},\ref{eq:T}) are defined by the Lagrangian of  MFDM model~\ref{eq:MFDM}, the complete set of  Feynman rules for which is given in  HEPMDB.

Recalling  that $M_+=M_{D'}$  from  Eq.~\ref{eq:T} 
it follows that 
\begin{equation}
	T \equiv 0 \ \ .
\end{equation}
This important feature of MFDM takes place because one of the
down parts of the vector-like doublet, corresponding to the 
neutral Majorana  fermion, does not mix and has the same mass as the charged  fermion. For the same reason 
\begin{equation}
	U \equiv 0 \ \ 
\end{equation}
for MFDM.
One should also note that for the expressions for  $S$, $T$ and $U$ observables 
both $div$ and  $\ln(\mu^2)$  terms cancel out as expected,
confirming the consistency and correctness of our evaluation.
}

%%%%%%%%%%%%%%%%%%%%%%%%%%%%%%%%%%%%%%
\section{Process \ \ $\pmb{e^+e^-\to Z\to DD_2\to DDZ}$}\label{secDDA}
%%%%%%%%%%%%%%%%%%%%%%%

One more process leading to production of $D$-odd particles at ILC
is also observable at $M_{D_2}+M_D<\sqrt{s}$ (in particular, at $\frac{\sqrt{s}}{2}>M_+>M_{D_2}$):
 \be
e^+e^-\to Z\to DD_2\to DDZ.
 \label{DDZproc}
 \ee
 This process has a clear signature in the modes suitable for observation:
\bes\label{DDAZ}
\be
\boxed{\mbox{\begin{minipage}{0.85\textwidth}
The \epe \ or $\mu^+\mu^-$ \ pair  with large \
$\MET$ and large $M(\MET)$ + {\large\it nothing}. The
effective mass of this dilepton is $\le M_Z$,
its energy is typically less than $\frac{\sqrt{s}}{2}$. \end{minipage} }}
\label{DDAZA}\ee
\be
\boxed{\mbox{\begin{minipage}{0.8\textwidth} A quark dijet with large \
$\MET$ and large $M(\MET)$ + {\large\it nothing}. The
effective mass of this dijet is $\le M_Z$,
its energy is typically less than $\frac{\sqrt{s}}{2}$. \end{minipage} }}
\label{DDAZB}\ee
\ees

At $M_{D_2}<M_+$ the BR for channel with signature \eqref{DDAZA} is 0.06, for the channel with signature \eqref{DDAZB} -- 0.7. We skip channel $Z\to \tau^+\tau^-$ with BR=0.03, 20\% of decays of $Z$ are invisible ($Z\to\nu\bar{\nu}$).

At $M_{D_2}>M_+$ BR's for processes with signature  \eqref{DDAZ} become less, since new decay channels $D_{2}\to D^\mp W^\pm\to DW^+ W^-$ are added with signature:
\be
\boxed{\mbox{\begin{minipage}{0.88\textwidth} $\epe\to DD_2\to DDW^+W^-$: Two quark dijets or dijet + single lepton or two leptons in one hemisphere with large \
$\MET$ and large $M(\MET)$ + {\large\it nothing}. The
effective mass of this system is $\le M_Z$,
its energy is typically less than $\frac{\sqrt{s}}{2}$. \end{minipage} }}
\label{DDAWW}\ee

The cross section of the process $\epe\to DD_2$ is model dependent. In the IDM it is determined unambiguously, in MSSM result depends on mixing angles and on the nature of fermions $D$ and $D_2$ (Dirac or Majorana).
In all considered cases at $\sqrt{s}>200$~GeV this cross section is smaller than $0.1\sigma_0$. Since the BR for events with signature \eqref{DDAZA} is 0.06, at the 500 fb$^{-1}$ luminosity the  number of events with this signature is of the order of  $10^3$
which is not enough for high precision measurements (but certain limits on the  masses can be obtained (cf. \cite{Lundstrom:2008ai-1,Lundstrom:2008ai-2,Lundstrom:2008ai-3} for LEP)).

Nevertheless we describe, for completeness, the energy distributions of $Z$
in this process. The obtained equations are similar to
 \eqref{rkinW}--\eqref{EPW} for new kinematics.

The $\gamma$-factor and velocity of
$D_2$ in c.m.s. for $e^+e^-$ are  \bear{c}
\gamma_{D_2}=\fr{s+M_{D_2}^2-M_D^2}{2\sqrt{s} M_{D_2}},\quad
\beta_{D_2}=\fr{\sqrt{(s^2-M_{D_2}^{2}-M_D^2)^2-4M_D^2M_{D_2}^{2}}}{s+M_{D_2}^2-M_D^2}.
\eear{Aprodcm}
For production of $Z$ with an effective mass $M^*_Z$ ($M^*_Z=M_Z$ at $M_{D_2}-M_D>M_Z$ and $M^*_Z\le M_{D_2}-M_D$ at $M_{D_2}-M_D<M_Z$)
in the rest frame of $D_2$
\be
E_Z^{D}\!=\!\fr{M_{D_2}^2 +M^{*2}_Z- M_{D}^2}{2M_{D_2}},\qquad p^{D}_Z\!=\!\fr{\sqrt{(M_{D_2}^2-M^{*2}_Z-M_D^2)^2-4M_D^2M^{*2}_Z}}{2M_{D_2}}.\label{rkin}
\ee

At $M_{D_2}-M_D>M_Z$ the $Z$-boson energy $E_Z$ lies within the interval with edges
 \be
E^{(-)}_{Z}\!=\!\gamma_{D_2}(E_Z^{D}\!-\!\beta_{D_2}
|\vb{p}|_Z^{D}),\;\;E^{(+)}_{Z}\!=\!\gamma_{D_2}(E_Z^{D}\!+\!\beta_{D_2}
|\vb{p}|^{D}_Z).\label{EPZ}
 \ee

At $M_{D_2}-M_D<M_Z$ similar equations are valid for each value of $M^*_Z$.
Absolute upper and lower edges of the energy distribution of $Z$
are reached at $M^*_Z=0$:
\be
E^{(\pm)}_{Z}=\gamma_{D_2}(1\pm \beta_{D_2})(M_{D_2}^2-M_D^2)/(2M_{D_2})\,.
\label{EPZoff}\ee
The peak in the energy distribution of dilepton appears at $M^*_Z=M_{D_2}-M_D$:
\be
E_{Z}=\gamma_{D_2}(M_{D_2}-M_D)\,.\label{EPZpeak}
\ee

%{\bf Masses $\pmb{M_D}$ and $\pmb{M_{D_2}}$.} 
At first sight,
measurement of kinematical edges of the dilepton spectrum \eqref{EPZ} (at $M_{D_2}-M_D>M_Z$) gives two equations for $M_D$ and $M_{D_2}$, allowing for determination of these masses. At $M_{D_2}-M_D<M_Z$, the same procedure can be performed separately for each value of the effective mass of dilepton \cite{Gin10}. In the latter case, the absolute edges of the dilepton energy spectrum \eqref{EPZoff} and the position of the peak in this spectrum \eqref{EPZpeak} could be also used for measuring $M_D$ and $M_{D_2}$.

In any case, the upper edge in the dijet energy spectrum $E^{(+)}_{Z}$ \eqref{EPZ}, \eqref{EPZoff} (signature \eqref{DDAZ}) gives one equation, necessary to find $M_{D_2}$ and $M_D$. In principle, necessary additional information gives position of lower edge in the dilepton energy spectrum $E^{(-)}_Z$. However, as it was noted above, the anticipated number of events with signature \eqref{DDAZA} looks insufficient for obtaining precise results. Together with good results for $M_D$ and $M_+$, one can hope to find an accurate value of $M_{D_2}$.

%%%%%%%%%%%%%%%%%%%%%%%%%%%%%%%%%%%%
%%%%%%%%%%%%%%%%%%%%%%%%%%%%%%%%%%%%

\section{Derivations} \label{D}

\subsection{$E_\mu$ end-point: $E_\mu^{max}$} \label{D2}
	In this section we present the derivation for the maximum muon energy, $E_\mu^{max}$, which is achieved when mass of virtual $W$ boson $M_W^*$ reaches its maximum value of $M_+-M_D$.	
	We start with the muon energy in the laboratory frame:
\begin{equation}
E_\mu=\gamma_{W}(1+c_1\beta_{W})(M^{(*)}_W/2),\\
\label{eq:Emu_appendixB1}
\end{equation}
where $c_1$ is $cos\theta_1$ of the escape angle of $\mu$ relative to the direction of the $W$ in the laboratory frame. We then substitute the $\gamma_{W}$ and $\beta_{W}$ variables for the edge, given by:
\begin{equation}
	\gamma_{W} = E_W /M^{(*)}_W=E(1-M_D/M_+)/M^{(*)}_W, \end{equation}
\begin{equation}
	\beta_{W}=\sqrt{1-M^{(*)2}_W/E_{W}^2}=\sqrt{1-\frac{M^{(*)2}_W}{E^2(1-M_D/M_+)^2}}\\
\end{equation}
into Eq.\eqref{eq:Emu_appendixB1}, which gives an $E_\mu$ for the off-shell $W$ boson case:
\begin{equation}
	E_\mu= \frac{E(1-M_D/M_+)}{M^*_W}\bigg(1+ c_1\sqrt{1-\frac{M^{*2}_W}{E^2(1-M_D/M_+)^2}}\bigg)(M^*_W/2).\\
	\label{eq:Emu_B1sub}
\end{equation}
By substituting $M^{(*)}_W=M_+-M_D$ for the maximum value of $M_W^*$ into Eq.\eqref{eq:Emu_B1sub} and setting $c_1$ to $+1$ corresponding to the maximum in $E_\mu$, this gives the maximum edge in muon energy,\\

\begin{equation}
	E_\mu^{max}= E\frac{(1-M_D/M_+)}{M_+-M_D}\bigg(1+ \sqrt{1-\frac{(M_+-M_D)^2}{E^2(1-\frac{M_D}{M_+})^2}}\bigg)\bigg(\frac{M_+-M_D}{2}\bigg).\\
\end{equation}
Simplifying this down to:
\begin{equation}
	E_\mu^{max}=\frac{E(1-\frac{M_D}{M+})}{2}\bigg(1+\sqrt{1-\bigg(\frac{M_+-M_D}{E(1-\frac{M_D}{M_+})}\bigg)^2}\bigg),\\
\end{equation}
%	$E_\mu=\frac{E(1-\frac{M_D}{M+})}{2}\bigg(1+c_1\sqrt{1-\bigg(\frac{M_+(1-\frac{M_D}{M_+})}{E(1-\frac{M_D}{M_+})}\bigg)^2}\bigg)$\\
it follows that:
%	$E_\mu=\frac{E(1-\frac{M_D}{M+})}{2}\bigg(1+c_1\sqrt{1-\bigg(\frac{M_+}{E}\bigg)^2}\bigg)$\\
\begin{equation}
	E_\mu^{max}=\frac{E(1-\frac{M_D}{M+})}{2}(1+\beta_{+}),\\
\end{equation}
where $\beta_{+}=\sqrt{1-\bigg(\frac{M_+}{E}\bigg)^2}$. %This $E_\mu^{max}$ gives the maximum muon energy edge for the off-shell $W$ boson case.\\
%\begin{equation}
%	E_\mu^{max}=E(1+\beta_{+})(1-M_D/M_+)/2.\\
%\end{equation}

%\begin{center}
%$E_\mu=\frac{E}{4}\bigg( 1+\frac{M_W^2-M_D^2}{M_+^2}+c \frac{\sqrt{(M_+^2-M_W^2-M_D^2)^2-4M_D^2M_W^2}}{M_+^2}\sqrt{1-\frac{M_+^2}{E^2}} \bigg) +  \frac{c_1}{2}\sqrt{\frac{E^2}{4}\bigg( 1+\frac{M_W^2-M_D^2}{M_+^2}+c\frac{\sqrt{(M_+^2-M_W^2-M_D^2)^2-4M_D^2M_W^2}}{M_+^2}\sqrt{1-\frac{M_+^2}{E^2}} \bigg)^2 -M_W^2}$
%\end{center}

\subsection{$E_\mu^{\pm}$ derivations} \label{D1}
In this section we present the derivation for the upper$(+)$ and lower$(-)$ kinks of the muon energy distribution $E_\mu$, defined as $E_\mu^{\pm}$.
We start with the muon energy in the laboratory frame:
\begin{equation}
	E_\mu=\gamma_{W}(1+c_1\beta_{W})\bigg(\frac{M_W^{(*)}}{2}\bigg)\\
	\label{eq:Emu_appb2}
\end{equation}
where we can substitute $\gamma^{(\pm)}_{W}$ and $\beta^{(\pm)}_{W}$ variables in terms of the upper and lower kinks of $E_W$, defined as:
\begin{equation}
	\gamma^{(\pm)}_{W}=\frac{E_W^{(\pm)}}{M_W^{(*)}},
\end{equation}	
\begin{equation}
 \beta^{(\pm)}_{W}=\sqrt{1-\bigg(\frac{M_W^{(*)}}{E_W^{(\pm)}}\bigg)^2}.\\
\end{equation}
We substitute these into $E_\mu$,  Eq.\eqref{eq:Emu_appb2}, and set $c_1$ to $\pm 1$ to give the maximum and minimum muon energy kinks $E^{(\pm)}_\mu$ in terms of $E_W^{(\pm)}$:
\begin{equation}
	E^{(\pm)}_\mu=\frac{E_W^{(\pm)}}{M_W^{(*)}}\bigg(1\pm\sqrt{1-\bigg(\frac{M_W^{(*)}}{E_W^{(\pm)}}\bigg)^2}\bigg)\bigg(\frac{M_W^{(*)}}{2}\bigg).\\
\end{equation}
After simplifying this down, this gives:
%\begin{equation}
%	$E_\mu=\frac{E_W^{(\pm)}}{2}\bigg(1+c_1\sqrt{1-\bigg(\frac{M_W^{(*)}}{E_W^{(\pm)}}\bigg)^2}\bigg)$\\
	
%	$E_\mu=\frac{E_W^{(\pm)}}{2}\bigg(1+c_1\sqrt{\frac{1}{E_W^{(\pm) 2}}(E_W^{(\pm) 2}-M_W^{(*)2})}\bigg)$\\
	
%	$E_\mu=\frac{E_W^{(\pm)}}{2}+c_1\frac{E_W^{(\pm)}}{2}\sqrt{\frac{1}{E_W^{(\pm) 2}}(E_W^{(\pm) 2}-M_W^{(*)2})}$\\
	
	%$E_\mu=\frac{E^{\pm}}{2}+c_1\frac{1}{2}\sqrt{(E^{\pm 2}-0)})$\\
	
	%$E_\mu=\frac{E^{\pm}}{2}+c_1\frac{E^{\pm}}{2}$.\\
\begin{equation}	
	E^{(\pm)}_\mu=\frac{E^{(\pm)}_W}{2}\pm\frac{1}{2}\sqrt{E^{(\pm) 2}_W-M_W^{(*)2}}.\\
\end{equation}
%for the corresponding maximum or minimum kinks in the muon energy distribution.
%:
%\begin{center}
%	$E^{(\pm)}_\mu=\frac{E_W^{(\pm)}}{2}\pm\frac{1}{2}\sqrt{E_W^{(\pm) 2}-M_W^{(*)2}}$.\\
	%$E_\mu^{\pm}=\frac{E^{\pm}}{2}\pm \frac{E^{\pm}}{2}$.\\
%\end{center}

%This gives a maximum muon energy of:\\
%\begin{center}
%$E_\mu^{+}=\frac{E^{+}}{2}+\frac{E^{+}}{2}=E^{+}$\\
%\end{center}

%and minimum muon energy of:\\
%\begin{center}
%$E_\mu^{-}=\frac{E^{-}}{2}-\frac{E^{-}}{2}=0$.\\
%\end{center}

\subsection{Simultaneous equations procedure for finding $M_+$ and $M_D$} \label{D3}
In this section we present the derivation for the DM masses $M_+$ and $M_D$, as a function of the muon energy upper and lower bounds $E_\mu^\pm$, that can be determined independent of each other.
Equations \eqref{EPW} and \eqref{Emuin} are used to give two simultaneous equations:\\
\begin{equation}
\begin{split}
\fr{4E_\mu^{(+)2}+M_W^2}{4E_\mu^{(+)}} &=\fr{E}{M_+}  \Bigg(\fr{M_+^2 +M_W^2- M_{D}^2}{2M_+} \\&\ \ \ \ \ \ + \sqrt{1-\fr{M_+^2}{E^2}}\fr{\sqrt{M_+^4+M_W^4+M_D^4-2M_+^{2}M_W^2-2M_+^{2}M_D^2-2M_W^{2}M_D^2}}{2M_+}\Bigg),\\
\end{split}
\label{eq:SimEq1_B3}
\end{equation}
\begin{equation}
\begin{split}
\fr{4E_\mu^{(-)2}+M_W^2}{4E_\mu^{(-)}}&=\fr{E}{M_+}
\Bigg(\fr{M_+^2 +M_W^2- M_{D}^2}{2M_+} \\&\ \ \ \ \ \ - \sqrt{1-\fr{M_+^2}{E^2}}\fr{\sqrt{M_+^4+M_W^4+M_D^4-2M_+^{2}M_W^2-2M_+^{2}M_D^2-2M_W^{2}M_D^2}}{2M_+} \Bigg).\\
\end{split}
\end{equation}
Performing the simultaneous equations procedure gives the equation of $M_D$ in terms of $M_+$:\\
\begin{equation}
M_D^2=M_W^2-M_+^2\left[\fr{1}{E}\left(\fr{4E_\mu^{(+)2}+M_W^2}{4E_\mu^{(+)}}+\fr{4E_\mu^{(-)2}+M_W^2}{4E_\mu^{(-)}}\right)-1\right]
\end{equation}
and substituting this onto the first simultaneous equation \eqref{eq:SimEq1_B3} results in the polynomial of $M_+$:\\
\begin{equation}
-M_+^4(\alpha+\beta)^2+4M_+^2E^2(\alpha\beta+M_W^2)-4M_W^2E^4=0
\end{equation}
where:
\begin{equation}
\alpha=\fr{4E_\mu^{(+)2}+M_W^2}{4E_\mu^{(+)}},     
   \ \ \ \ \beta=\fr{4E_\mu^{(-)2}+M_W^2}{4E_\mu^{(-)}}.
\end{equation}
This gives 4 roots for $M_+$:\\
\begin{equation}
M_+=\pm \sqrt{2}\sqrt{\fr{-\sqrt{E^4(\alpha^2-M_W^2)(\beta^2-M_W^2)}+\alpha \beta E^2 +E^2M_W^2}{(\alpha+\beta)^2}}\\
\end{equation}
\begin{equation}
M_+=\pm \sqrt{2}\sqrt{\fr{\sqrt{E^4(\alpha^2-M_W^2)(\beta^2-M_W^2)}+\alpha \beta E^2 +E^2M_W^2}{(\alpha+\beta)^2}}.\\
\end{equation}

Two of these roots will be positive and the top equation will correspond to the physical mass
of $D^{\pm}$.
By rearranging the equation of $M_D$ in terms of $M_+$ we  obtain the the following equation
for $M_D$:\\
\begin{equation}
-\left(\fr{M_W^2-M_D^2}{\alpha+\beta-E}\right)^2(\alpha+\beta)^2+4\fr{M_W^2-M_D^2}{\alpha+\beta-E}E(\alpha\beta+M_W^2)-4M_W^2E^2=0\\
\end{equation}
which gives two real and two complex roots for $M_D$. Out of the two real roots, one is positive and gives
the physical mass for $D$.

\newpage
\bibliography{bib}

\providecommand{\href}[2]{#2}\begingroup\raggedright\begin{thebibliography}{100}

\bibitem{Jungman:1995df}
G.~Jungman, M.~Kamionkowski and K.~Griest, {\it {Supersymmetric dark matter}},
  {\em Phys. Rept.} {\bf 267} (1996) 195--373
  [\href{http://arXiv.org/abs/hep-ph/9506380}{{\tt hep-ph/9506380}}].

\bibitem{Ellwanger:2009dp}
U.~Ellwanger, C.~Hugonie and A.~M. Teixeira, {\it {The Next-to-Minimal
  Supersymmetric Standard Model}},  {\em Phys. Rept.} {\bf 496} (2010) 1--77
  [\href{http://arXiv.org/abs/0910.1785}{{\tt 0910.1785}}].

\bibitem{Giudice:2004tc}
G.~Giudice and A.~Romanino, {\it {Split supersymmetry}},  {\em Nucl. Phys. B}
  {\bf 699} (2004) 65--89 [\href{http://arXiv.org/abs/hep-ph/0406088}{{\tt
  hep-ph/0406088}}]. [Erratum: Nucl.Phys.B 706, 487--487 (2005)].

\bibitem{Dodelson:1993je}
S.~Dodelson and L.~M. Widrow, {\it {Sterile-neutrinos as dark matter}},  {\em
  Phys. Rev. Lett.} {\bf 72} (1994) 17--20
  [\href{http://arXiv.org/abs/hep-ph/9303287}{{\tt hep-ph/9303287}}].

\bibitem{Cirelli:2005uq}
M.~Cirelli, N.~Fornengo and A.~Strumia, {\it {Minimal dark matter}},  {\em
  Nucl. Phys. B} {\bf 753} (2006) 178--194
  [\href{http://arXiv.org/abs/hep-ph/0512090}{{\tt hep-ph/0512090}}].

\bibitem{Kim:2008hd}
J.~E. Kim and G.~Carosi, {\it {Axions and the Strong CP Problem}},  {\em Rev.
  Mod. Phys.} {\bf 82} (2010) 557--602
  [\href{http://arXiv.org/abs/0807.3125}{{\tt 0807.3125}}]. [Erratum:
  Rev.Mod.Phys. 91, 049902 (2019)].

\bibitem{Cheng:2002ej}
H.-C. Cheng, J.~L. Feng and K.~T. Matchev, {\it {Kaluza-Klein dark matter}},
  {\em Phys. Rev. Lett.} {\bf 89} (2002) 211301
  [\href{http://arXiv.org/abs/hep-ph/0207125}{{\tt hep-ph/0207125}}].

\bibitem{Hooper:2007qk}
D.~Hooper and S.~Profumo, {\it {Dark Matter and Collider Phenomenology of
  Universal Extra Dimensions}},  {\em Phys. Rept.} {\bf 453} (2007) 29--115
  [\href{http://arXiv.org/abs/hep-ph/0701197}{{\tt hep-ph/0701197}}].

\bibitem{Boehm:2003hm}
C.~Boehm and P.~Fayet, {\it {Scalar dark matter candidates}},  {\em Nucl. Phys.
  B} {\bf 683} (2004) 219--263 [\href{http://arXiv.org/abs/hep-ph/0305261}{{\tt
  hep-ph/0305261}}].

\bibitem{Branco:2011iw}
G.~Branco, P.~Ferreira, L.~Lavoura, M.~Rebelo, M.~Sher and J.~P. Silva, {\it
  {Theory and phenomenology of two-Higgs-doublet models}},  {\em Phys. Rept.}
  {\bf 516} (2012) 1--102 [\href{http://arXiv.org/abs/1106.0034}{{\tt
  1106.0034}}].

\bibitem{Belyaev:2016lok}
A.~Belyaev, G.~Cacciapaglia, I.~P. Ivanov, F.~Rojas-Abatte and M.~Thomas, {\it
  {Anatomy of the Inert Two Higgs Doublet Model in the light of the LHC and
  non-LHC Dark Matter Searches}},  {\em Phys. Rev.} {\bf D97} (2018), no.~3
  035011 [\href{http://arXiv.org/abs/1612.00511}{{\tt 1612.00511}}].
%%CITATION = ARXIV:1612.00511;%%

\bibitem{Christensen:2013sea}
N.~D. Christensen and D.~Salmon, {\it {New method for the spin determination of
  dark matter}},  {\em Phys. Rev.} {\bf D90} (2014), no.~1 014025
  [\href{http://arXiv.org/abs/1311.6465}{{\tt 1311.6465}}].
%%CITATION = ARXIV:1311.6465;%%

\bibitem{Belyaev:2016pxe}
A.~Belyaev, L.~Panizzi, A.~Pukhov and M.~Thomas, {\it {Dark Matter
  characterization at the LHC in the Effective Field Theory approach}},  {\em
  JHEP} {\bf 04} (2017) 110 [\href{http://arXiv.org/abs/1610.07545}{{\tt
  1610.07545}}].
%%CITATION = ARXIV:1610.07545;%%

\bibitem{Asano:2011aj}
M.~Asano, T.~Saito, T.~Suehara, K.~Fujii, R.~S. Hundi, H.~Itoh, S.~Matsumoto,
  N.~Okada, Y.~Takubo and H.~Yamamoto, {\it {Discrimination of New Physics
  Models with the International Linear Collider}},  {\em Phys. Rev.} {\bf D84}
  (2011) 115003 [\href{http://arXiv.org/abs/1106.1932}{{\tt 1106.1932}}].
%%CITATION = ARXIV:1106.1932;%%

\bibitem{Christensen:2014yya}
N.~D. Christensen, T.~Han, Z.~Qian, J.~Sayre, J.~Song and Stefanus, {\it
  {Determining the Dark Matter Particle Mass through Antler Topology Processes
  at Lepton Colliders}},  {\em Phys. Rev.} {\bf D90} (2014) 114029
  [\href{http://arXiv.org/abs/1404.6258}{{\tt 1404.6258}}].
%%CITATION = ARXIV:1404.6258;%%

\bibitem{Burns:2008va}
M.~Burns, K.~Kong, K.~T. Matchev and M.~Park, {\it {Using Subsystem MT2 for
  Complete Mass Determinations in Decay Chains with Missing Energy at Hadron
  Colliders}},  {\em JHEP} {\bf 03} (2009) 143
  [\href{http://arXiv.org/abs/0810.5576}{{\tt 0810.5576}}].

\bibitem{Sirunyan:2017jix}
{\bf CMS} Collaboration, A.~M. Sirunyan {\em et.~al.}, {\it {Search for new
  physics in final states with an energetic jet or a hadronically decaying $W$
  or $Z$ boson and transverse momentum imbalance at $\sqrt{s}=13\text{ }\text{
  }\mathrm{TeV}$}},  {\em Phys. Rev.} {\bf D97} (2018), no.~9 092005
  [\href{http://arXiv.org/abs/1712.02345}{{\tt 1712.02345}}].
%%CITATION = ARXIV:1712.02345;%%

\bibitem{Aaboud:2017phn}
{\bf ATLAS} Collaboration, M.~Aaboud {\em et.~al.}, {\it {Search for dark
  matter and other new phenomena in events with an energetic jet and large
  missing transverse momentum using the ATLAS detector}},  {\em JHEP} {\bf 01}
  (2018) 126 [\href{http://arXiv.org/abs/1711.03301}{{\tt 1711.03301}}].
%%CITATION = ARXIV:1711.03301;%%

\bibitem{Sirunyan:2017hci}
{\bf CMS} Collaboration, A.~M. Sirunyan {\em et.~al.}, {\it {Search for dark
  matter produced with an energetic jet or a hadronically decaying W or Z boson
  at $ \sqrt{s}=13 $ TeV}},  {\em JHEP} {\bf 07} (2017) 014
  [\href{http://arXiv.org/abs/1703.01651}{{\tt 1703.01651}}].
%%CITATION = ARXIV:1703.01651;%%

\bibitem{Sirunyan:2017onm}
{\bf CMS} Collaboration, A.~M. Sirunyan {\em et.~al.}, {\it {Search for dark
  matter and unparticles in events with a Z boson and missing transverse
  momentum in proton-proton collisions at $ \sqrt{s}=13 $ TeV}},  {\em JHEP}
  {\bf 03} (2017) 061 [\href{http://arXiv.org/abs/1701.02042}{{\tt
  1701.02042}}]. [Erratum: JHEP09,106(2017)].
%%CITATION = ARXIV:1701.02042;%%

\bibitem{Aaboud:2018xdl}
{\bf ATLAS} Collaboration, M.~Aaboud {\em et.~al.}, {\it {Search for dark
  matter in events with a hadronically decaying vector boson and missing
  transverse momentum in $pp$ collisions at $\sqrt{s} = 13$ TeV with the ATLAS
  detector}},  {\em JHEP} {\bf 10} (2018) 180
  [\href{http://arXiv.org/abs/1807.11471}{{\tt 1807.11471}}].
%%CITATION = ARXIV:1807.11471;%%

\bibitem{Basalaev:2017cpw}
{\bf ATLAS} Collaboration, A.~Basalaev, {\it {Search for dark matter particle
  candidates produced in association with a Z boson in pp collisions at a
  center-of-mass energy of 13 TeV with the ATLAS detector}},  {\em J. Phys.
  Conf. Ser.} {\bf 934} (2017), no.~1 012024.
%%CITATION = 00462,934,012024;%%

\bibitem{Aaboud:2017bja}
{\bf ATLAS} Collaboration, M.~Aaboud {\em et.~al.}, {\it {Search for an
  invisibly decaying Higgs boson or dark matter candidates produced in
  association with a $Z$ boson in $pp$ collisions at $\sqrt{s} =$ 13 TeV with
  the ATLAS detector}},  {\em Phys. Lett.} {\bf B776} (2018) 318--337
  [\href{http://arXiv.org/abs/1708.09624}{{\tt 1708.09624}}].
%%CITATION = ARXIV:1708.09624;%%

\bibitem{Aaboud:2017dor}
{\bf ATLAS} Collaboration, M.~Aaboud {\em et.~al.}, {\it {Search for dark
  matter at $\sqrt{s}=13$ TeV in final states containing an energetic photon
  and large missing transverse momentum with the ATLAS detector}},  {\em Eur.
  Phys. J.} {\bf C77} (2017), no.~6 393
  [\href{http://arXiv.org/abs/1704.03848}{{\tt 1704.03848}}].
%%CITATION = ARXIV:1704.03848;%%

\bibitem{Sirunyan:2018fpy}
{\bf CMS} Collaboration, A.~M. Sirunyan {\em et.~al.}, {\it {Search for dark
  matter produced in association with a Higgs boson decaying to $\gamma\gamma$
  or $\tau^+\tau^-$ at $\sqrt{s} =$ 13 TeV}},  {\em Submitted to: JHEP} (2018)
  [\href{http://arXiv.org/abs/1806.04771}{{\tt 1806.04771}}].
%%CITATION = ARXIV:1806.04771;%%

\bibitem{Sirunyan:2017hnk}
{\bf CMS} Collaboration, A.~M. Sirunyan {\em et.~al.}, {\it {Search for
  associated production of dark matter with a Higgs boson decaying to $
  \mathrm{b}\overline{\mathrm{b}} $ or $\gamma \gamma$ at $ \sqrt{s}=13$ TeV}},
   {\em JHEP} {\bf 10} (2017) 180 [\href{http://arXiv.org/abs/1703.05236}{{\tt
  1703.05236}}].
%%CITATION = ARXIV:1703.05236;%%

\bibitem{ATLAS:2018bvd}
{\bf ATLAS} Collaboration, T.~A. collaboration, {\it {Search for Dark Matter
  Produced in Association with a Higgs Boson decaying to $b\bar{b}$ at
  $\sqrt{s}= 13\,$TeV with the ATLAS Detector using 79.8$\,$fb$^{-1}$ of
  proton-proton collision data}},  2018.
%%CITATION = ATLAS-CONF-2018-039;%%

\bibitem{Aaboud:2017uak}
{\bf ATLAS} Collaboration, M.~Aaboud {\em et.~al.}, {\it {Search for dark
  matter in association with a Higgs boson decaying to two photons at
  $\sqrt{s}$ = 13 TeV with the ATLAS detector}},  {\em Phys. Rev.} {\bf D96}
  (2017), no.~11 112004 [\href{http://arXiv.org/abs/1706.03948}{{\tt
  1706.03948}}].
%%CITATION = ARXIV:1706.03948;%%

\bibitem{Aad:2015yga}
{\bf ATLAS} Collaboration, G.~Aad {\em et.~al.}, {\it {Search for Dark Matter
  in Events with Missing Transverse Momentum and a Higgs Boson Decaying to Two
  Photons in $pp$ Collisions at $\sqrt{s}=8$ TeV with the ATLAS Detector}},
  {\em Phys. Rev. Lett.} {\bf 115} (2015), no.~13 131801
  [\href{http://arXiv.org/abs/1506.01081}{{\tt 1506.01081}}].
%%CITATION = ARXIV:1506.01081;%%

\bibitem{Sirunyan:2018dub}
{\bf CMS} Collaboration, A.~M. Sirunyan {\em et.~al.}, {\it {Search for dark
  matter particles produced in association with a top quark pair at $\sqrt{s}
  =$ 13 TeV}},  {\em Submitted to: Phys. Rev. Lett.} (2018)
  [\href{http://arXiv.org/abs/1807.06522}{{\tt 1807.06522}}].
%%CITATION = ARXIV:1807.06522;%%

\bibitem{Sirunyan:2018gka}
{\bf CMS} Collaboration, A.~M. Sirunyan {\em et.~al.}, {\it {Search for dark
  matter in events with energetic, hadronically decaying top quarks and missing
  transverse momentum at $ \sqrt{s}=13 $ TeV}},  {\em JHEP} {\bf 06} (2018) 027
  [\href{http://arXiv.org/abs/1801.08427}{{\tt 1801.08427}}].
%%CITATION = ARXIV:1801.08427;%%

\bibitem{Aaboud:2017rzf}
{\bf ATLAS} Collaboration, M.~Aaboud {\em et.~al.}, {\it {Search for dark
  matter produced in association with bottom or top quarks in $\sqrt{s}=13$ TeV
  pp collisions with the ATLAS detector}},  {\em Eur. Phys. J.} {\bf C78}
  (2018), no.~1 18 [\href{http://arXiv.org/abs/1710.11412}{{\tt 1710.11412}}].
%%CITATION = ARXIV:1710.11412;%%

\bibitem{Sirunyan:2018owy}
{\bf CMS} Collaboration, A.~M. Sirunyan {\em et.~al.}, {\it {Search for
  invisible decays of a Higgs boson produced through vector boson fusion in
  proton-proton collisions at $\sqrt{s} =$ 13 TeV}},  {\em Phys. Lett. B} {\bf
  793} (2019) 520--551 [\href{http://arXiv.org/abs/1809.05937}{{\tt
  1809.05937}}].

\bibitem{Aaboud:2019rtt}
{\bf ATLAS} Collaboration, M.~Aaboud {\em et.~al.}, {\it {Combination of
  searches for invisible Higgs boson decays with the ATLAS experiment}},  {\em
  Phys. Rev. Lett.} {\bf 122} (2019), no.~23 231801
  [\href{http://arXiv.org/abs/1904.05105}{{\tt 1904.05105}}].

\bibitem{Sirunyan:2018xlo}
{\bf CMS} Collaboration, A.~M. Sirunyan {\em et.~al.}, {\it {Search for narrow
  and broad dijet resonances in proton-proton collisions at $\sqrt{s}=$ 13 TeV
  and constraints on dark matter mediators and other new particles}},
  \href{http://arXiv.org/abs/1806.00843}{{\tt 1806.00843}}.
%%CITATION = ARXIV:1806.00843;%%

\bibitem{Sirunyan:2018wcm}
{\bf CMS} Collaboration, A.~M. Sirunyan {\em et.~al.}, {\it {Search for new
  physics in dijet angular distributions using proton-proton collisions at
  $\sqrt{s}=$ 13 TeV and constraints on dark matter and other models}},
  \href{http://arXiv.org/abs/1803.08030}{{\tt 1803.08030}}.
%%CITATION = ARXIV:1803.08030;%%

\bibitem{Sirunyan:2016iap}
{\bf CMS} Collaboration, A.~M. Sirunyan {\em et.~al.}, {\it {Search for dijet
  resonances in proton–proton collisions at $\sqrt{s}$ = 13 TeV and
  constraints on dark matter and other models}},  {\em Phys. Lett.} {\bf B769}
  (2017) 520--542 [\href{http://arXiv.org/abs/1611.03568}{{\tt 1611.03568}}].
  [Erratum: Phys. Lett.B772,882(2017)].
%%CITATION = ARXIV:1611.03568;%%

\bibitem{Ellis:1997wva}
J.~R. Ellis, T.~Falk, K.~A. Olive and M.~Schmitt, {\it {Constraints on
  neutralino dark matter from LEP-2 and cosmology}},  {\em Phys. Lett. B} {\bf
  413} (1997) 355--364 [\href{http://arXiv.org/abs/hep-ph/9705444}{{\tt
  hep-ph/9705444}}].

\bibitem{Aaboud:2017mpt}
{\bf ATLAS} Collaboration, M.~Aaboud {\em et.~al.}, {\it {Search for long-lived
  charginos based on a disappearing-track signature in pp collisions at $
  \sqrt{s}=13 $ TeV with the ATLAS detector}},  {\em JHEP} {\bf 06} (2018) 022
  [\href{http://arXiv.org/abs/1712.02118}{{\tt 1712.02118}}].

\bibitem{Sirunyan:2018ldc}
{\bf CMS} Collaboration, A.~M. Sirunyan {\em et.~al.}, {\it {Search for
  disappearing tracks as a signature of new long-lived particles in
  proton-proton collisions at $\sqrt{s} =$ 13 TeV}},  {\em JHEP} {\bf 08}
  (2018) 016 [\href{http://arXiv.org/abs/1804.07321}{{\tt 1804.07321}}].

\bibitem{Aaboud:2017iio}
{\bf ATLAS} Collaboration, M.~Aaboud {\em et.~al.}, {\it {Search for
  long-lived, massive particles in events with displaced vertices and missing
  transverse momentum in $\sqrt{s}$ = 13 TeV $pp$ collisions with the ATLAS
  detector}},  {\em Phys. Rev. D} {\bf 97} (2018), no.~5 052012
  [\href{http://arXiv.org/abs/1710.04901}{{\tt 1710.04901}}].

\bibitem{Sirunyan:2018pwn}
{\bf CMS} Collaboration, A.~M. Sirunyan {\em et.~al.}, {\it {Search for
  long-lived particles with displaced vertices in multijet events in
  proton-proton collisions at $\sqrt{s}= $13 TeV}},  {\em Phys. Rev. D} {\bf
  98} (2018), no.~9 092011 [\href{http://arXiv.org/abs/1808.03078}{{\tt
  1808.03078}}].

\bibitem{Barducci:2015ffa}
D.~Barducci, A.~Belyaev, A.~K.~M. Bharucha, W.~Porod and V.~Sanz, {\it
  {Uncovering Natural Supersymmetry via the interplay between the LHC and
  Direct Dark Matter Detection}},  {\em JHEP} {\bf 07} (2015) 066
  [\href{http://arXiv.org/abs/1504.02472}{{\tt 1504.02472}}].
%%CITATION = ARXIV:1504.02472;%%

\bibitem{Athron:2017yua}
{\bf GAMBIT} Collaboration, P.~Athron {\em et.~al.}, {\it {A global fit of the
  MSSM with GAMBIT}},  {\em Eur. Phys. J.} {\bf C77} (2017), no.~12 879
  [\href{http://arXiv.org/abs/1705.07917}{{\tt 1705.07917}}].
%%CITATION = ARXIV:1705.07917;%%

\bibitem{Athron:2018vxy}
{\bf GAMBIT} Collaboration, P.~Athron {\em et.~al.}, {\it {Combined collider
  constraints on neutralinos and charginos}},
  \href{http://arXiv.org/abs/1809.02097}{{\tt 1809.02097}}.
%%CITATION = ARXIV:1809.02097;%%

\bibitem{Ginzburg:2014ora}
I.~F. Ginzburg, {\it {Measuring mass and spin of Dark Matter particles with the
  aid energy spectra of single lepton and dijet at the $e^+e^-$ Linear
  Collider}},  {\em J. Mod. Phys.} {\bf 5} (2014) 1036--1049
  [\href{http://arXiv.org/abs/1410.0869}{{\tt 1410.0869}}].

\bibitem{inert-1}
N.~G. Deshpande and E.~Ma, {\it Pattern of symmetry breaking with two higgs
  doublets},  {\em Phys. Rev. D} {\bf 18} (Oct, 1978) 2574--2576.

\bibitem{inert-2}
R.~Barbieri, L.~J. Hall and V.~S. Rychkov, {\it {Improved naturalness with a
  heavy Higgs: An Alternative road to LHC physics}},  {\em Phys. Rev. D} {\bf
  74} (2006) 015007 [\href{http://arXiv.org/abs/hep-ph/0603188}{{\tt
  hep-ph/0603188}}].

\bibitem{inert-3}
I.~F. Ginzburg, K.~A. Kanishev, M.~Krawczyk and D.~Sokolowska, {\it {Evolution
  of Universe to the present inert phase}},  {\em Phys. Rev. D} {\bf 82} (2010)
  123533 [\href{http://arXiv.org/abs/1009.4593}{{\tt 1009.4593}}].

\bibitem{inert-4}
M.~Gustafsson, S.~Rydbeck, L.~Lopez-Honorez and E.~Lundstrom, {\it {Status of
  the Inert Doublet Model and the Role of multileptons at the LHC}},  {\em
  Phys. Rev. D} {\bf 86} (2012) 075019
  [\href{http://arXiv.org/abs/1206.6316}{{\tt 1206.6316}}].

\bibitem{Lundstrom:2008ai-1}
M.~Espírito-Santo, K.~Hultqvist, P.~Johansson and A.~Lipniacka, {\it Search
  for neutralino pair production at $\sqrt{s}$ from 192 to 208 gev},  2003.

\bibitem{Lundstrom:2008ai-2}
E.~Lundström, M.~Gustafsson and J.~Edsjö, {\it Inert doublet model and lep ii
  limits},  {\em Physical Review D} {\bf 79} (Feb, 2009).

\bibitem{Lundstrom:2008ai-3}
M.~Aoki, S.~Kanemura and H.~Yokoya, {\it Reconstruction of inert doublet
  scalars at the international linear collider},  {\em Physics Letters B} {\bf
  725} (Oct, 2013) 302–309.

\bibitem{Ma:2006km}
E.~Ma, {\it {Verifiable radiative seesaw mechanism of neutrino mass and dark
  matter}},  {\em Phys.Rev.} {\bf D73} (2006) 077301
  [\href{http://arXiv.org/abs/hep-ph/0601225}{{\tt hep-ph/0601225}}].
%%CITATION = HEP-PH/0601225;%%

\bibitem{Barbieri:2006dq}
R.~Barbieri, L.~J. Hall and V.~S. Rychkov, {\it {Improved naturalness with a
  heavy Higgs: An Alternative road to LHC physics}},  {\em Phys.Rev.} {\bf D74}
  (2006) 015007 [\href{http://arXiv.org/abs/hep-ph/0603188}{{\tt
  hep-ph/0603188}}].
%%CITATION = HEP-PH/0603188;%%

\bibitem{LopezHonorez:2006gr}
L.~Lopez~Honorez, E.~Nezri, J.~F. Oliver and M.~H. Tytgat, {\it {The Inert
  Doublet Model: An Archetype for Dark Matter}},  {\em JCAP} {\bf 0702} (2007)
  028 [\href{http://arXiv.org/abs/hep-ph/0612275}{{\tt hep-ph/0612275}}].
%%CITATION = HEP-PH/0612275;%%

\bibitem{Arina:2009um}
C.~Arina, F.-S. Ling and M.~H.~G. Tytgat, {\it {IDM and iDM or The Inert
  Doublet Model and Inelastic Dark Matter}},  {\em JCAP} {\bf 0910} (2009) 018
  [\href{http://arXiv.org/abs/0907.0430}{{\tt 0907.0430}}].
%%CITATION = ARXIV:0907.0430;%%

\bibitem{Nezri:2009jd}
E.~Nezri, M.~H.~G. Tytgat and G.~Vertongen, {\it {e+ and anti-p from inert
  doublet model dark matter}},  {\em JCAP} {\bf 0904} (2009) 014
  [\href{http://arXiv.org/abs/0901.2556}{{\tt 0901.2556}}].
%%CITATION = ARXIV:0901.2556;%%

\bibitem{Miao:2010rg}
X.~Miao, S.~Su and B.~Thomas, {\it {Trilepton Signals in the Inert Doublet
  Model}},  {\em Phys. Rev.} {\bf D82} (2010) 035009
  [\href{http://arXiv.org/abs/1005.0090}{{\tt 1005.0090}}].
%%CITATION = ARXIV:1005.0090;%%

\bibitem{Gustafsson:2012aj}
M.~Gustafsson, S.~Rydbeck, L.~Lopez-Honorez and E.~Lundstrom, {\it {Status of
  the Inert Doublet Model and the Role of multileptons at the LHC}},  {\em
  Phys. Rev.} {\bf D86} (2012) 075019
  [\href{http://arXiv.org/abs/1206.6316}{{\tt 1206.6316}}].
%%CITATION = ARXIV:1206.6316;%%

\bibitem{Arhrib:2012ia}
A.~Arhrib, R.~Benbrik and N.~Gaur, {\it {$H\to \gamma \gamma$ in Inert Higgs
  Doublet Model}},  {\em Phys. Rev.} {\bf D85} (2012) 095021
  [\href{http://arXiv.org/abs/1201.2644}{{\tt 1201.2644}}].
%%CITATION = ARXIV:1201.2644;%%

\bibitem{Swiezewska:2012eh}
B.~Swiezewska and M.~Krawczyk, {\it {Diphoton rate in the inert doublet model
  with a 125 GeV Higgs boson}},  {\em Phys. Rev.} {\bf D88} (2013), no.~3
  035019 [\href{http://arXiv.org/abs/1212.4100}{{\tt 1212.4100}}].
%%CITATION = ARXIV:1212.4100;%%

\bibitem{Goudelis:2013uca}
A.~Goudelis, B.~Herrmann and O.~StÃ¥l, {\it {Dark matter in the Inert Doublet
  Model after the discovery of a Higgs-like boson at the LHC}},  {\em JHEP}
  {\bf 1309} (2013) 106 [\href{http://arXiv.org/abs/1303.3010}{{\tt
  1303.3010}}].
%%CITATION = ARXIV:1303.3010;%%

\bibitem{Arhrib:2013ela}
A.~Arhrib, Y.-L.~S. Tsai, Q.~Yuan and T.-C. Yuan, {\it {An Updated Analysis of
  Inert Higgs Doublet Model in light of the Recent Results from LUX, PLANCK,
  AMS-02 and LHC}},  {\em JCAP} {\bf 1406} (2014) 030
  [\href{http://arXiv.org/abs/1310.0358}{{\tt 1310.0358}}].
%%CITATION = ARXIV:1310.0358;%%

\bibitem{Krawczyk:2013jta}
M.~Krawczyk, D.~Sokolowska, P.~Swaczyna and B.~Swiezewska, {\it {Constraining
  Inert Dark Matter by $R_{\gamma\gamma}$ and WMAP data}},  {\em JHEP} {\bf 09}
  (2013) 055 [\href{http://arXiv.org/abs/1305.6266}{{\tt 1305.6266}}].
%%CITATION = ARXIV:1305.6266;%%

\bibitem{Krawczyk:2013pea}
M.~Krawczyk, D.~Sokolowska, P.~Swaczyna and B.~Swiezewska, {\it {Higgs $\to
  \gamma \gamma $, $Z\gamma $ in the Inert Doublet Model}},  {\em Acta Phys.
  Polon.} {\bf B44} (2013), no.~11 2163--2170
  [\href{http://arXiv.org/abs/1309.7880}{{\tt 1309.7880}}].
%%CITATION = ARXIV:1309.7880;%%

\bibitem{Ilnicka:2015jba}
A.~Ilnicka, M.~Krawczyk and T.~Robens, {\it {Inert Doublet Model in light of
  LHC Run I and astrophysical data}},  {\em Phys. Rev.} {\bf D93} (2016), no.~5
  055026 [\href{http://arXiv.org/abs/1508.01671}{{\tt 1508.01671}}].
%%CITATION = ARXIV:1508.01671;%%

\bibitem{Diaz:2015pyv}
M.~A. Diaz, B.~Koch and S.~Urrutia-Quiroga, {\it {Constraints to Dark Matter
  from Inert Higgs Doublet Model}},  {\em Adv. High Energy Phys.} {\bf 2016}
  (2016) 8278375 [\href{http://arXiv.org/abs/1511.04429}{{\tt 1511.04429}}].
%%CITATION = ARXIV:1511.04429;%%

\bibitem{Modak:2015uda}
K.~P. Modak and D.~Majumdar, {\it {Confronting Galactic and Extragalactic
  $\gamma$-rays Observed by Fermi-lat With Annihilating Dark Matter in an Inert
  Higgs Doublet Model}},  {\em Astrophys. J. Suppl.} {\bf 219} (2015), no.~2 37
  [\href{http://arXiv.org/abs/1502.05682}{{\tt 1502.05682}}].
%%CITATION = ARXIV:1502.05682;%%

\bibitem{Queiroz:2015utg}
F.~S. Queiroz and C.~E. Yaguna, {\it {The CTA aims at the Inert Doublet
  Model}},  {\em JCAP} {\bf 1602} (2016), no.~02 038
  [\href{http://arXiv.org/abs/1511.05967}{{\tt 1511.05967}}].
%%CITATION = ARXIV:1511.05967;%%

\bibitem{Garcia-Cely:2015khw}
C.~Garcia-Cely, M.~Gustafsson and A.~Ibarra, {\it {Probing the Inert Doublet
  Dark Matter Model with Cherenkov Telescopes}},  {\em JCAP} {\bf 1602} (2016),
  no.~02 043 [\href{http://arXiv.org/abs/1512.02801}{{\tt 1512.02801}}].
%%CITATION = ARXIV:1512.02801;%%

\bibitem{Hashemi:2016wup}
M.~Hashemi and S.~Najjari, {\it {Observability of Inert Scalars at the LHC}},
  {\em Eur. Phys. J.} {\bf C77} (2017), no.~9 592
  [\href{http://arXiv.org/abs/1611.07827}{{\tt 1611.07827}}].
%%CITATION = ARXIV:1611.07827;%%

\bibitem{Poulose:2016lvz}
P.~Poulose, S.~Sahoo and K.~Sridhar, {\it {Exploring the Inert Doublet Model
  through the dijet plus missing transverse energy channel at the LHC}},  {\em
  Phys. Lett.} {\bf B765} (2017) 300--306
  [\href{http://arXiv.org/abs/1604.03045}{{\tt 1604.03045}}].
%%CITATION = ARXIV:1604.03045;%%

\bibitem{Alves:2016bib}
A.~Alves, D.~A. Camargo, A.~G. Dias, R.~Longas, C.~C. Nishi and F.~S. Queiroz,
  {\it {Collider and Dark Matter Searches in the Inert Doublet Model from
  Peccei-Quinn Symmetry}},  {\em JHEP} {\bf 10} (2016) 015
  [\href{http://arXiv.org/abs/1606.07086}{{\tt 1606.07086}}].
%%CITATION = ARXIV:1606.07086;%%

\bibitem{Datta:2016nfz}
A.~Datta, N.~Ganguly, N.~Khan and S.~Rakshit, {\it {Exploring collider
  signatures of the inert Higgs doublet model}},  {\em Phys. Rev.} {\bf D95}
  (2017), no.~1 015017 [\href{http://arXiv.org/abs/1610.00648}{{\tt
  1610.00648}}].
%%CITATION = ARXIV:1610.00648;%%

\bibitem{Aprile:2018dbl}
{\bf XENON} Collaboration, E.~Aprile {\em et.~al.}, {\it {Dark Matter Search
  Results from a One Ton-Year Exposure of XENON1T}},  {\em Phys. Rev. Lett.}
  {\bf 121} (2018), no.~11 111302 [\href{http://arXiv.org/abs/1805.12562}{{\tt
  1805.12562}}].
%%CITATION = ARXIV:1805.12562;%%

\bibitem{Enberg:2007rp}
R.~Enberg, P.~J. Fox, L.~J. Hall, A.~Y. Papaioannou and M.~Papucci, {\it {LHC
  and dark matter signals of improved naturalness}},  {\em JHEP} {\bf 11}
  (2007) 014 [\href{http://arXiv.org/abs/0706.0918}{{\tt 0706.0918}}].
%%CITATION = ARXIV:0706.0918;%%

\bibitem{Cohen:2011ec}
T.~Cohen, J.~Kearney, A.~Pierce and D.~Tucker-Smith, {\it {Singlet-Doublet Dark
  Matter}},  {\em Phys. Rev.} {\bf D85} (2012) 075003
  [\href{http://arXiv.org/abs/1109.2604}{{\tt 1109.2604}}].
%%CITATION = ARXIV:1109.2604;%%

\bibitem{CALCHEP}
A.~Belyaev, C.~N. D. and A.~Pukhov, {\it {CalcHEP 3.4 for collider physics
  within and beyond the Standard Model}},  {\em Comput. Phys. Commun.} {\bf
  184} (2013) 1729 [\href{http://arXiv.org/abs/1207.6082}{{\tt 1207.6082}}].
%%CITATION = ARXIV:1207.6082;%%

\bibitem{Semenov:2014rea}
A.~Semenov, {\it {LanHEP \textemdash{} A package for automatic generation of
  Feynman rules from the Lagrangian. Version 3.2}},  {\em Comput. Phys.
  Commun.} {\bf 201} (2016) 167--170
  [\href{http://arXiv.org/abs/1412.5016}{{\tt 1412.5016}}].

\bibitem{HEPMDB}
https://hepmdb.soton.ac.uk, {\it {High Energy Physics Models DataBase
  (HEPMDB)}},  {webpage}, 2011.

\bibitem{micromegas}
G.~Belanger, F.~Boudjema, A.~Pukhov and A.~Semenov, {\it {micrOMEGAs: A Tool
  for dark matter studies}},  {\em Nuovo Cim.} {\bf C033N2} (2010) 111--116
  [\href{http://arXiv.org/abs/1005.4133}{{\tt 1005.4133}}].
%%CITATION = ARXIV:1005.4133;%%

\bibitem{checkmate}
D.~Dercks, N.~Desai, J.~S. Kim, K.~Rolbiecki, J.~Tattersall and T.~Weber, {\it
  {CheckMATE 2: From the model to the limit}},  {\em Comput. Phys. Commun.}
  {\bf 221} (2017) 383--418 [\href{http://arXiv.org/abs/1611.09856}{{\tt
  1611.09856}}].
%%CITATION = ARXIV:1611.09856;%%

\bibitem{CMS:2017fdz}
{\bf CMS} Collaboration, A.~M. Sirunyan {\em et.~al.}, {\it {Search for
  electroweak production of charginos and neutralinos in multilepton final
  states in proton-proton collisions at $\sqrt{s}=$ 13 TeV}},  {\em JHEP} {\bf
  03} (2018) 166 [\href{http://arXiv.org/abs/1709.05406}{{\tt 1709.05406}}].
%%CITATION = ARXIV:1709.05406;%%

\bibitem{Peskin:1991sw}
M.~E. Peskin and T.~Takeuchi, {\it {Estimation of oblique electroweak
  corrections}},  {\em Phys. Rev. D} {\bf 46} (1992) 381--409.

\bibitem{Workman:2022}
{\bf Particle Data Group} Collaboration, R.~Workman {\em et.~al.}, ``{Review of
  Particle Physics}.''
\newblock (2022),\url{https://pdg.lbl.gov/2022/}.

\bibitem{Veltman:1977kh}
M.~J.~G. Veltman, {\it {Limit on Mass Differences in the Weinberg Model}},
  {\em Nucl. Phys. B} {\bf 123} (1977) 89--99.

\bibitem{Chanowitz:1978uj}
M.~S. Chanowitz, M.~A. Furman and I.~Hinchliffe, {\it {Weak Interactions of
  Ultraheavy Fermions}},  {\em Phys. Lett. B} {\bf 78} (1978) 285.

\bibitem{ILD}
{\bf Linear Collider ILD Concept Group -} Collaboration, T.~Abe {\em et.~al.},
  {\it {The International Large Detector: Letter of Intent}},
  \href{http://arXiv.org/abs/1006.3396}{{\tt 1006.3396}}.
%%CITATION = ARXIV:1006.3396;%%

\bibitem{Sjostrand:2014zea}
T.~Sjöstrand, S.~Ask, J.~R. Christiansen, R.~Corke, N.~Desai, P.~Ilten,
  S.~Mrenna, S.~Prestel, C.~O. Rasmussen and P.~Z. Skands, {\it {An
  Introduction to PYTHIA 8.2}},  {\em Comput. Phys. Commun.} {\bf 191} (2015)
  159--177 [\href{http://arXiv.org/abs/1410.3012}{{\tt 1410.3012}}].
%%CITATION = ARXIV:1410.3012;%%

\bibitem{deFavereau:2013fsa}
{\bf DELPHES 3} Collaboration, J.~de~Favereau, C.~Delaere, P.~Demin,
  A.~Giammanco, V.~Lemaître, A.~Mertens and M.~Selvaggi, {\it {DELPHES 3, A
  modular framework for fast simulation of a generic collider experiment}},
  {\em JHEP} {\bf 02} (2014) 057 [\href{http://arXiv.org/abs/1307.6346}{{\tt
  1307.6346}}].
%%CITATION = ARXIV:1307.6346;%%

\bibitem{Cacciari:2008gp}
M.~Cacciari, G.~P. Salam and G.~Soyez, {\it {The anti-$k_t$ jet clustering
  algorithm}},  {\em JHEP} {\bf 04} (2008) 063
  [\href{http://arXiv.org/abs/0802.1189}{{\tt 0802.1189}}].

\bibitem{PDG}
{\bf Particle Data Group} Collaboration, C.~Patrignani {\em et.~al.}, {\it
  {Review of Particle Physics}},  {\em Chin. Phys.} {\bf C40} (2016), no.~10
  100001.
%%CITATION = CHPHD,C40,100001;%%

\bibitem{TESLA-1}
{\bf ECFA/DESY LC Physics Working Group} Collaboration, J.~A. Aguilar-Saavedra
  {\em et.~al.}, {\it {TESLA: The Superconducting electron positron linear
  collider with an integrated x-ray laser laboratory. Technical design report.
  Part 3. Physics at an e+ e- linear collider}},
  \href{http://arXiv.org/abs/hep-ph/0106315}{{\tt hep-ph/0106315}}.
%%CITATION = HEP-PH/0106315;%%

\bibitem{TESLA-2}
H.~Baer, T.~%Barklow, K.~Fujii, Y.~Gao, A.~Hoang, S.~Kanemura, J.~List, H.~E.
  Logan, A.~Nomerotski, M.~Perelstein, M.~E. Peskin, R.~Pöschl, J.~Reuter,
  S.~Riemann, A.~Savoy-Navarro, G.~Servant, T.~M.~P. Tait and J.~Yu, {\it {The
  International Linear Collider Technical Design Report - Volume 2: Physics}},
  \href{http://arXiv.org/abs/1306.6352}{{\tt 1306.6352}}.

\bibitem{WILC-1}
Y.~Li and A.~Nomerotski, {\it {Chargino and Neutralino Masses at ILC}},  in
  {\em {International Linear Collider Workshop}}, July, 2010.
\newblock \href{http://arXiv.org/abs/1007.0698}{{\tt 1007.0698}}.

\bibitem{WILC-2}
M.~Asano, T.~Saito, T.~Suehara, K.~Fujii, R.~S. Hundi, H.~Itoh, S.~Matsumoto,
  N.~Okada, Y.~Takubo and H.~Yamamoto, {\it {Discrimination of New Physics
  Models with the International Linear Collider}},  {\em Phys. Rev. D} {\bf 84}
  (2011) 115003 [\href{http://arXiv.org/abs/1106.1932}{{\tt 1106.1932}}].

\bibitem{ILDConceptGroup:2020sfq}
{\bf ILD Concept Group} Collaboration, H.~Abramowicz {\em et.~al.}, {\it
  {International Large Detector: Interim Design Report}},
  \href{http://arXiv.org/abs/2003.01116}{{\tt 2003.01116}}.

\bibitem{selectron-1}
G.~Moortgat-Pick, {\it Lhc/ilc interplay in susy searches},  {\em Journal of
  Physics: Conference Series} {\bf 110} (May, 2008) 072027.

\bibitem{selectron-2}
A.~Freitas, H.~U. Martyn, U.~Nauenberg and P.~M. Zerwas, {\it Sleptons: Masses,
  mixings, couplings},  2004.

\bibitem{selectron-3}
J.~A. Conley, H.~K. Dreiner and P.~Wienemann, {\it Measuring a light neutralino
  mass at the ilc: Testing the mssm neutralino cold dark matter model},  {\em
  Physical Review D} {\bf 83} (Mar, 2011).

\bibitem{Skrzypek:1991}
M.~Skrzypek and S.~Jadach, {\it {Exact and approximate solutions for the
  electron nonsinglet structure function in QED}},  {\em Zeitschrift fuer
  Physik C} {\bf 49(4)} (1991) 577--584.

\bibitem{Chen:1991wd}
P.~Chen, {\it {Differential luminosity under multi - photon beamstrahlung}},
  {\em Phys. Rev. D} {\bf 46} (1992) 1186--1191.

\bibitem{Bian:2015zha}
L.~Bian, J.~Shu and Y.~Zhang, {\it {Prospects for Triple Gauge Coupling
  Measurements at Future Lepton Colliders and the 14 TeV LHC}},  {\em JHEP}
  {\bf 09} (2015) 206 [\href{http://arXiv.org/abs/1507.02238}{{\tt
  1507.02238}}].

\bibitem{DeBlas:2019qco}
J.~De~Blas, G.~Durieux, C.~Grojean, J.~Gu and A.~Paul, {\it {On the future of
  Higgs, electroweak and diboson measurements at lepton colliders}},  {\em
  JHEP} {\bf 12} (2019) 117 [\href{http://arXiv.org/abs/1907.04311}{{\tt
  1907.04311}}].

\bibitem{Barbieri:2004qk}
R.~Barbieri, A.~Pomarol, R.~Rattazzi and A.~Strumia, {\it {Electroweak symmetry
  breaking after LEP-1 and LEP-2}},  {\em Nucl. Phys. B} {\bf 703} (2004)
  127--146 [\href{http://arXiv.org/abs/hep-ph/0405040}{{\tt hep-ph/0405040}}].

\bibitem{Cynolter:2008ea}
G.~Cynolter and E.~Lendvai, {\it {Electroweak Precision Constraints on
  Vector-like Fermions}},  {\em Eur. Phys. J. C} {\bf 58} (2008) 463--469
  [\href{http://arXiv.org/abs/0804.4080}{{\tt 0804.4080}}].

\bibitem{Gin10}
I.~F. Ginzburg, {\it {Simple and robust method for search Dark Matter particles
  and measuring their properties at ILC in various models of DM}},  {\em PoS}
  {\bf QFTHEP2010} (2010) 028 [\href{http://arXiv.org/abs/1010.5579}{{\tt
  1010.5579}}].

\end{thebibliography}\endgroup
\bibliographystyle{bib}

\end{document}